\documentclass[aps,showpacs,twocolumn,
prd,superscriptaddress,nofootinbib]{revtex4-1}

\usepackage{amsmath}
\usepackage{amsfonts}
\usepackage{amssymb}
\usepackage{latexsym}
\usepackage{graphicx}
\usepackage{bm}
\usepackage{color}
\usepackage{enumerate}
\usepackage{tabularx}
\usepackage{braket}

\usepackage{graphicx}
\usepackage[caption=false]{subfig}

\usepackage{color}
\usepackage[usenames,dvipsnames,svgnames,table]{xcolor}
\usepackage[colorlinks=true,
            linkcolor=YellowOrange,
            urlcolor=RoyalBlue,
            citecolor=RedViolet]{hyperref}

\newcommand{\be}{\begin{equation}}
\newcommand{\ee}{\end{equation}}
\newcommand\ud{{\mathrm{d}}}

\newcommand\calO{{\mathcal{O}}}

\newcommand\calF{{\mathcal{F}}}
\newcommand\calT{{\mathcal{T}}}
\newcommand\calD{{\mathcal{D}}}
\newcommand\calA{{\mathcal{A}}}

\newcommand{\ov}[1]{\overline{#1}}

\newcommand{\nn}{\nonumber}
\newcommand{\hatk}{k}
\newcommand{\Hz}{\,\mathrm{Hz}}
\newcommand{\yr}{\,\mathrm{yr}}
\newcommand{\sinc}{\,\mathrm{sinc}}
\newcommand{\Msol}{M_{\odot}}
\newcommand{\Mchirp}{M_{c}}
\newcommand{\tf}{t_{f}}
\newcommand{\Tf}{T_{f}}
\newcommand{\tfd}{t_{f}^{d}}
\newcommand{\tfSPA}{t_{f}^{\rm SPA}}
\newcommand{\TfSPA}{T_{f}^{\rm SPA}}

\newcolumntype{C}[1]{>{\centering\arraybackslash}p{#1}}
\newcolumntype{L}[1]{>{\raggedright\arraybackslash}p{#1}}

\begin{document}

\title{Fourier-domain modulations and delays of gravitational-wave signals}

\author{Sylvain Marsat}
\affiliation{Max Planck Institute for Gravitational Physics (Albert Einstein Institute), Am M\"uhlenberg 1, Potsdam-Golm, 14476, Germany}
\author{John G. Baker}
\affiliation{Gravitational Astrophysics Laboratory, NASA Goddard Space Flight Center, 8800 Greenbelt Rd., Greenbelt, MD 20771, USA}
\affiliation{Joint Space-Science Institute, University of Maryland, College Park, MD 20742, USA}

\date{\today}

\begin{abstract}

We present a Fourier-domain approach to modulations and delays of gravitational wave signals, a problem which arises in two different contexts. For space-based detectors like LISA, the orbital motion of the detector introduces a time-dependency in the response of the detector, consisting of both a modulation and a varying delay. In the context of signals from precessing spinning binary systems, a useful tool for building models of the waveform consists in representing the signal as a time-dependent rotation of a quasi-non-precessing waveform. In both cases, being able to compute transfer functions for these effects directly in the Fourier domain may enable performance gains for data analysis applications by using fast frequency-domain waveforms. Our results generalize previous approaches based on the stationary phase approximation for inspiral signals, extending them by including delays and computing corrections beyond the leading order, while being applicable to the broader class of inspiral-merger-ringdown signals. In the LISA case, we find that a leading-order treatment is accurate for high-mass and low-mass signals that are chirping fast enough, with errors consistently reduced by the corrections we derived. By contrast, low-mass binary black holes, if far away from merger and slowly-chirping, cannot be handled by this formalism and we develop another approach for these systems. In the case of precessing binaries, we explore the merger-ringdown range for a handful of cases, using a simple model for the post-merger precession. We find that deviations from leading order can give large fractional errors, while affecting mainly subdominant modes and giving rise to a limited unfaithfulness in the full waveform. Including higher-order corrections consistently reduces the unfaithfulness, and we further develop an alternative approach to accurately represent post-merger features.

\end{abstract}


\maketitle


\section{Introduction}
\label{sec:intro}

With the unprecedented recent gravitational-wave detections of coalescencing binary black holes and binary neutron stars, announced by the LIGO-Virgo collaboration~\cite{LIGO-theevent-2016, LIGO-christmasevent-2016, LIGO-O1BBH-2016, LIGO-BNSevent-2017}, gravitational-wave astronomy has entered its observational era. As LIGO prepares for even more sensitive observation runs, and with the recent expansion of the ground-based detectors network with Virgo~\cite{Virgo} (and eventually also KAGRA~\cite{KAGRA} and LIGO-India~\cite{INDIGO}), observations of such compact object coalescences are expected at an ever-increasing rate.

Moreover, the European Space Agency has recently selected the Laser Interferometer Space Antenna (LISA)~\cite{LISA17} to realize the ``Gravitational Universe'' science theme~\cite{elisa13} as the 3rd large space mission of its Cosmic Vision program, with a tentative launch around 2034. A technology demonstrator, LISA Pathfinder, has tested with great success some of the key mission technologies~\cite{LISAPathfinder2016, LISAPathfinder2018}. LISA will be able to detect and characterize, among several important gravitational wave source targets, comparable-mass binary black hole coalescences from cosmological distances over a wide range of masses. These will range from high-redshift observations of supermassive black hole binaries with $M\sim 10^{7} \Msol$ down to the observation of LIGO-type sources with $M\sim 10-100 \Msol$~\cite{Sesana16}.

Data analysis for gravitational-wave observations of compact binary coalescences require accurate models (or templates) for the signals, both for ensuring efficient detections of signals that may be buried in instrumental noise, and to extract the physical parameters of the source in a subsequent analysis. Bayesian analysis for parameter estimation of gravitational-wave signals, as was performed for the LIGO detections~\cite{LIGO-theeventPE-2016,LIGO-O1BBH-2016}, may require millions of evaluations of the likelihood function to sample the posterior probability distribution. The greater sensitivity of LISA and other future instruments will require further increases in the accuracy and computational efficiency of signal templates.

The GW community is making progress assembling higher-fidelity tools for these coming challenges. State-of-the-art IMR templates combine information from the perturbative results of post-Newtonian (PN) theory covering the inspiral (see e.g.~\cite{BlanchetLiving}) and from numerical relativity (NR) simulations covering the end of the inspiral and the merger-ringdown phase (see e.g.~\cite{Pfeiffer12}).  Approaches to template construction include phenomenological templates postulating an analytic ansatz for the Fourier-domain amplitude and phase~\cite{Husa+15, Khan+15,Hannam+13}, and the Effective-One-Body (EOB) approach~\cite{BD99,Taracchini+13, Pan+13, Bohe+16} incorporating PN and NR information. Where needed, Reduced Order Models (ROM), also called surrogate models, have been developed to considerably speed up waveform generation, without losing accuracy~\cite{Field+13, Puerrer14, Blackman+17a}. Put together, these tools provide efficient non-precessing IMR Fourier-domain waveforms, while recent progress has been made for precessing systems as well~\cite{Hannam+13, Chatziioannou+17, Blackman+17b}. Importantly for this work, the resulting waveforms can be represented by an amplitude and phase for each mode, with only a few hundred samples~\cite{Puerrer14}.

A complete representation of spin effects across parameter space in fast IMR templates still remains a frontier of gravitational wave signal modelling. In presence of misaligned spins, the system will endure precession of the orbital plane as it evolves, leading to modulations of the signal as seen by the observer~\cite{Apostolatos+94, Kidder95}. The presence of six degrees of freedom for the spins increases the dimensionality of the problem.

A promising approach to modeling the effect of precession on the emitted waveform, as proposed in~\cite{BCV03b, BCPTV05, Schmidt+10, OShaughnessy+11, Boyle+11}, is to decompose precessing waveforms by performing a time-dependent rotation, following the precession of the orbital plane. The resulting waveform in the rotated frame can then be modelled by a non-precessing waveform, an approximation which is used both in the construction of the inspiral part of precessing EOB waveforms~\cite{Pan+13} and in the construction of precessing phenomenological waveforms~\cite{Hannam+13}. To follow this modelling approach and efficiently create Fourier-domain waveforms, one needs to understand how to translate the time-domain modulations created by the frame rotation into a Fourier-domain transfer function.

Beyond a fast representation of the incident gravitational wave, observational analyses also require transforming the signals through some instrumental response. For short duration mergers such as LIGO  has detected, this can be treated by a simple multiplier and a fixed timeshift between detectors.  For future instruments though, the instrumental response will be more complicated.

Whereas LIGO and Virgo are typically sensitive to chirping binaries for a minute or less, LISA signals may accumulate over months or years. The response of a LISA-type instrument is thus time-dependent~\cite{Cutler97}. The motion and change of orientation of the detector constellation along its orbit lead to significant time variability in the form of a modulation and a varying delay. These effects then convey information about the localization of the gravitational-wave source in the sky. Direct time-domain implementation of the detector response is straightforward~\cite{Vallisneri04, Petiteau+08, CR02, RCP04}, but at a high computational cost for parameter-estimation analyses. To leverage the performance of state-of-the-art Fourier-domain IMR templates~\cite{BTB16,Khan+15}, we must efficiently process the signals through the time-dependent response of the dectector while staying in the Fourier domain.

The purpose of this paper is to introduce a formalism for efficiently processing signals through a time-domain modulation and delay within the Fourier domain, while retaining the compactness of a Fourier-domain amplitude and phase representation of the signals. This will allow us to address both the issue of the Fourier-domain response of the LISA instrument, as well as the issue of Fourier-domain precession modulation for IMR signals from precessing binaries.

In previous works focused on gravitational-wave inspirals, the Stationary Phase Approximation (SPA) (see e.g.~\cite{Thorne300, CF94}) has been often used for this purpose. While the SPA is a common approximation to compute the Fourier transform of non-precessing signals during the inspiraling phase, it is not applicable for IMR waveforms.

In the case of precessing binaries, applying the SPA directly to the modulated signal is prone to pathologies. In Refs.~\cite{KCY13,KCY14}, a formalism (called shifted uniform asymptotics or SUA) was introduced to go beyond the SPA and compute more accurately the modulation in the Fourier domain; however, this formalism still relies on the SPA for the underlying precessing-frame signal, and is as such limited to inspiraling signals. The simplified treatment of the precession response in the phenomenological waveforms of Ref.~\cite{Hannam+13} takes another approach, treating the precession modulation in the frequency domain by directly associating the Fourier frequency with the post-Newtonain orbital frequency. As we will explain, this corresponds to the zeroth-order approximation of the SUA.

In the case of the LISA response, the SPA provides a natural map from time-domain (for the orbit) to Fourier-domain (for the signal)~\cite{Cutler97}, and was used by many previous studies with inspiral waveforms. Ref.~\cite{Klein+15} included the orbital motion of the detector in the SUA treatment of precessing inspiral signals, within a low-frequency approximation of the constellation response. Consistently extending these previous approaches to the merger-ringdown part of the signals, while including the delays in the full LISA response at all frequencies, assessing and understanding the errors made along the way, is part of the objectives of this paper.

We seek to overcome two limitations in such previous approaches.  The first is that  SPA-based methods are not applicable to IMR waveforms. Second, there is often no clear way to improve the accuracy of these methods beyond the intuitive leading order treatment in order to meet the high-accuracy needs of future detectors. Our approach exploits separation of time-scales approximations, based on a general treatment directly in the Fourier domain of slowly varying delays and amplitude modulations for chirping waveforms.

The plan of the paper is as follows. In Sec.~\ref{sec:motivation}, we provide a general presentation of the problem of Fourier-domain modulation and introduce the relevant timescales for both the response of LISA-type detectors and the modulation of precessing signals. In Sec.~\ref{sec:formalism}, we present our general formalism, give its leading order approximation as well as higher-order corrections, introduce new timescales based on the Fourier-domain signal, and refine the previous results for both the quadratic-in-phase corrections and the treatment of the delays. We then apply our formalism to the response of the LISA detector in Sec.~\ref{sec:LISA}, and to the case of signals from precessing binaries in Sec~\ref{sec:precession}. We discuss and summarize our results in Sec.~\ref{sec:discussion}.


\section{GW signals in the frequency domain}
\label{sec:motivation}


In our presentation, we will consider a formal signal processing problem encompassing the challenges posed by both the LISA response and the precession modulations. Given a signal $h(t)$, we apply a time-varying delay $d(t)$ to the waveform followed by a multiplicative modulation function $F(t)$,
\be
\label{eq:delay-mod-defs}
	h_{d}(t) = h(t+d(t)) \,, \quad s(t) = F(t)h_{d}(t) \,.
\ee
We then seek an efficient way to compute the Fourier transform $\tilde{s}(f)$, expressed by means of a Fourier-domain transfer function $\calT$ such that
\be\label{eq:deftransfer}
	\tilde{s}(f) \equiv \calT(f) \tilde{h}(f) \,.
\ee

As is well known, GW signals decomposed in spin-weighted spherical harmonic components are smoothly varying functions of frequency in amplitude/phase form. An important consequence is that these signal components can be accurately represented in terms of a relatively coarsely sampled frequency grid. This property will also extend to the transfer functions, allowing one to keep a compact representation of the full signals. (See Appendix~\ref{app:notation} for details of our notation and conventions for Fourier transforms and spherical harmonics.)

In the case of precessing binaries, the delays will be absent and the modulation functions will be the time-dependent Wigner coefficients applied to rotate the waveform from a precessing orbital frame, in which waveform modes exhibit smooth amplitude and phase variation, to an inertial frame where the observations take place. In the case of a LISA-type detector, the signal $h(t)$ will simply be the waveform in a fixed heliocentric frame, and the delays will come from the motion of each detector against the wave front, while the modulation will represent the time-variation in the detector orientation.

In full generality computing the transfer function $\calT(f)$ would require a convolution, a costly (discretized) integral over the full frequency domain for each value of $f$. For our context though, there are some properties of $h(t)$, $d(t)$ and $F(t)$ which we can exploit for a more efficient computation. In particular, we will be able to exploit the separation of the different timescales in the problem.

First, the gravitational waveforms present a clear separation between the timescales of orbital motion and radiation-reaction. This leads to a general feature of GW signals from compact binaries, during the inspiral phase but also for black hole mergers, that the signal is relatively localized in time-frequency. The localization is particularly clear during the inspiral phase, where the SPA (see Sec.~\ref{subsec:SPA} below) provides an unambiguous time-to-frequency correspondence, however we will find that the applicability of the SPA will not be a limiting factor of our approach.

Second, the delay and modulation functions we consider are much more slowly varying than the GW signal. The relevant timescales are either the precession timescale for precessing binaries or the fixed annual-orbital motion timescale for the LISA response. This means that the modulation and delay have relatively compact support in the Fourier domain, hence the convolution with the signal will be localized in frequency, which justifies writing its output as a transfer function as in~\eqref{eq:deftransfer}.

Together, these observations lead us to expect that, for a given $f$, only a limited range of times should be relevant in $d(t)$ and $F(t)$ so that we may expect to find a treatment for the transfer function that would be local in time for the modulation and delay. This general idea has already been applied for inspiral signals in the limit of an extremely slowly varying $d(t)$ and $F(t)$, where intuitively we should be able to simply evaluate them at the time given by the time-to-frequency correspondence of the SPA.

Furthermore, a natural quantitative criterion for the applicability of this idea is given by the comparison of the radiation-reaction timescale of the signal with the modulation timescale. In other words, the change in frequency of the signal over a characteristic time of the modulation should be large for the separation of timescales to work. However, as we will see below, for this problem the dimensional analysis falls short of the full picture: the separation of timescales can be affected by frequency-dependent dimensionless factors in presence of delays.

Our objective is to find an approximate treatment for $\calT(f)$ which allows us to exploit these properties without relying on unnecessary limiting assumptions for the signal (like the limitation of the SPA to inspiral signals), which is extensible to the high-accuracies which will be required by LISA and other future gravitational-wave instruments, while being computationally efficient and widely applicable to GW analysis. In preparation for developing our formalism we first review the salient features of our application problems.

\subsection{Instrumental modulations and delays for LISA-type detectors}
\label{subsec:modulationLISA}

The response of a detector of the LISA type to an incident gravitational wave can be written in two different, equivalent forms, in terms of phase or frequency measurements. Here we will work with the second representation, which will prove more convenient for our purposes. Moreover, various notation and conventions have been used in the literature to label the spacecraft and describe their orbits. We refer the reader to~\cite{Vallisneri04} for a comparative account on these various conventions. In this work, we will keep close to the conventions of~\cite{Vallisneri04}, which were also used in the Mock LISA Data Challenges (MLDC)~\cite{MLDC4}.

We use a coordinate system centered on the solar system barycenter (SSB), and represent the center of the constellation by the vector $p_{0}$. We introduce the notation $\hatk$ for the propagation vector of the gravitational wave, which we denote by $h_{ij}^{\rm TT}(t)$ in transverse-traceless matrix form, as measured at the SSB (thus, at position $p$, $h_{ij}^{\rm TT}(t, p) = h_{ij}^{\rm TT}(t - \hatk \cdot p)$). We denote by $p_{A}$ ($A=1,2,3$) the position of the individual spacecraft and $n_{l}$ the unit vectors of the three links, with the convention that $n_{3}$ points from 1 to 2.

Written in terms of the fractional laser frequency shifts between two spacecraft, the elementary response of the detector reads~\cite{EW75, RCP04, Vallisneri04}
\begin{align}\label{eq:yslr}
	y_{slr} &\equiv \frac{\nu_{r} - \nu_{s}}{\nu} \nn\\
	&= \frac{1}{2} \frac{n_{l}^{i}n_{l}^{j}}{1 - \hatk\cdot n_{l}} \left[ h_{ij}^{\rm TT}(t - L - \hatk\cdot p_{s}) - h_{ij}^{\rm TT}(t - \hatk\cdot p_{r}) \right] \,.
\end{align}
We will use a rigid instantaneous model, approximating the geometry of the constellation by a moving equilateral triangle (neglecting the flexing of the arms induced by corrections in the orbits), and evaluating all geometric factors in~\eqref{eq:yslr} at a single time $t$ (neglecting point-ahead corrections). The additional delay $L$ (we use $c=1$) in the first term represents the light propagation time along the arm between spacecraft $s$ and $r$.

The first- and second-generation TDI observables are then built as combinations of these basic building blocks, evaluated at delayed times. Since in our rigid approximation these delays will take a simple form in Fourier domain, we will focus in this paper on the $y_{slr}$ observables. Sec.~\ref{subsec:modelLISA} provides more details on the response of LISA-type detectors and on the approximations that enter the derivation of the basic response~\eqref{eq:yslr} above.

The structure of~\eqref{eq:yslr} is such that one can build the full signal from individual contributions of the form
\be
	s(t) = F(t) h(t + d(t)) \,.
\ee
Here, $h(t)$ represents one of the individual modes building the full gravitational wave signal (see Eq.~\eqref{eq:defmodes}), $d(t)$ represents the time-varying delays of the form $-\hatk\cdot p_{A}(t)$, and $F(t)$ incorporates all the relevant geometric prefactors.

For the LISA response, the functions $F(t)$ and $d(t)$ vary on a timescale of one year, with a frequency $f_{0} = 1/\mathrm{yr} \simeq 3.169\times10^{-8} \mathrm{Hz}$ (we will also use $\Omega_{0} = 2\pi f_{0}$). When neglecting small non-periodic orbital perturbations (such as those caused by the influence of the Earth and of the other planets on the orbits), both $F(t)$ and $d(t)$ are periodic. It will also be useful to separate the delays in two types of terms: the first, $d_{0} = -\hatk\cdot p_{0}$, relates the waveform at the SSB to the waveform at the center of the constellation, whereas the second, $d_{L}$, represents the various delays between the spacecraft of the constellation. Concretely, we assume the basic reference design parameters of the LISA mission\cite{LISA17} recently selected by the European Space Agency(ESA), with orbit radius $R=1\,\mathrm{au}$ and armlength $L=2.5\times10^{6}\mathrm{km}$, and with $d_{0} \sim R/c \simeq 500\mathrm{s}$ and $d_{L} \sim L/c \simeq 8\mathrm{s}$.


\subsection{Precession modulation for spinning binaries}
\label{subsec:modulationPrec}

For spinning compact objects, angular momentum interactions typically lead to the precession of the orbit~\cite{Apostolatos+94, Kidder95}, which can have large effects on the waveform. In particular, it breaks the planar symmetry of the gravitational wave emission, causing modulations that are especially important for systems that are observed edge-on.

A number of authors~\cite{BCV03b, BCPTV05, Schmidt+10, OShaughnessy+11, Boyle+11} have suggested that the effect of the precession can be modeled, to a good approximation, by a time-dependent rotation of a effectively non-precessing waveform. This allows for a modelling approach where one separately models a precessing frame following the evolution of the plane of the orbit, and approximates the waveform in this precessing frame by using a effective non-precessing model. Different prescriptions have been proposed for the construction of a precessing frame from the waveform itself~\cite{Schmidt+10, OShaughnessy+11, Boyle+11}.

If $(\alpha, \beta, \gamma)$ are the Euler angles relating the precessing frame to the inertial frame in the $(z,y,z)$ convention, the modes in the inertial frame $h_{\ell m}^{\rm I}$ are then related to the modes in the precessing frame $h_{\ell m}^{\rm P}$ by~\cite{Goldberg+67}
\be\label{eq:wignerrotintro}
	h_{\ell m}^{\rm I} = \sum\limits_{m=-\ell}^{\ell} \calD^{\ell *}_{mm'} (\alpha,\beta,\gamma) h_{\ell m'}^{\rm P} \,,
\ee
where the $\calD^{\ell}_{mm'}$ are Wigner matrices (see App.~\ref{app:wigner}).

If we make the assumption that the precessing-frame waveform $h^{\rm P}$ is approximated by a non-precessing model that provides us with a smooth Fourier-domain amplitude and phase, then the problem reduces to computing the Fourier transform of the signal
\be\label{eq:defmodulationprec}
	s(t) = F(t) h(t) \,,
\ee
where the modulation function $F(t)$ is given by a Wigner matrix and depends on time through the Euler angles $(\alpha, \beta, \gamma)(t)$.

In Eq.~\eqref{eq:defmodulationprec} above, the modulation function $F$ has time variations on the precessional timescale, which evolves throughout the inspiral. In the limit of low frequencies, we will see in Sec.~\ref{sec:precession} below that, although the precession and orbital timescales become more and more separated, the decrease in the chirping rate gives raise to a corrective contribution that does not vanish in this limit. We will also explore  the application of our formalism to a precessing-frame decomposition of the waveform extended through the merger and ringdown phase.


\subsection{Stationary phase approximation}
\label{subsec:SPA}

As a preliminary step, we recall here an approximation widely used for similar purposes in treating gravitational waves signals emitted by inspiraling binaries, the Stationary Phase Approximation (thereafter SPA, and also sometimes called the steepest-descent method). It applies in general to chirping signals, and we refer the reader to~\cite{FC93, CF94} for details. One first writes the time-domain signal in amplitude and phase form as $h(t) = a(t) e^{-2i\varphi(t)}$, where for gravitational wave signals $\varphi$ will correspond to the orbital phase, with the orbital frequency being $\omega = \dot{\varphi}$. In order to keep close to the notation used in the gravitational wave literature, we introduced a factor of 2 in the phase of the wave, which is appropriate for the dominant 22 harmonic of the signal. The approximation then applies to signals verifying the conditions\footnote{Note that the third condition is not always written explicitly, in particular not in Refs.~\cite{FC93, CF94}. It becomes important if we generalize $a$ to be an envelope function incorporating a modulation.}
\be\label{eq:conditionsSPA}
	\left| \frac{\dot{a}/a}{\omega} \right| \ll 1\,, \quad \left|\frac{\dot{\omega}}{\omega^{2}} \right| \ll 1\,, \quad \left| \frac{(\dot{a}/a)^{2}}{\dot{\omega}} \right| \ll 1 \,.
\ee
Since the integral~\eqref{eq:defFT} defining the Fourier transform is rapidly oscillatory unless the term $2\pi f t$ cancels the evolution of $-2\varphi(t)$, its support is well centered around the point of stationary phase. This determines the time-to-frequency correspondence in the stationary phase approximation, and leads to the definition of this time as an implicit function of frequency by the relation
\be\label{eq:deftfSPA}
	\omega(\tfSPA) = \pi  f \,.
\ee
For a chirping signal of increasing phase, $\omega>0$ and $\dot{\omega}>0$, there is a unique point of stationary phase, located in the positive frequency range $f>0$. Using the conditions above, one can formally expand the signal around $\tfSPA$ to quadratic order in time, according to
\begin{align}
	\tilde{h}_{\rm SPA} (f) &\simeq a(\tfSPA) \exp\left[2i\pi f \tfSPA-2i\varphi(\tfSPA) \right] \nn\\
	& \qquad \cdot \int \ud t \, e^{-i \dot{\omega} (\tfSPA) (t-\tfSPA)^{2}} \,.
\end{align}
Note that to be able to treat the amplitude as a constant in the integral above, we used the third condition in~\eqref{eq:conditionsSPA}. The resulting complex Gaussian integral yields\footnote{The expressions given here are valid for the dominant mode $h_{22}$ of the waveform. They can be generalized to other modes $h_{\ell m}$, $m\neq 0$ with phase $e^{-im\varphi}$ as follows: $t_{f}^{\rm SPA}$ is now such that $m\omega = 2\pi f$, $A$ acquires a factor $\sqrt{2/m}$, in $\Psi$ the term $2\varphi$ becomes $m\varphi$.}
\begin{subequations}
\begin{align}
	\tilde{h}_{\rm SPA}(f) &= A_{\rm SPA}(f) e^{-i\Psi_{\rm SPA}(f)} \,, \\
	A_{\rm SPA}(f) &= a(\tfSPA) \sqrt{\frac{\pi}{\dot{\omega}(\tfSPA)}} \,, \label{eq:ASPA} \\
	\Psi_{\rm SPA}(f) &= 2\varphi(\tfSPA) - 2\pi f \tfSPA + \frac{\pi}{4} \,. \label{eq:PsiSPA}
\end{align}
\end{subequations}
Ref.~\cite{Droz+99} evaluated the first correction to this approximation, within the context of post-Newtonian signals, and found that it can be considered as a term of the fifth post-Newtonian order, beyond the accuracy level of our current best models~\cite{BlanchetLiving}.

To understand the separation of timescales in the problem, it will be useful to have at hand the leading-order scaling laws for an inspiral (labeled the Newtonian order in the PN language). Although inaccurate for the purpose of waveform modelling, these leading-order estimates will give useful orders of magnitude of the relevant timescales in the inspiral. For a binary with masses $m_{1}, m_{2}$, we define the total mass $M=m_{1}+m_{2}$ and the symmetric mass ratio $\nu = m_{1}m_{2}/M^{2}$. Introducing a time of coalescence $t_{c}$, the relations between the orbital frequency and phase and the time to coalescence $t_{c} - t$ are then given by
\begin{subequations}\label{eq:omegaphiN}
\begin{align}
	\omega(t) &= \left[ \frac{256\nu}{5c^{5}} (GM)^{5/3} (t_{c}-t) \right]^{-3/8} \,, \\
	\varphi(t) &= -\left[ \frac{c^{3}}{5 G M \nu^{3/5}} (t_{c}-t) \right]^{5/8} \,.
\end{align}
\end{subequations}
As for the leading-order time-domain amplitude of the $22$ mode, we have~\cite{BlanchetLiving}
\be\label{eq:a22N}
	a_{22}^{\rm N} (t) = \frac{2 G M \nu v^{2}}{D c^{2}} \sqrt{\frac{16 \pi}{5}} \,,
\ee
where we set $v = (G M\omega/c^{3})^{1/3}$ and where $D$ is the luminosity distance to the observer. For this Newtonian inspiral, applying the SPA gives for the $22$ mode $\tilde{h}(f) = A_{\rm N}(f)e^{-i\Psi_{\rm N}(f)}$ with
\begin{subequations}\label{eq:SPAN}
\begin{align}
	A_{N}(f) &= \frac{G^{2}M^{2} \pi}{Dc^{5}} \sqrt{\frac{2\nu}{3}} v^{-7/2}\,, \label{eq:ASPAN}\\
	\Psi_{\rm SPA}^{\rm N}(f) &= \phi_{0} - 2\pi f t_{0} - \frac{3}{128\nu v^{5}} \,, \label{eq:PsiSPAN}
\end{align}
\end{subequations}
where $v=(G M \pi f/c^{3})^{1/3}$ according to the SPA correspondence~\eqref{eq:deftfSPA}, and where $t_{0}, \phi_{0}$ are constants\footnote{For modes $h_{\ell m}$ with $m\neq 0$, the phase $\Psi$ acquires a factor $m/2$ and has to be evaluated at $(2/m)^{1/3} v$.}. It is also customary to rewrite the above relations in terms of the chirp mass $\Mchirp \equiv M\nu^{3/5}$, which is the only mass combination characterizing the signal at the leading PN order.

In this Newtonian, low-frequency limit one can check that each of the combinations \eqref{eq:conditionsSPA} indeed vanish at $\calO{(v^5)}$. but as the system approaches merger, these condition are no-longer satisfied. Near and after the merger, the SPA treatment is not applicable. Nonetheless, the SPA has been a useful workhorse in many gravitational analyses involving frequency domain transfer functions of the form~\eqref{eq:deftransfer}. The usual approach is simply to replace any time dependencies appearing the transfer function using \eqref{eq:deftfSPA}. We will reference the SPA treatment, as a familiar touchstone, as we develop a more general formalism.


\section{Perturbative Fourier-domain approach to modulations and delays}
\label{sec:formalism}


In this section we develop a perturbative Fourier-domain formalism for treating time-delays and temporally multiplicative signal transformations, exploiting the separation of timescales in the problem.

\subsection{Fourier transform of a modulated and delayed signal}
\label{subsec:FTgeneral}

We begin by expression our delay and modulation function definitions in \eqref{eq:delay-mod-defs} using the Fourier transform (note our unusal convention~\eqref{eq:defFT}),
\begin{align}
h_{d}(t) & =  h(t+d(t)) \nonumber\\
&=\int \ud f \, e^{-2i\pi f (t+d(t))}\tilde{h}(f) \,,
\end{align}
and
\begin{align}
  \tilde{s}(f) &= \mathrm{FT} \left[ F h_{d}\right] (f) \nn \\
  &= \int \ud t \, e^{2i\pi f t} F(t)  \int \ud f' \, e^{-2i\pi f' (t+d(t))}\tilde{h}(f') \nn\\
	&= \int \ud f' \, \tilde{h}(f-f') \int \ud t \, e^{2i\pi f' t} e^{-2i\pi (f-f') d(t)} F(t) \,.
\end{align}
The last equation can then be rewritten as a generalized convolution integral with a frequency-dependent Kernel, according to
\be\label{eq:FDkernel}
	\tilde{s}(f) = \int \ud f' \, \tilde{h}(f-f') \tilde{G}(f-f',f') \,,
\ee
where we introduced the frequency-dependent function of time $G(f,t)$, and its Fourier transform in the auxiliary frequency $f'$, denoted by $\tilde{G}(f,f')$, as
\begin{subequations}\label{eq:defG}
\begin{align}
	G(f,t) &= e^{-2i\pi f d(t)} F(t) \,, \\
	\tilde{G}(f,f') &= \int \ud t \, e^{2i\pi f' t} G(f,t) \,.
\end{align}
\end{subequations}
In the absence of delays $d(t)$, as in the case of precessing binaries, the function $G$ loses its frequency-dependence and becomes a modulation in the form of a function of time $F(t)$, and the result~\eqref{eq:FDkernel} above reduces to the familiar convolution theorem for the Fourier transform.

A direct computation of the generalized convolution~\eqref{eq:FDkernel} using~\eqref{eq:defG} will generally be computationally demanding, but we can exploit the separation of timescales in the problem to seek an accurate but efficient approximation and to compute the transfer function $\calT(f)$ as in~\eqref{eq:deftransfer}.

For example, under appropriate conditions with slowly varying modulations and delays, we can expect that the Fourier transform $\tilde{G}(f,f')$ should have a compact support, limited to $f' \in [-f_{\rm max}, f_{\rm max}]$ with $f_{\rm max}$ a maximal frequency for the modulation, roughly the inverse of its characteristic timescale. This makes the convolution integral~\eqref{eq:FDkernel} localized in frequency, as the waveform is to be evaluated only at frequencies close to $f$. We can then approximate $\tilde{h}(f-f')$ by $\tilde{h}(f)$, with a simple Taylor expansion of their difference. Doing so, we will recover at leading order a locality in time: the response is approximately reduced to an evaluation of the modulation and delay at a representative signal-dependent time $\tf$, which will be equivalent to $\tfSPA$ for inspiral signals.

Now consider the limitations to this straightforward argument. The first is that we did not yet specify what $f_{\rm max}$ should be compared with, in order for the approximation to work. As shown in Sec.~\ref{subsec:SPA}, the Fourier-domain expressions for the amplitude and phases inspiral signals are steep power-laws. We will therefore need a quantitative criterion to ensure $\tilde{h}(f-f')$ does not vary too much on the range $f'\in [-f_{\rm max}, f_{\rm max}]$. Second, although clear when considering a fixed characteristic timescale for the modulation and delay (like in the LISA case), the above argument does not apply as such to modulations with a varying timescale (like in the case of precessing binaries, where the precession goes faster when arriving at merger). We will show below that what will be relevant is the characteristic timescale at the time $t_{f}$ associated to $f$.

In presence of delays, the above is also complicated by the additional delay phases in the signal Fourier transform. Depending on the frequency $f$, this can lead $G(f,t)$ to have faster variations than its nominal timescale ($1\mathrm{yr}$ for LISA). Thus, to assess the validity of our approximations we cannot limit ourselves to comparing the dimensionful timescales at play in the problem. As we will see in Sec.~\ref{subsec:lisafom}, we must also include in the analysis dimensionless factors of the form $2\pi f d$.


\subsection{Leading order: the local-in-frequency approximation}
\label{subsec:LLP}

\begin{figure}
  \centering
  \includegraphics[width=.98\linewidth]{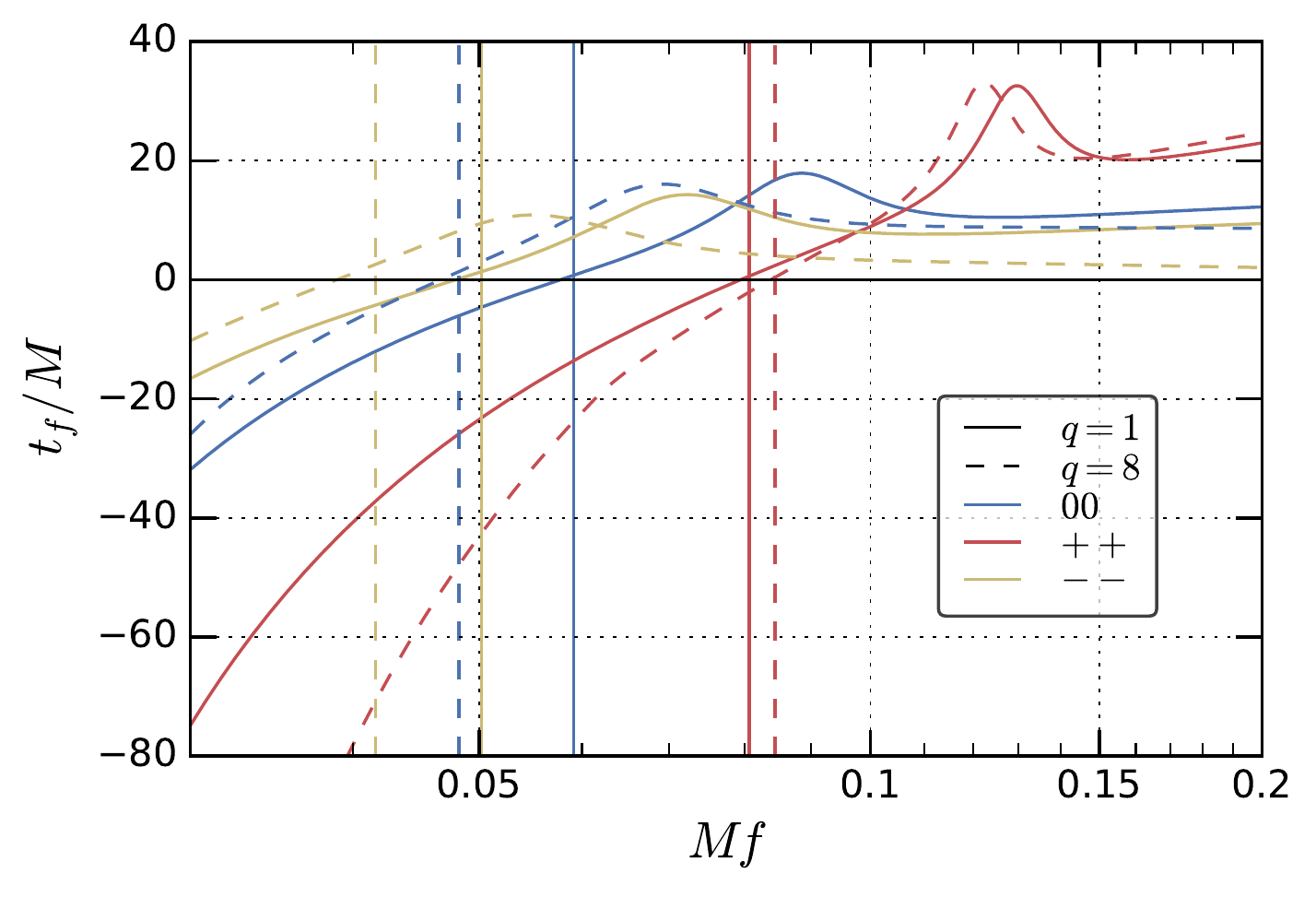}
  \caption{Time-to-frequency correspondence $t_{f}$, as defined directly from the phase of the Fourier-domain signal in~\eqref{eq:deftf}, in geometric units. We show only the high-frequency part of the signal, corresponding to the merger region. The PhenomD waveforms have been aligned such that the time-domain amplitude (obtained by an IFFT) peaks at $t=0$. The two aligned spins are equal, with components of $0.95$ ($++$, blue), $0$ ($00$, red), and $-0.95$ ($--$, yellow), for mass ratios $q=1$ (full line) and $q=8$ (dashed). The vertical lines shows the time-domain instantaneous frequency at the peak $\omega^{22}_{\rm peak}/(2\pi)$, and the fact that the curves $t_{f}$ do not pass exactly by the crossing with the $t=0$ horizontal line reflects the fact that the link between time-domain frequency and Fourier frequency $f$ is only approximate at merger. Note that $t_{f}$ increases to at most $\sim 30M$ after the peak, and is not monotonous.}
  \label{fig:tf}
\end{figure}

As a first step, we perform a formal leading-order expansion of the signal $\tilde{h}(f-f')$ around $f$. We use the amplitude/phase decomposition~\eqref{eq:defAPsi}, treat the Fourier-domain amplitude $A$ as a constant, expand the Fourier-domain phase $\Psi$ to the first order, and discard the $f'$ dependence in the first argument of $\tilde{G}(f-f', f')$.

For the signal, then, we have
\be
	\tilde{h}(f-f') \simeq A(f) \exp\left[ -i\left( \Psi(f) - f' \frac{\ud \Psi}{\ud f} \right) \right] \,,\label{eq:leadingorderwf}
\ee
Plugging this relation into~\eqref{eq:FDkernel}, we obtain
\begin{align}
	\tilde{s}(f) &\simeq \tilde{h}(f) \int \ud f' \, \exp\left[ i f' \frac{\ud \Psi}{\ud f} \right] \tilde{G}(f,f') \nn\\
	&= \tilde{h}(f) G\left( f, -\frac{1}{2\pi} \frac{\ud \Psi}{\ud f} \right) \,,\label{eq:leadingorderresponse}
\end{align}
which we can think of as a local evaluation of the kernel function $G(f,t)$ at a frequency-dependent effective time
\be\label{eq:deftf}
	\tf \equiv -\frac{1}{2\pi} \frac{\ud \Psi}{\ud f} \,.
\ee
It is worth noting that a shift in time of the time-domain signal will, by virtue of~\eqref{eq:shifttime}, be appropriately propagated to $t_{f}$. Because of the freedom of adding a linear term to $\Psi(f)$ by simply shifting the signal in time, no assumption can be made on the smallness of the first derivative of the phase, and this is really a leading order approximation.

The definition~\eqref{eq:deftf} is a straightforward generalization of the time-to-frequency correspondence at the heart of the SPA~\eqref{eq:deftfSPA}. Indeed, using~\eqref{eq:deftfSPA} one can verify that the derivative of the SPA phase $\Psi_{\rm SPA}$~\eqref{eq:PsiSPA} with respect to $f$ yields back $\tfSPA$, as
\be\label{eq:tfSPA}
	\tfSPA = -\frac{1}{2\pi} \frac{\ud \Psi_{\rm SPA}}{\ud f} \,.
\ee
However, $\tf$ refers only to the Fourier-domain waveform. We do not need to relate the frequency $f$ to a time-domain frequency like the orbital frequency $\omega$, and the definition is independent of the SPA being valid or not for the underlying signal $\tilde{h}$.

The main advantage of the time-of-frequency function~\eqref{eq:deftf} is that it extends naturally to the merger-ringdown part of the signals. As such, it is used in the PhenomD and PhenomHM waveform models~\cite{Khan+15, London+17}. Fig.~\ref{fig:tf} shows the behaviour of the time function $\tf$ around merger for six example waveforms, for mass ratios $q=1$ and $q=8$ and for aligned spin components $\chi=0.95,0.,-0.95$. In particular, one should note that $\tf$ is not monotonically increasing with frequency anymore after reaching in the high-frequency part of the waveform, corresponding to the ringdown. As long as the Fourier phase is differentiable $t_{f}$ is a well-defined function of $f$. While its non-monotonicity would forbid an unambiguous definition of a reciprocal frequency-of-time function $f(t)$, like that in the SPA, no such function will be needed in our treatment.

With \eqref{eq:leadingorderresponse} we have brought the modulated and delayed signal in to the form~\eqref{eq:deftransfer} with transfer function
\be\label{eq:transferlocal}
	\calT_{\rm local}(f) = G(f, \tf) = F(t_{f}) e^{-2i\pi f d(t_{f})}\,.
\ee
The interpretation of this approximation is straightforward: the signal is simply multiplied by the response function evaluated at the time $\tf$, the delay phase becoming the same linear phase contribution as one would have in~\eqref{eq:shifttime} with a time shift $d(t_{f})$ treated like a constant. The locality in frequency at $f$ for $\tilde{h}$ translates into a locality in time at $t_{f}$ for $F,d$.


\subsection{Taylor expansion in the Fourier domain}
\label{subsec:TaylorFD}

If the width of the kernel function $G(f-f',f')$ is not quite negligible compared to the scale of significant variations of $\tilde{h}(f)$ with $f$, it can be useful to extend our approach beyond the leading-order approximation. With the waveform represented in the amplitude and phase form~\eqref{eq:defAPsi}, the elements of~\eqref{eq:FDkernel} may be formally Taylor-expanded in the variable $f'$:
\begin{subequations}\label{eq:expandfprime}
\begin{align}
	\Psi(f-f') &= \Psi(f) + 2\pi f' \tf + \sum\limits_{p\geq 2} \frac{(-1)^{p}}{p!} {f'}^{p} \frac{\ud^{p} \Psi}{\ud f^{p}} \,, \label{eq:expandPsi}\\
	A(f-f') &= A(f)+A(f) \sum\limits_{q\geq 1} \frac{(-1)^{q}}{q!} {f'}^{q} \frac{1}{A}\frac{\ud^{q} A}{\ud f^{q}} \,, \label{eq:expandA}\\
	\tilde{G}(f-f', f') &=\tilde G(f,f')+ \sum\limits_{r\geq 1} \frac{(-1)^{r}}{r!} {f'}^{r} \frac{\partial^{r} }{\partial f^{r}}  \tilde{G}(f,f') \label{eq:expandG} \,,
\end{align}
\end{subequations}
using the definition of $t_{f}$ introduced in~\eqref{eq:deftf}.

The leading order transfer function~\eqref{eq:transferlocal} is obtained by leaving off all the terms in the sums.  In the following, we will consider the resulting transfer functions when keeping some of the next few terms in each of these expansions. Notice that we expand $\tilde{G}(f-f',f')$ in $f'$ only in its first argument, and that we can commute the $f$-derivatives of $G$ with the Fourier transform operation.

We can also formally expand the exponential of the phase corrections $\delta \Psi$ beyond the first two terms in~\eqref{eq:expandPsi} as
\be
	\exp\left[ -i\delta\Psi \right] = \sum_{j \geq 0}\frac{(-i\delta\Psi)^j}{j!} \,,
\label{eq:expandexp}
\ee
to obtain a pure $f'$-expansion. The resulting power series in $f'$ can then be recast as a temporal Taylor series, by applying the formal derivative rule
\allowdisplaybreaks
\be
	\int \ud f'\, {(-2i\pi f')}^{n} \frac{\partial^{m}}{\partial f^{m}} \tilde{G}(f,f') e^{-2i\pi f' \tf} = \frac{\partial^{m} }{\partial f^{m}} \frac{\partial^{n} }{\partial t^{n}} G (f,\tf) \,.
\ee
The fully expanded result is a rather cumbersome expression with multiple sums, that we will not use directly. Instead, it will be more instructive to separately consider the different expansions in~\eqref{eq:expandfprime}. Later, in Sec.~\ref{subsec:executivesummary} we will come back to combining our results together.

We first consider the effect of the higher-order corrections in~\eqref{eq:expandPsi}. We will find that the third and higher derivatives of the phase are always negligible for our purposes, and we will ignore them. Keeping only the first term of the sum in~\eqref{eq:expandPsi}, corresponding to the second derivative of $\Psi$, with just the leading terms from \eqref{eq:expandA} and \eqref{eq:expandG}, and expanding the phase exponential \eqref{eq:expandexp} so that the result can be cast as a Taylor series in time, we obtain straightforwardly
\be\label{eq:resulttaylorPsi}
	\calT_{\rm phase}(f) = \sum\limits_{p\geq 0} \frac{1}{p!} \left( \frac{i}{8\pi^{2}}\frac{\ud^{2} \Psi}{\ud f^{2}} \right)^{p} \left( \frac{\partial^{2p} }{\partial t^{2p}} G \right)(f, \tf) \,.
\ee
This correction to the transfer function is of particular interest to us. It will be quantitatively dominant over the other ones in most contexts, and it will be shown in Sec.~\ref{subsec:resumquadphase} that it generalizes the previous approach of~\cite{KCY14}. This result shows that the transfer function is signal-dependent, not only through the time-to-frequency correspondence $t_{f}$ but also through the second derivative of the phase $\Psi$.

Similarly, applying the expansion of the Fourier-domain amplitude~\eqref{eq:expandA} while preserving only the leading order terms in \eqref{eq:expandPsi} and \eqref{eq:expandG} gives
\be\label{eq:resulttaylorA}
	\calT_{\rm amp}(f) = \sum\limits_{p\geq 0} \frac{1}{(2i\pi)^{p}p!} \frac{1}{A} \frac{\ud^{p} A}{\ud f ^{p}}  \left( \frac{\partial^{p} }{\partial t^{p}} G \right)(f,\tf) \,.
\ee

Lastly, expanding only the frequency-dependence of $G$ as in~\eqref{eq:expandG} yields
\be\label{eq:resulttaylordelay}
	\calT_{\rm delay}(f) = \sum\limits_{p\geq 0} \frac{1}{(2i\pi)^{p}p!} \left( \frac{\partial^{p} }{\partial f^{p}} \frac{\partial^{p} }{\partial t^{p}} G \right)(f,\tf) \,,
\ee
which interestingly looks like a Taylor expansion of $G$, but this time with joint derivatives in frequency and time. This last expansion, taken separately from the other corrections, is signal-independent as it only depends on the kernel function $G$ and not on $A$, $\Psi$.

As will be explained in Sec.~\ref{subsec:executivesummary}, we will also use these formal Taylor expansions to build error measures (constructed as the magnitude of the first term ignored in the series, see~\eqref{eq:deffom}) designed to estimate if these three types of corrections are important to take into account.


\subsection{Signal-dependent timescales}
\label{subsec:timescales}

\begin{figure}
  \centering
  \includegraphics[width=.98\linewidth]{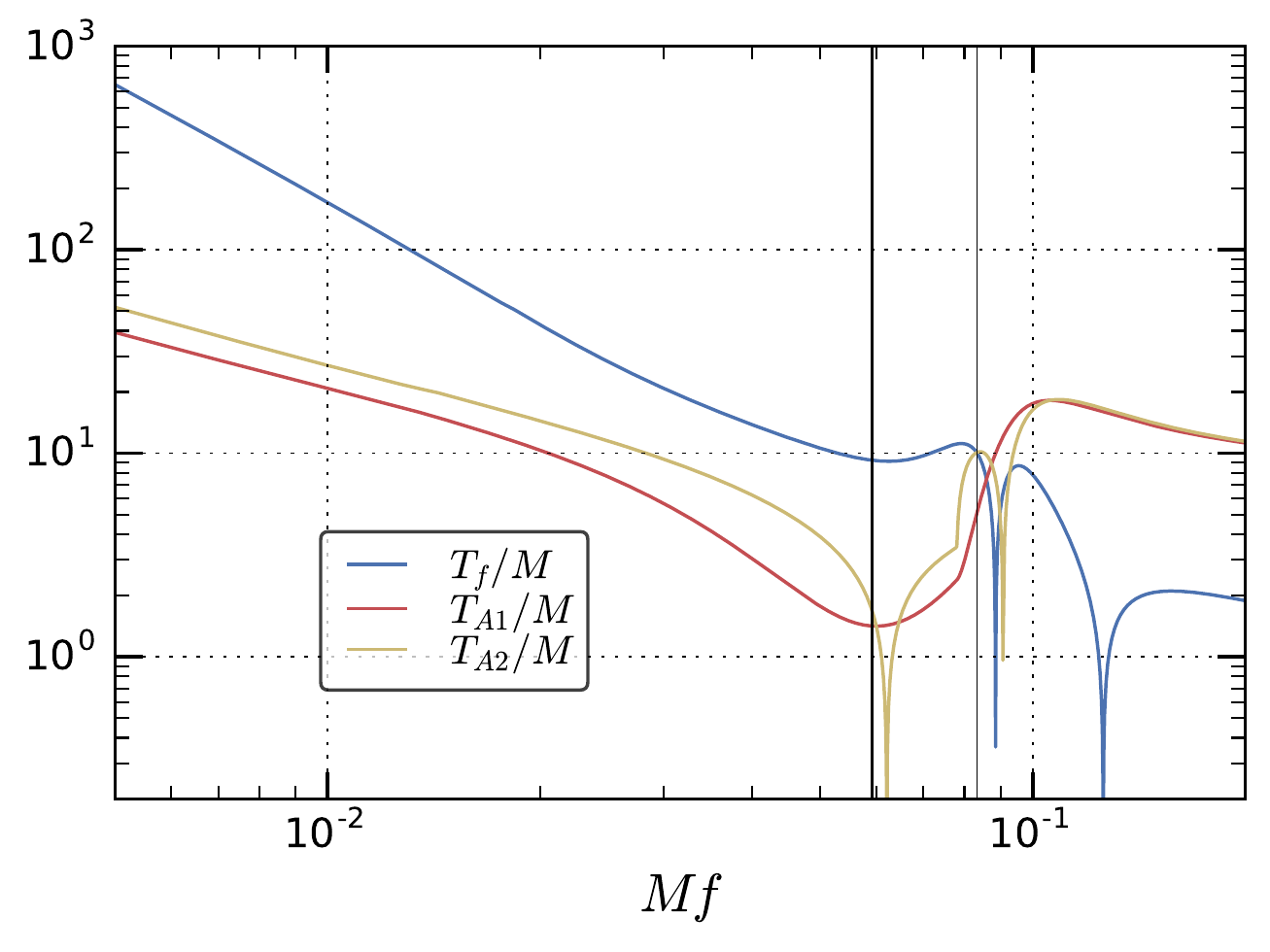}
  \caption{Fourier-domain amplitude and phase timescales, as defined in~\eqref{eq:defTf} and~\eqref{eq:defTA}, in geometric units for an equal-mass, non-spinning system. The hierarchy of these timescales is roughly the same for higher mass ratio and higher spin systems. The time-domain frequency at merger and the ringdown frequency are represented by the thick and thin vertical lines. The derivatives from which these timescales are built encounter zero-crossings and change sign in the high-frequency range.}
  \label{fig:TfTA}
\end{figure}

When considering the impact on the transfer function from including the next higher-order phase term, that is the difference between~\eqref{eq:resulttaylorPsi} and~\eqref{eq:transferlocal}, it is natural to define a new timescale, as a function of frequency,
\be\label{eq:defTf}
	\Tf^{2} = \frac{1}{4\pi^{2}}\left| \frac{\ud^{2}\Psi}{\ud f^{2}} \right| \,.
\ee
For the correction to be small, this timescale should be small compared to the time-scale of variations in the kernel $G(f,t)$ encoding the modulation and delay. We can use this notation to rewrite~\eqref{eq:resulttaylorPsi} as
\begin{align}\label{eq:resultdffPsiTf}
	 \calT_{\rm phase}(f) &= \sum\limits_{p\geq 0} \frac{(-i\epsilon)^{p}}{2^{p}p!} \Tf^{2p} \left( \frac{\partial^{2p} }{\partial t^{2p}} G \right)(f, \tf) \,,
\end{align}
where $\epsilon = -\mathrm{sgn}(\ud^{2}\Psi/\ud f^{2} )$ is $1$ in the inspiral.

We can obtain a straighforward physical interpretation of this timescale $\Tf$ by considering inspiral signals for which the SPA is valid. In that case
\be
	\left(\Tf^{\rm SPA} \right)^{2} = -\frac{1}{4\pi^{2}}  \frac{\ud^{2} \Psi_{\rm SPA}}{\ud f^{2}} \,,
\ee
where $\ud^{2}\Psi/\ud f^{2} < 0$ in the SPA with our sign conventions.
Then, taking two derivatives of~\eqref{eq:PsiSPA}, we find
\be\label{eq:TfSPA}
	\Tf^{\rm SPA} \equiv \frac{1}{\sqrt{2\dot{\omega}(\tfSPA)}} \,.
\ee
Thus, when the SPA applies, $\Tf$ corresponds to the radiation-reaction timescale: the shorter this timescale, the faster the binary chirps to higher frequencies on its quasi-circular inspiral.

However, in the same way that the definition~\eqref{eq:deftf} for the time-of-frequency function $t_{f}$ generalizes the SPA definition~\eqref{eq:deftfSPA}, the definition~\eqref{eq:defTf} only refers to the phase of the Fourier-domain signal and does not require introducing a time-domain frequency like $\omega$. This defintion thus extends naturally to the merger-ringdown part of the signal. In this part of the signal, its physical interpretation as the timescale of radiation reaction is obscured, and the second derivative $\ud^{2}\Psi/\ud f^{2}$ can go through zero and change sign, as shown in Fig.~\ref{fig:TfTA}. We include an absolute value in the definition~\eqref{eq:defTf} to allow for this possibility, and keep track of the sign via $\epsilon$.

Next, we consider the impact on the transfer function~\eqref{eq:resulttaylorA} from amplitude corrections beyond leading order. This series also leads to the natural introduction of a set of another set of timescales related to the successive derivatives of the amplitude. In an analogous manner to the definition~\eqref{eq:defTf} of the timescale $\Tf$, we can define
\be\label{eq:defTA}
	\left( T_{Ap} \right)^{p} \equiv \frac{1}{(2 \pi)^{p}} \frac{1}{A(f)} \left| \frac{\ud^{p} A}{\ud f^{p}} \right| \,,
\ee
where we included an absolute value to accomodate the possible sign changes in the right-hand side. With this notation, \eqref{eq:resulttaylorA} becomes simply
\be\label{eq:resultATA}
	\calT_{\rm amp}(f) = \sum\limits_{p\geq 0} \frac{1}{p!} (T_{Ap})^{p}  \left( \partial_{t}^{p} G \right) (f,\tf) \,,
\ee
Although the above is written for a generic $p\geq 0$, in practice only the first few of these timescales will be relevant. In this paper we will use only the first two, $T_{A1}$ and $T_{A2}$.

By contrast, the impact of higher-order terms in the kernel function on the transfer function~\eqref{eq:resulttaylordelay} is signal-independent. It does not lead to the introduction of new timescales since the coupled time and frequency derivatives are dimensionless. We will see however in Sec.~\ref{subsec:lisafom} that treating delays does require taking into account dimensionless factors of the type $2\pi f d$.

We can obtain useful estimates for these timescales from the leading order post-Newtonian expressions~\eqref{eq:SPAN}, valid for the dominant harmonic $h_{22}$. The leading-order radiation-reaction and amplitude timescales are:
\begin{subequations}\label{eq:timescalesN}
\begin{align}
	\Tf^{\rm N} &= \frac{1}{8} \sqrt{\frac{5}{3\nu}} \frac{G M}{c^{3}} v^{-11/2} \,, \label{eq:TfN}\\
	T_{A1}^{\rm N} &= \frac{7}{6} \frac{1}{2\pi f}\,, \quad T_{A2}^{\rm N} = \frac{\sqrt{91}}{6} \frac{1}{2\pi f} \,. \label{eq:TA1N-TA2N}
\end{align}
\end{subequations}
One can check explicitly that this expression for $\Tf$ agrees with~\eqref{eq:TfSPA}. With a simple power-law amplitude as in~\eqref{eq:ASPAN}, all higher-order amplitude timescales are also simply proportional to $1/f$ and differ only by their numerical factor. For higher harmonics $h_{\ell m}$ with $m\neq 0$, as discussed in Sec.~\ref{subsec:SPA}, $T_{f}$ acquires a factor $(m/2)^{4/3}$. The amplitude of higher harmonics starts at a higher PN order~\cite{BlanchetLiving}, differing from~\eqref{eq:a22N} by a different constant and an additional scaling $v^{\kappa_{\ell m}}$ with $\kappa_{\ell m} = \ell - 2 + (\ell + m \; \mathrm{mod} \; 2)$, which changes the constant in the amplitude timescale as $7/6 \rightarrow (7-2\kappa_{\ell m})/6$ in $T_{A1}^{\rm N}$ and $\sqrt{91}/6 \rightarrow \sqrt{(7-2\kappa_{\ell m})(13-2\kappa_{\ell m})}/6$ in $T_{A2}^{\rm N}$.

We show in Fig.~\ref{fig:TfTA} the timescales $\Tf$, $T_{A1}$, $T_{A2}$ for an equal-mass and non-spinning system. In the inspiral, they follow the scalings~\eqref{eq:timescalesN} and $\Tf$ is much larger than the other two. For frequencies above the merger frequency, $\Tf$ can go through zero while the amplitude-related timescales become comparable or larger.

In the following, we will compare these signal-dependent timescales to the timescales present in the modulations and delays. In the case of signals from precessing binaries, the precession timescale decreases as the system gets closer to merger, as will be discussed in details in Sec.~\ref{subsec:sizecorrPrec} below. In the case of the response of a LISA-like detectors, the modulation and delay evolve with a fixed timescale of one year, as will be detailed in Sec.~\ref{subsec:lisafom}.


\subsection{Quadratic term in the phase and relation to the SUA}
\label{subsec:resumquadphase}

If we restrict to the case of a pure undelayed modulation, $d=0$ and $G(f,t) = F(t)$, we can relate our result~\eqref{eq:resultdffPsiTf} for the transfer function including up to quadratic phase terms, with the treatment of Ref.~\cite{KCY13, KCY14}, where the authors extend the SPA in a formalism called the Shifted Uniform Asymptotic expansion (SUA).

The main intermediate result of~\cite{KCY14}, their equation~(34), reads exactly like~\eqref{eq:resultdffPsiTf} with $\epsilon=1$ and with the identifications $\tilde{H}_{corr}(f)\rightarrow \calT(f)$ for the transfer function, $T\rightarrow \Tf$ for the radiation-reaction timescale and $e^{-i\delta\phi} \rightarrow F$ for the modulation function (which is restricted in their framework to a phase, amplitudes being treated jointly with the amplitude of the signal). Thus, our treatment gives a straightforward rederivation of the result of~\cite{KCY14} which corresponds in our framework to the approximation~\eqref{eq:expandPsi}, where in the Fourier-domain convolution the phase of the signal is expanded to quadratic order and the amplitude is not expanded. The main difference is that the approach of~\cite{KCY14} still relies on the SPA being valid for the underlying signal.

The authors of~\cite{KCY14} then proposed a resummation scheme for~\eqref{eq:resultdffPsiTf}, using finite differences for the derivatives. Indeed, the result~\eqref{eq:resultdffPsiTf} looks like a symmetrized Taylor expansion, except for the factors $i^{p}$ and $1/p!$ instead of $1/(2p)!$. Truncating the sum at some finite order $N$, one can write (following~\cite{KCY14})
\be\label{eq:stencilresult}
	\calT_{\rm phase}(f) \simeq \sum\limits_{p = 0}^{N} \frac{(-i\epsilon\Tf^{2})^{p}}{2^{p}p!} \partial_{t}^{2p}F(\tf) \simeq \calF^{N}_{\Tf, \epsilon}[F] (\tf) \,,
\ee
with the operator $\calF_{T, \epsilon}^{N}$ defined as
\be\label{eq:stencilfresnel}
	\calF_{T, \epsilon}^{N}[F] (t) \equiv \frac{1}{2}\sum\limits_{k=0}^{N} a_{N,k}^{\epsilon} \left( F(t + kT) + F(t - k T) \right) \,,
\ee
where the complex coefficients $a_{N,k}^{\epsilon}$ are a solution of the $N+1$-dimensional linear system~\cite{KCY14}
\be\label{eq:stencilsystem}
	(-i\epsilon)^{p} (2p-1)!! = \sum\limits_{k=0}^{N} a_{N,k}^{\epsilon} k^{2p} \quad \text{for } p=0,\dots,N \,.
\ee
In practice, $N$ will be most often less than $5$, although we will consider in one case $N=10$ and $20$. In~\cite{KCY14}, only the case $\epsilon=1$ was needed, as inspiral signals were considered. The two solutions for the stencil coefficients in the two cases $\epsilon = \pm 1$ are simply related by a complex conjugation. Explicit expressions for the stencil coefficients $a_{N,k}^{\epsilon}$ are given in App.~\ref{app:stencil} for the first values of $N$.

An immediate advantage of this reformulation is its improved numerical stability. In waveform modelling applications, it can be bery hard to control high-order numerical derivatives of the modulation. Here, one simply evaluates the original smooth modulation function at shifted times. In the following, we will adopt this implementation for our quadratic-in-phase treatment.


\subsection{Quadratic phase corrections as an integral transform}
\label{subsec:fresneltransform}

We can give an alternative interpretation of the quadratic-phase expansion~\eqref{eq:resultdffPsiTf} and of Sec.~\ref{subsec:resumquadphase} in terms of an integral transform. For simplicity of notation, here we keep to the case of a pure modulation $F$ with no delays, $d=0$. We will reintroduce the delays below in Sec.~\ref{subsec:delays}. To obtain~\eqref{eq:resultdffPsiTf}, we expanded the phase exponential using~\eqref{eq:expandexp}. If instead we do not expand this factor, we have
\be\label{eq:integralquadphase}
	\tilde{s}(f)	\simeq \tilde{h}(f) \int \ud t\, F(t) \int\ud f'\, e^{2i\pi f' (t-\tf)} \exp\left[ 2i\pi^{2} \epsilon{f'}^{2} \Tf^{2} \right] \,,
\ee
where we recall that $\epsilon = -\mathrm{sgn}(\ud ^{2} \Psi/\ud f^{2})$. The integral over $f'$ is a simple complex Gaussian integral which can be carried out explicitly. The result is a complex Gaussian integral over time which is analogous to a Fresnel transform of the function $F$. For this Fresnel transform, we introduce the notation
\be\label{eq:defFresnel}
	\calF_{\tau}[F](t_{0}) \equiv \frac{e^{i\frac{\pi}{4}}}{\sqrt{2\pi} \tau} \int \ud t \, \exp\left[ - \frac{i}{2} \left( \frac{t-t_{0}}{\tau} \right)^{2}\right] F(t) \,,
\ee
together with the additional notation
\be\label{eq:Fresnelsign}
	\calF_{\tau, \epsilon}[F](t_{0}) \equiv
\begin{cases}
	 \calF_{\tau}[F](t_{0}) &\text{ if } \epsilon=1 \\
	 \calF_{\tau}[F^{*}](t_{0})^{*} &\text{ if } \epsilon=-1
\end{cases}
\ee
to accomodate for the possible sign change represented by $\epsilon$. The integral~\eqref{eq:integralquadphase} then gives for the transfer function
\be\label{eq:resultFresnel}
	\calT_{\rm phase}(f) = \calF_{\Tf, \epsilon}[F](\tf) \,.
\ee

The Fresnel transform~\eqref{eq:defFresnel} is localized, in the sense that the part of the integral that is centered around $t_{0}$ contributes predominantly, due to the cancelling oscillations far from $t_{0}$. The parameter $\tau$ determines how local the transform is. In the limit $\tau\rightarrow 0$, fast oscillations away from the central value $t_{0}$ will cancel out, leading to the integral taking the value $F(t_{0})$. For large values of $\tau$, by contrast, the integral~\eqref{eq:defFresnel} has an extended support. Note also that only the part of the function $F$ that is symmetric about $t_{0}$ contributes to the integral in~\eqref{eq:defFresnel}.

In our result~\eqref{eq:resultFresnel}, both the scale $\Tf$ and the central time $\tf$ are functions of the frequency $f$. Since we have seen that $\Tf$ can be interpreted as the radiation reaction timescale in the SPA regime, this means that a faster-chirping signal ($\Tf$ small) will have a Fresnel transform that is more focused, whereas a slower-chirping signal ($\Tf$ large) will have a Fresnel transform that is more extended. The Fresnel width must then be compared with how fast the function $F(t)$ in the integrand is varying. In Sec.~\ref{subsec:executivesummary}, we will build an estimate for the magnitude of these phase corrections by comparing the radiation-reaction timescale to the timescale of variation of the modulation.

Thus, the previous result~\eqref{eq:stencilresult}-\eqref{eq:stencilfresnel} can be rephrased as a quadrature rule, and the stencil $\calF_{T, \epsilon}^{N}$ is an quadrature approximation of the Fresnel transfrom $\calF_{T, \epsilon}$. If one allows for polynomial integrands\footnote{Note that such integrals with a polynomial integrand are formally divergent. One can regularize them for instance by introducing a small imaginary part in $\tau$ that is sent it to $0$ at the end of the computation.} in~\eqref{eq:defFresnel}, using the stencil~\eqref{eq:stencilfresnel} amounts to building a quadrature rule for the particular choice of nodes $\tf \pm k \Tf$, which is exact (with a regularization) if $F$ is a symmetric polynomial of degree $\leq 2N$. As a verification, performing a formal Taylor expansion in time of $F(t)$ around $\tf$ in the integral~\eqref{eq:defFresnel} and integrating term by term yields back~\eqref{eq:resultdffPsiTf}.

Note that the choice of a stencil with quadrature nodes $\tf \pm k\Tf$, even if natural, is by no means unique, and other choices would have led to different stencils. The formulation of the result~\eqref{eq:resultdffPsiTf} as a Fresnel transform~\eqref{eq:resultFresnel} opens the way for future investigations of different numerical approaches to the problem. For practical computations, in this work we will rely on the approximations $\calF_{T, \epsilon}^{N}[F] (t)$ in \eqref{eq:stencilfresnel}, introduced by~\cite{KCY14}.


\subsection{Response treatment including delays}
\label{subsec:delays}

In the case of a LISA-type detector response, the presence of the delay $d(t)$ is responsible for the frequency-dependence of the kernel $G$ introduced in~\eqref{eq:defG}. As an improvement over the Taylor expansion of the kernel function in~\eqref{eq:resulttaylordelay}, in this section we will propose a different approach based on a change of variable.

As a first step, we consider the transfer function when we keep only the leading order of the expansion~\eqref{eq:leadingorderwf}, leaving off additional amplitude and phase terms in~\eqref{eq:resulttaylorA} and~\eqref{eq:resulttaylorPsi}, but treating the kernel function nonperturbatively. The response then depends on the signal only by the time-to-frequency correspondence and reads
\begin{equation}
	\tilde{s}(f) = \tilde{h}(f) \int \ud t \, F(t) e^{-2i\pi f d(t)} \int \ud f' \, e^{2i\pi f' (t+d(t) - t_{f})} \,,
\end{equation}
with $\tf$ defined in~\eqref{eq:deftf}. This motivates a change of variable to the delayed time function $t_{d}:t \mapsto t+d(t)$. Assuming the delay does not vary too quickly, we can also define the reciprocal function $t_{d}^{-1}$, and a modified time-to-frequency correspondence $\tfd = t_{d}^{-1}(\tf)$, defined implicitly by
\be
	\left. (t + d(t) - t_{f})\right|_{t=t_{f}^{d}} = 0 \,.
\ee
The above integral gives then for the transfer function
\begin{align}\label{eq:delaycorrleading}
	\calT(f) &= \int \ud t \, F(t) e^{-2i\pi f d(t)} \delta(t + d(t) - t_{f}) \nn \\
	&= F(t_{f}^{d}) \frac{e^{-2i\pi f d(t_{f}^{d})}}{1+\dot{d}(t_{f}^{d})} \,.
\end{align}
This differs from~\eqref{eq:transferlocal} both by the replacement $\tf \rightarrow t_{f}^{d}$ and by the extra denominator.

For the proposed LISA configuration~\cite{LISA17}, we have for the orbital delays the scaling $d_{0}\sim R/c \simeq 500s$, and for the constellation delays the scaling $d_{L}\sim L/c \simeq 8s$ (ignoring the dependence on angular factors). Since the motion of the constellation is anually periodic, with a frequency $\Omega_{0} \simeq 2 \times 10^{-7}\mathrm{rad}.s^{-1}$, we have $\dot{d}_{0} \sim \Omega_{0} R/c \simeq 10^{-4}$ and $\dot{d}_{L} \sim \Omega_{0} L/c \simeq 1.7 \times 10^{-6}$. The smallness of the dimensionless quantity $\dot{d} \ll 1$ (and of its subsequent derivatives) will allow us to treat it perturbatively with a very good approximation, and shows also that the function $t_{d}$ is univalued and that there is no ambiguity in defining the reciprocal $t_{d}^{-1}$.

By treating $\dot{d}$ as a perturbation and keeping only first-order terms, we obtain for the delayed time reciprocal function
\begin{align}
	t_{d}^{-1}(t) &\simeq t-d(t) (1-\dot{d}(t)) \,,\nn\\
	d(t_{f}^{d}) &\simeq d(t_{f}) ( 1 - \dot{d}(t_{f})) \,.
\end{align}
Now, the most relevant correction in~\eqref{eq:delaycorrleading} comes from the phase factor at high frequencies, where the factors $2\pi f d_{0}$ and $2\pi f d_{L}$ give a magnification reaching respectively $3.10^{3}$ and $10^{2}$ at $1\Hz$. Ignoring the other corrections, we thus arrive at the following form for the dominant delay correction in the transfer function:
\be
	\calT(f) \simeq F(t_{f})\exp\left[ -2i\pi f d(t_{f}) (1-\dot{d}(t_{f})) \right] \,.
\ee
This first correction beyond the leading order is signal-independent and affects purely the phase of the output signal.

Next, we consider the case where the quadratic phase correction is kept as well, as in Sec.~\ref{subsec:fresneltransform}, and where we keep all the first-order terms in $\dot{d}$ (neglecting its higher derivatives). We can write
\begin{widetext}
\begin{align}
	\tilde{s}(f) &\simeq \tilde{h}(f) \int \ud t \, F(t) e^{-2i\pi f d(t)} \int \ud f' \, \exp\left[ 2i\pi \epsilon \Tf^{2} f'^{2} + 2i\pi f' (t+d(t) - \tf) \right] \nn\\
	&\simeq \tilde{h}(f) \frac{e^{i\epsilon\frac{\pi}{4}}}{\sqrt{2\pi}\Tf} \int \ud \tau \, \frac{F(\tau - d(\tau))}{1+\dot{d}(\tau)} e^{-2i\pi f d(\tau)(1-\dot{d}(\tau))}\exp\left[ -\frac{i\epsilon}{2} \frac{(\tau - \tf)^{2}}{\Tf^{2}} \right] \,,
\end{align}
where we used a change of variable $\tau = t_{d}(t)$. We see that the result can again be expressed as a Fresnel transform.

Finally, when considering amplitude corrections as well, as in~\eqref{eq:resulttaylorA}, additional powers of $f'$ can be translated as time derivatives with respect to the variable $\tau$ after performing the change of variables. This produces the result:
\be\label{eq:transferfinal}
	\calT(f) = \sum\limits_{k \geq 0} \frac{(-i)^{k}}{k!} (T_{Ak})^{k} \calF_{\Tf, \epsilon} \left[ \frac{\ud^{k}}{\ud \tau^{k}} \left( \frac{F(\tau - d(\tau))}{1+\dot{d}(\tau)} e^{-2i\pi f d(\tau)(1-\dot{d}(\tau))} \right) \right] (\tf) \,.
\ee
\end{widetext}

\subsection{Summary of the formalism}\label{subsec:executivesummary}

In this Section, we gather our previous results for the convenience of the reader, and explain how we will use them in practice. For a modulation $F$ and delay $d$, so that $s(t) = F(t) h(t+d(t))$, we obtained the transfer function $\calT (f) = \tilde{s}(f)/\tilde{h}(f)$ given in~\eqref{eq:transferfinal}.

In this result, the Fresnel transform $\calF^{\epsilon}_{\Tf}$ can in turn be approximated by the stencil $\calF^{N}_{\Tf, \epsilon}$ using the formula~\eqref{eq:stencilfresnel}. The timescales $\Tf$ and $T_{Ak}$ were defined in~\eqref{eq:defTf} and~\eqref{eq:defTA}, and the generalized time-to-frequency function $t_{f}$ was given in~\eqref{eq:deftf}. The delays $d$ are present only in the LISA context, while in the context of precessing binaries we only have to consider the modulation function $F$. In practice, only the first few of the terms in the series expansion are relevant. We will investigate several orders of approximation, combining corrections from the phase, amplitude and delays, and explore which ones are relevant for a given level of accuracy. We will use symbols of the form $\{N | A | d\}$ to indicate the order of the stencil in~\eqref{eq:stencilfresnel}, the maximal order of the amplitude correction included, and the inclusion or not of the delay corrections at first order in $\dot{d}$ as derived in Sec.~\ref{subsec:delays}.

In the LISA context, due to the smallness of the corrections we will only go up to $k=1$ in~\eqref{eq:transferfinal}, and in computing the remaining time derivatives in~\eqref{eq:transferfinal} we will neglect the second and higher derivatives. We will also use $F(\tau - d(\tau)) \simeq (F - d \dot{F})(\tau)$. This gives concretely:
\begin{widetext}
\begin{subequations}\label{eq:summaryNAd}
\begin{align}
	\{N | A:0 | d:0\}&: \; \calT(f) = \calF^{N}_{\Tf, \epsilon} \left[ F e^{-2i\pi f d} \right] (t_{f}) \,, \\
	\{N | A:1 | d:0\}&: \; \calT(f) = \calF^{N}_{\Tf, \epsilon} \left[ \left(F - i T_{A1} \left( \dot{F} - 2i\pi f \dot{d}\right) \right) e^{-2i\pi f d} \right] (t_{f}) \,, \\
	\{N | A:0 | d:1\}&: \; \calT(f) = \calF^{N}_{\Tf, \epsilon} \left[ \frac{F - d\dot{F}}{1+\dot{d}} e^{-2i\pi f d (1-\dot{d})} \right] (t_{f})\,, \\
	\{N | A:1 | d:1\}&: \; \calT(f) = \calF^{N}_{\Tf, \epsilon} \left[ \frac{1}{1+\dot{d}}\left(  F - d\dot{F} - i T_{A1} \left( \dot{F} - 2i\pi f \dot{d} \right) \right) e^{-2i\pi f d (1 - \dot{d})} \right] (t_{f})\,.
\end{align}
\end{subequations}
\end{widetext}

In the context of precessing binaries, the delays are absent and we will go up to $k=2$ in~\eqref{eq:transferfinal}. The tranfer function at different orders of approximation will be:
\begin{subequations}\label{eq:summaryNA}
\begin{align}
	\{N | A:0\}&: \; \calT(f) = \calF^{N}_{\Tf, \epsilon} \left[ F \right] (t_{f}) \,, \\
	\{N | A:1\}&: \; \calT(f) = \calF^{N}_{\Tf, \epsilon} \left[ F - i T_{A1} \dot{F} \right] (t_{f}) \,, \\
	\{N | A:2\}&: \; \calT(f) = \calF^{N}_{\Tf, \epsilon} \left[ F - i T_{A1} \dot{F} - \frac{1}{2} (T_{A2})^{2} \ddot{F} \right] (t_{f}) \,.
\end{align}
\end{subequations}

In the following, it will be convenient to introduce error estimates built from the Taylor-like series~\eqref{eq:resulttaylorPsi},~\eqref{eq:resulttaylorA} and~\eqref{eq:resulttaylordelay}. We simply define these error estimates at a certain level of approximation as the magnitude of the first term ignored in the original Taylor series. To give these quantities a relative meaning, we divide by the leading term. Thus we define, with $G(f,t) = F(t) e^{-2i\pi f d(t)}$,
\begin{subequations}\label{eq:deffom}
\begin{align}
	\epsilon_{\Psi 2} &\equiv \frac{1}{2} \Tf^{2} \left| \frac{1}{G}\partial_{tt}G \right| \,, \\
	\epsilon_{A 1} &\equiv T_{A1} \left| \frac{1}{G} \partial_{t} G \right| \,, \\
	\epsilon_{A 2} &\equiv \frac{1}{2} T_{A2}^{2} \left| \frac{1}{G} \partial_{tt}G \right| \,, \\
	\epsilon_{d} &\equiv \frac{1}{2\pi} \left| \frac{1}{G} \partial_{tf} G \right| \,,
\end{align}
\end{subequations}
where the function $G$ and its derivatives are evaluated at $(f, t_{f})$. For $\epsilon \ll 1$, the perturbative approach applies and $\epsilon$ can be used as an estimate for the magnitude of the effect. Reaching $\epsilon \sim 1$ will indicate a breakdown of the perturbative approach.

The physical interpretation of these error measures is clear: when the modulation and delay obey the simple scaling $\partial_{t}^{n} \rightarrow \Omega^{n}$, with $\Omega_{0}$ a characteristic frequency, the phase and amplitude error estimates are simply ratios of timescales. For instance, with this scaling $\epsilon_{\Psi 2} \sim T_{f}^{2}\Omega^{2}/2$, so that the approximation will work well when the radiation-reaction timescale is shorter than the characteristic timescale of the modulation. However, as we will show in Sec.~\ref{subsec:lisafom} this simple picture will need to be refined in the presence of delays, due to the presence of additional dimensionless factors of the form $2\pi f d$, which can be larger than 1.

On top of this perturbative formalism, in the LISA case we will also develop another approach exploiting the periodicity of the modulation and delays (see Sec.~\ref{subsec:comblisa}), while in the case of precessing binaries we will use a trigonometric polynomial approach to represent the merger-ringdown part of the signal (see Sec.~\ref{subsec:trigopoly})


\section{Application to the response of LISA-type detectors}
\label{sec:LISA}

In this Section, we apply the formalism of Sec.~\ref{sec:formalism} to the Fourier-domain response of a LISA-like detector, and assess the accuracy of our approach at various levels of approximation. We also propose an alternative approach for slowly-chirping signals.


\subsection{The response model}
\label{subsec:modelLISA}

We begin by detailing the model that we use for the response of a LISA-like detector, together with the assumptions used and their limitations. Since our aim is to assess the accuracy of our direct Fourier-domain treatment of the response, we can focus on the gravitational-wave contribution to the basic single-link observables. We therefore use a somewhat simplified model for the time-domain response~\cite{Krolak+04}, ignoring corrections that would be crucial from the point of view of noise cancellations, keeping in mind that the response model can be enriched later without affecting the conclusions of the present analysis.

The frequency-shift response for a single link~\eqref{eq:yslr} was derived in Ref.~\cite{EW75} (see also~\cite{CR02, RCP04, Finn08, Cornish09}). Several assumptions enter the result as written in~\eqref{eq:yslr}: (i) effects of the order $v/c$ are neglected, including for instance the special relativistic Doppler effect created by the relative speeds of the spacecraft on their orbits (ii) the propagation is assumed to take place in a flat spacetime, perturbed only by the gravitational wave; thus the gravitational redshift as well as the deflection of light created by the gravitational potential of the Sun is ignored (iii) all geometric factors are evaluated at a single time, whereas one should consider the beam as propagating from the position of the first spacecraft at the time of emission to the position of the second scapecraft at the time of reception, leading to a point-ahead effect.

Additionally, we limit ourselves to a rigid model for the orbits of the constellation, namely we assume that the constellation remains in an equilateral configuration with fixed armlengths. These simplified orbits neglect (iv) effects of order $e^{2}$ from the eccentricity of the individual Keplerian orbits, (v) the effect of gravitational perturbations coming from other celestial bodies, such as the Earth, the quadrupole of the Sun, and the other planets. Note that although we can neglect all the effects (i)-(v) for our present study, keeping track of these corrections is crucial for the purpose of laser noise cancellations, and led to the development of new generations of TDI observables~\cite{Tintoliving}.

It is natural to split the response~\eqref{eq:yslr} into two steps: first the orbital delay related to the orbit around the Sun of the whole constellation, and then the constellation response. The baselines for the delays are indeed very different in the two cases. Geometrical projection factors aside, we have for the orbit around the Sun $R=1\text{au}=1.5\times 10^{8} \text{km}$, while the detector armlength is $L=2.5\times 10^{6}\text{km}$ (in the configuration proposed in~\cite{LISA17}). It is natural to define two transfer frequencies for the two relevant length scales for the delays, defined such that a wavelength fits within this length scale, i.e. $2\pi f d = 1$. This gives
\begin{subequations}\label{eq:transferfrequencies}
\begin{align}
	f_{R} &= 3.2\times10^{-4}\Hz \,,\\
	f_{L} &= 1.9\times 10^{-2}\Hz \,.
\end{align}
\end{subequations}
The LISA response will behave qualitatively differently on the three frequency bands $f \leq f_{R}$, $f_{R} \leq f \leq f_{L}$ and $f_{L} \leq f$.

The first stage of the response, the orbital delay, consists simply in applying the varying time delay to bring the wavefront sampling point from the SSB reference to the center of the LISA triangular constellation, common to all $y_{slr}$ observables. For $h^{\rm TT}$ the transverse-traceless gravitational waveform in matrix form, we write this orbital time delay as
\be\label{eq:defresponse0}
	h_{0}^{\rm TT} (t) = h^{\rm TT}(t-\hatk\cdot p_{0}) \,,
\ee
with $p_{0}$ the position of the constellation center, which follows the Earth orbit around the Sun. The second stage of the response calculation comprises the remaining, constellation-centered response. For the single-link contribution to the response, for the laser link from spacecraft $s$ to spacecraft $r$ along a path in direction $n_l$, we write
\begin{align}\label{eq:defresponseL}
	y_{slr} &= \frac{1}{2} \frac{1}{1 - \hatk\cdot n_{l}} \nn\\
	& \cdot n_{l}\cdot \left[ h_{0}^{\rm TT}(t - L - \hatk\cdot p^{L}_{s}) - h_{0}^{\rm TT}(t - \hatk\cdot p^{L}_{r}) \right] \cdot n_{l}\,,
\end{align}
where we reference the positions of the spacecraft relative to the center of the constellation, $p^{L}_{A} \equiv p_{A} - p_{0}$.

As described in App.~\ref{app:notation} We will decompose the full signal in the contributions of the individual spin-weighted spherical modes $h_{\ell m}$, whose Fourier transforms are assumed to have a smooth amplitude and phase. First, we define the matrices $P_{+},P_{\times}$ such that, in the sense of matrices,
\be
	h^{\rm TT} = h_{+}P_{+} + h_{\times}P_{\times} \,.
\ee
We focus only on positive frequencies. Assuming that the approximation~\eqref{eq:zeronegativef} applies, we consider a single mode contribution, $h=h_{\ell m}$ with $m>0$. For each given mode we define a complex matrix $P_{\ell m}$ incorporating the spin-weighted spherical harmonic constant factor as
\be
	P_{\ell m} =
	\begin{cases}
	\frac{1}{2} {}_{-2}Y_{\ell m} \left( P_{+} + i P_{\times} \right) \text{ for } m>0\,,\\
	\frac{1}{2} {}_{-2}Y_{\ell m}^{*} \left( P_{+} - i P_{\times} \right) \text{ for } m<0\,.
	\end{cases}
\ee

We now turn to the transformation of~\eqref{eq:defresponse0} and~\eqref{eq:defresponseL} to the Fourier domain. Applying a pure delay as in~\eqref{eq:defresponse0} translates into
\be\label{eq:G0}
	G_{0}(f, t) = e^{-2i\pi f d_{0}(t)} \,,
\ee
with $d_{0} = -\hatk \cdot p_{0}$ the delay associated to the orbit around the Sun. For the leading-order response~\eqref{eq:transferlocal}, this gives a Fourier-domain transfer function common to all modes, that is a pure phase factor, proportional to the frequency but also $t_{f}$-dependent:
\be\label{eq:transfer0local}
	\calT_{0}^{\rm local}(f) = G_{0}(f, \tf)\,.
\ee
If $(\lambda, \beta)$ are the ecliptic longitude and latitude of the source in the sky, and if the orbital phase is set by convention to $0$ at $t=0$, the orbital delay has the simple expression
\be\label{eq:delay0}
	d_{0}(t) = -R \cos\beta \cos\left(\Omega_{0}t - \lambda\right)\,.
\ee

For the constellation part of response~\eqref{eq:defresponseL}, treated separately from the delay~\eqref{eq:delay0}, we write
\begin{align}\label{eq:decomposeGslr}
	F_{slr}^{L}(t) &= \frac{1}{2} \frac{1}{1 - \hatk\cdot n_{l}(t)} n_{l}(t) \cdot P_{\ell m} \cdot n_{l} (t) \,,\nn\\
	d_{s}(t) &= - k\cdot p_{s}^{L}(t) \,, \quad d_{r}(t) = - k\cdot p_{r}^{L}(t) \,,\nn\\
	G_{slr}^{L}(f,t) &=  F_{slr}^{L}(t) \left( e^{-2i\pi f (d_{s,L}(t) + L)} - e^{-2i\pi f d_{r}(t)} \right) \,.
\end{align}
The superscript $L$ indicates that the orbital delay~\eqref{eq:delay0} is not included. Since we also assume the rigid approximation for the constellation, where the armlengths are fixed, a particular simplification occurs when combining these individual delays, thanks to the relation $p^{L}_{r} - p^{L}_{s} =  L n_{l}$:
\begin{align}\label{eq:GslrL}
	G_{slr}^{L}(f,t) &= \frac{i \pi f L}{2} \sinc \left[ \pi f L\left(1-\hatk\cdot n_{l} \right) \right] \nn\\
	& \quad \cdot \exp\left[ i \pi f \left( L + \hatk\cdot \left( p_{1}^{L} + p_{2}^{L} \right) \right) \right]  n_{l} \cdot P_{\ell m} \cdot n_{l} \,,
\end{align}
with all time-dependent vectors evaluated at $t$. This expression is well known as describing the frequency-dependency in the LISA response~\cite{Larson+99, Cornish01, CR02, RCP04}. In the local approximation~\eqref{eq:transferlocal}, the Fourier-domain transfer function then reads
\begin{align}\label{eq:transferLlocal}
	\calT_{slr}^{L, \mathrm{local}}(f) &= G_{slr}^{L}(f, \tf) \,.
\end{align}
For plotting purposes, we will also define
\be\label{eq:transferLenvelope}
	\overline{\calT}_{slr}^{L} (f) = \frac{i \pi f L}{2} n_{l} \cdot P_{\ell m} \cdot n_{l} (\tf)
\ee
which will serve as an estimate for the enveloppe function of the response, devoid of the zero-crossings at high frequencies of the $\sinc$ term in~\eqref{eq:GslrL}.

Note that, if the corrections of Sec.~\ref{subsec:delays}, for non-negligible $\dot d$, are included for the constellation delays, the transfer function will not have this simple form anymore, as $\dot{d}$ will have a different velocity-dependent expression for the sending and receiving spacecraft. One must then separately handle $d_{s}$ and $d_{r}$ in~\eqref{eq:decomposeGslr} to compute the corrections.

The orbital response~\eqref{eq:transfer0local}-\eqref{eq:delay0} takes a simple analytic form, but the phase contribution of this delay is significant across most of the frequency band and can be large for $f \gg f_{R}$.

The constellation response~\eqref{eq:GslrL}-\eqref{eq:transferLlocal} can be interpreted as the Fourier-domain translation of a discrete derivative taken on the waveform. The leading factor in~\eqref{eq:GslrL} shows that the amplitude of the response is proportional to $f$ in the low-frequency limit $f\ll f_{L}$, where the other factors are essentially unity. For $f\gtrsim f_{L}$, the $\sinc$ and the phase of the exponential generate additional structure in the response, including zero-crossings when the projected armlength is an integer number of wavelengths. From~\eqref{eq:transferLlocal}, an expansion for small $f\ll f_{L}$ yields back a Fourier-domain analog of the low-frequency approximation of the response~\cite{Cutler97, RCP04}, which is equivalent to having two LIGO-type interferometers turned by $\pi/4$ and set in motion.

For analysis of the response we need a concrete set of gravitational waveforms. We will use the PhenomD model~\cite{Khan+15,Husa+15}, which provides Fourier-domain inspiral-merger-ringdown waveforms for aligned spins. We refer to App.~\ref{app:precLISA} for a brief discussion of the prospects for applying our formalism for the LISA response to precessing Fourier-domain waveforms.


\subsection{Estimates for the magnitude of higher-order corrections}
\label{subsec:lisafom}

\begin{figure}
  \centering
  \includegraphics[width=.99\linewidth]{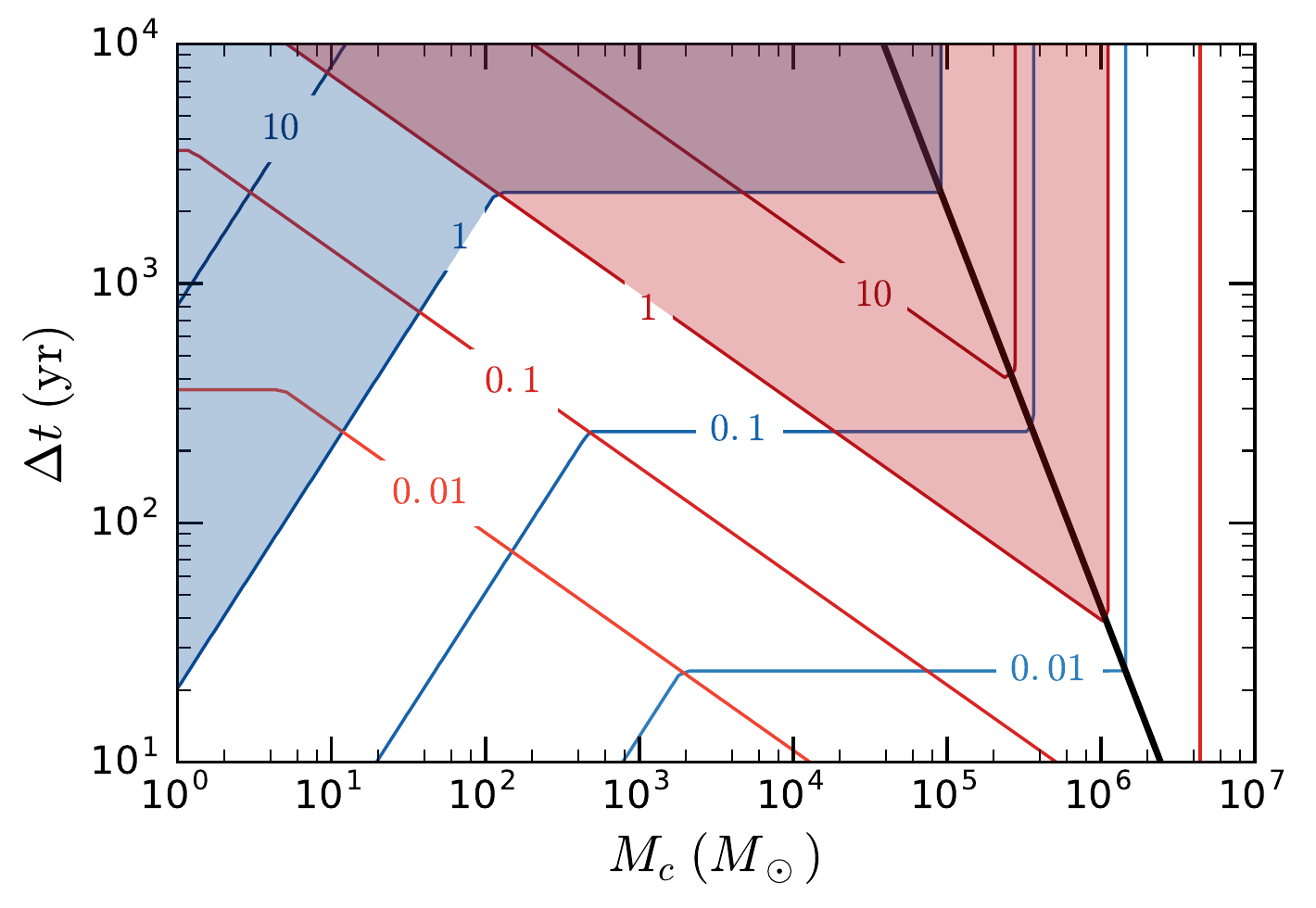}
  \caption{Contour levels for the analytical estimate of the error measure $\epsilon_{\Psi 2}$ at the starting frequency, as a function of chirp mass $\Mchirp$ and time to coalescence $\Delta t$. Blue corresponds to the orbital response and red to the constellation response. The colored shaded areas indicate regions where $\epsilon_{\Psi 2} \geq 1$, where the perturbative formalism is expected to break down. In the region to the right of the black line, $f_{\rm start}$ given in~\eqref{eq:fstartN} is lower than the lowest in-band frequency $f_{\rm min} = 10^{-5}\Hz$, so that the signal starts at $f_{\rm min}$ and $\epsilon_{\Psi 2}$ becomes independent of $\Delta t$.}
  \label{fig:lisafomPsiMcDeltat}
\end{figure}

\begin{figure*}
  \centering
  \includegraphics[width=.98\linewidth]{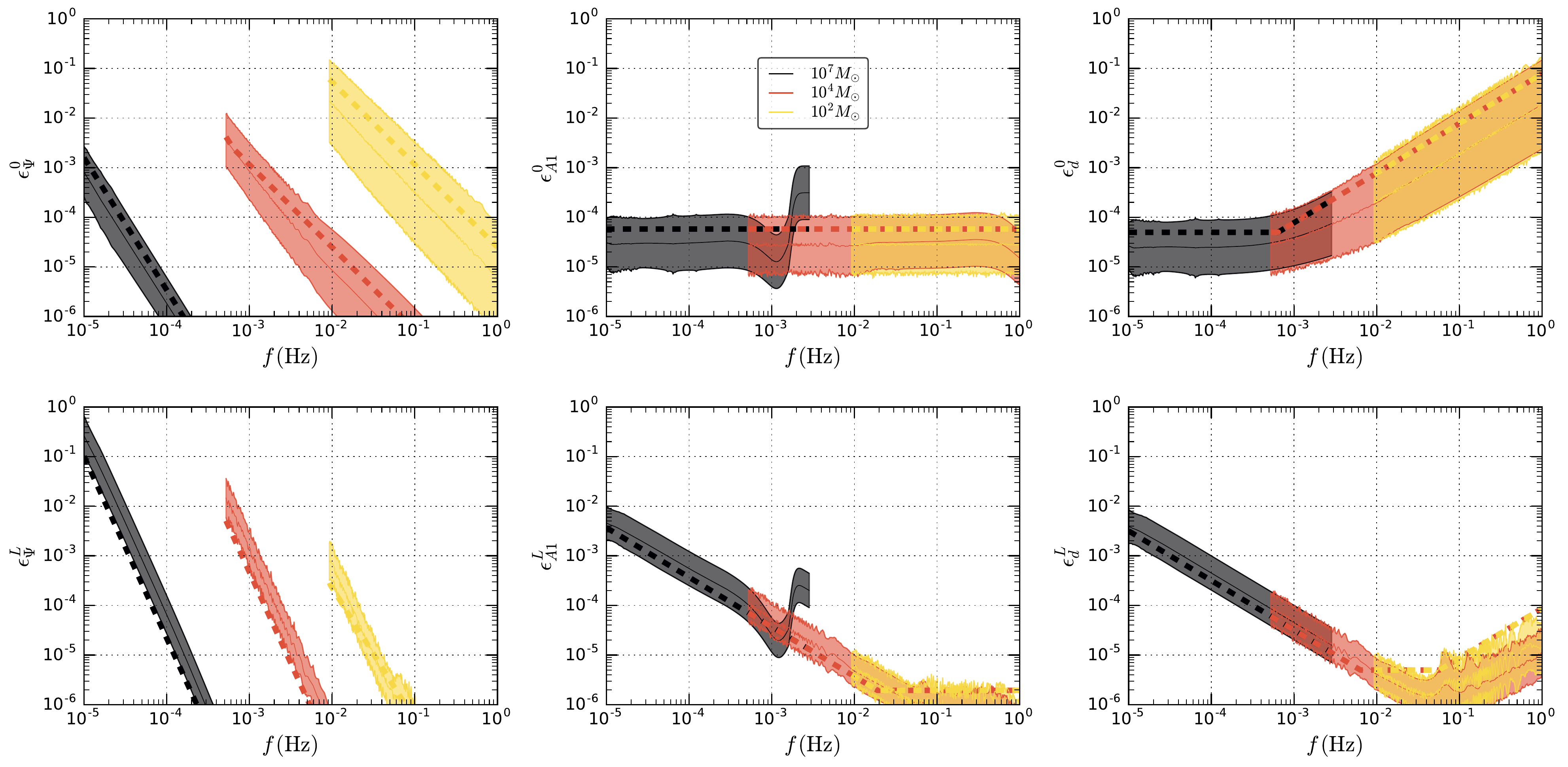}
  \caption{Error estimates as defined in~\eqref{eq:deffom}, for an equal-mass, non-spinning system and for total masses $M=10^{7} \Msol$, $10^{4} \Msol$ and $10^{2} \Msol$. The top row corresponds to the orbital delay part of the response~\eqref{eq:defresponse0}, and the lower row shows the LISA-centered constellation response~\eqref{eq:defresponseL}. The error measures $\epsilon_{\Psi 2}$ for the phase corrections, $\epsilon_{A1}$ for the amplitude, $\epsilon_{d}$ for the delay are shown from left to right. The central line and interval are the mean and $1\sigma$ standard deviation of the logarithm of $\epsilon$ computed with numerical derivatives over 400 random values for the position in the sky, inclination and polarization. The starting frequency is set by an observation time of $\Delta t = 10 \text{yrs}$ before merger. We overlay in dashed the analytical estimates obtained from from~\eqref{eq:timescalesNfstart} and~\eqref{eq:estimatederivorb}-\eqref{eq:estimatederivconst}.}
  \label{fig:fomLISA}
\end{figure*}

Using the approximate error measures $\epsilon$ introduced in~\eqref{eq:deffom}, we will now estimate, for each type of correction (phase, amplitude, delay), the size of errors in the transfer function.

To obtain an order-of-magnitude estimate for the error measures $\epsilon$~\eqref{eq:deffom}, we will use the Newtonian-order expressions~\eqref{eq:timescalesN} for the signal-dependent timescales $\Tf$ and $T_{A1}$. Higher-order amplitude terms beyond the first one in~\eqref{eq:resulttaylorA}, as well as phase terms beyond the second derivative in~\eqref{eq:expandPsi}, will be negligible and we will ignore them in the following. It is useful to separate the orbital response~\eqref{eq:transfer0local} and the constellation response~\eqref{eq:transferLlocal}, as the different baseline of the delays (orbital radius $R$ or armlength $L$) as well as the presence of a time-varying prefactor $F(t)$ both affect the result. When estimating the magnitude of the relevant derivatives of $G$, we must also take into account dimensionless delay factors of the type $2\pi f d$.

We start with the orbital response, which takes the form of a pure delay $G_{0}(f, t) = e^{-2 i \pi f d_{0}(t)}$, and obtain
\begin{subequations}
\begin{align}
	\frac{1}{G_{0}} \partial_{t} G_{0} &= -2i\pi f \dot{d}_{0}\,,\\
	\frac{1}{G_{0}} \partial_{tt} G_{0} &= -2i\pi f \ddot{d}_{0} - 4\pi^{2} f^{2} \dot{d}_{0}^{2} \,,\\
	\frac{1}{G_{0}} \partial_{tf} G_{0} &= -2 i \pi \dot{d}_{0} - 4\pi^{2} f d_{0} \dot{d}_{0} \,.
\end{align}
\end{subequations}

Since the one-arm constellation response~\eqref{eq:decomposeGslr}-\eqref{eq:transferLlocal} is analogous to a discrete time derivative of the signal, it is appropriate to keep explicit an overall factor $f$ reflecting this structure. Hence we write symbolically $G_{L}(f,t) \sim f F(t) e^{-2i \pi f d_{L}(t)}$, where $d_{L}$ represents a delay term and $F(t)$ represents the rest of the geometric factors in~\eqref{eq:transferLlocal}, for which we momentarily ignore the $f$-dependence. This gives
\begin{subequations}
\begin{align}
	\frac{1}{G_{L}} \partial_{t} G_{L} &\sim -2i\pi f \dot{d}_{L} + \frac{\dot{F}}{F}\,,\\
	\frac{1}{G_{L}} \partial_{tt} G_{L} &\sim -2i\pi f \ddot{d}_{L} - 4\pi^{2} f^{2} \dot{d}_{L}^{2} - 2i\pi f \dot{d}_{L} \frac{\dot{F}}{F} - \frac{\dot{F}^{2}}{F^{2}} + \frac{\ddot{F}}{F} \,,\\
	\frac{1}{G_{L}} \partial_{tf} G_{L} &\sim -4 i \pi \dot{d}_{L} - 2i\pi d_{L} \frac{\dot{F}}{F} - 4\pi^{2} f d_{L} \dot{d}_{L} + \frac{1}{f}\frac{\dot{F}}{F}\,.
\end{align}
\end{subequations}

In order to obtain simple scalings for the error estimates, we will make the replacements $\partial_{t}^{n} d \sim \Omega_{0}^{n} d$ as well as $\partial_{t}^{n} F \sim \Omega_{0}^{n} F$. These scalings are only approximate as different orientation angles can lead to significant variations. To represent the average of the geometric projection factor of the gravitational wave propagation vector on the plane of the orbit, we will take $d_{0} \sim R/2 $. For the constellation delays, we simply take $d_{L} \sim L$ as there projection effects can lead to variations in both ways. The resulting estimates for the magnitude of these derivatives are
\begin{subequations}\label{eq:estimatederivorb}
\begin{align}
	\left| \frac{1}{G_{0}} \partial_{t} G_{0} \right| &\sim \pi f \Omega_{0} R\,,\\
	\left| \frac{1}{G_{0}} \partial_{tt} G_{0} \right| &\sim \text{max} \left[ \pi f \Omega_{0}^{2} R, \pi^{2} f^{2} \Omega_{0}^{2} R^{2}\right] \,,\\
	\left| \frac{1}{G_{0}} \partial_{tf} G_{0} \right| &\sim \text{max} \left[ \pi  \Omega_{0} R, \pi^{2} f \Omega_{0} R^{2} \right] \,,
\end{align}
\end{subequations}
for the orbital response and
\begin{subequations}\label{eq:estimatederivconst}
\begin{align}
	\left| \frac{1}{G_{L}} \partial_{t} G_{L} \right| &\sim \text{max} \left[ \Omega_{0}, 2 \pi f \Omega_{0} L \right] \,,\\
	\left| \frac{1}{G_{L}} \partial_{tt} G_{L} \right| &\sim \text{max} \left[ \Omega_{0}^{2}, 4 \pi f \Omega_{0}^{2} L, 4\pi^{2} f^{2} \Omega_{0}^{2} L^{2} \right] \,,\\
	\left| \frac{1}{G_{L}} \partial_{tf} G_{L} \right| &\sim \text{max} \left[ \Omega_{0}/f, 6 \pi \Omega_{0} L, 4\pi^{2} f \Omega_{0}^{2} L^{2} \right] \,,
\end{align}
\end{subequations}
for the constellation response. In the presence of different terms, we simply take the maximum of their norms.

An important point in~\eqref{eq:estimatederivorb} and~\eqref{eq:estimatederivconst} above is the presence of delay factors in the forms of powers of $2\pi f d$, so that we cannot use the simple replacement $\partial_{t}^{n}G \rightarrow \Omega_{0}^{n}G$. Thus, the suitability of the leading-order treatment cannot be estimated by a mere separation between the signal timescales and the annual timescale for the response, but will also depend on the frequency being above or below the transfer frequencies $f_{R}$ and $f_{L}$ defined in~\eqref{eq:transferfrequencies}. The error measure $\epsilon_{d}$ does not depend on signal-dependent timescales and can be directly read off the estimates above, with $\epsilon_{d} = |\partial_{tf}G/(2\pi G)|$. For the error measures $\epsilon_{\Psi 2}$ and $\epsilon_{A1}$, we will combine the above derivatives with the Newtonian timescales given in~\eqref{eq:timescalesN} for the inspiral phase of the signal.

The starting frequency of the signal will play an important role. As a function of the time remaining before merger $\Delta t$, from the Newtonian relation~\eqref{eq:omegaphiN} we have (using the chirp mass $\Mchirp = M \nu^{3/5}$):
\be\label{eq:fstartN}
	f_{\rm start} = 1.75\times 10^{-5} \Hz \left( \frac{\Mchirp}{10^{6}M_{\odot}} \right)^{-5/8} \left( \frac{\Delta t}{10 \yr} \right)^{-3/8} \,.
\ee
This gives in turn for the Newtonian timescales~\eqref{eq:timescalesN}
\begin{subequations}\label{eq:timescalesNfstart}
\begin{align}
	\Tf^{\rm N} &= 8.78\times10^{-2}\yr \left( \frac{\Mchirp}{10^{6}M_{\odot}} \right)^{5/16} \left( \frac{\Delta t}{10 \yr} \right)^{11/16} \,, \\
	T_{A1}^{\rm N} &= 3.37\times10^{-4}\yr \left( \frac{\Mchirp}{10^{6}M_{\odot}} \right)^{5/8} \left( \frac{\Delta t}{10 \yr} \right)^{3/8} \,.
\end{align}
\end{subequations}
Note that if we think of LISA as effectively insensitive below some minimal frequency $f_{\rm min}$, then for sufficiently high $\Mchirp$ this point of entry in the sensitive band will mark the beginning of the signal, obviating the relevance of $f_{\rm start}$. For $f_{\rm min} = 10^{-5}\Hz$ and $\Delta t \leq 10\yr$, this is the case for $\Mchirp \geq 2.45\times 10^{6}\Msol $. Thus, our final analytical estimates for the $\epsilon$ error measures are built by inserting~\eqref{eq:timescalesNfstart} and~\eqref{eq:estimatederivorb}-\eqref{eq:estimatederivconst} in~\eqref{eq:deffom}, while ensuring the frequency cut $10^{-5}\Hz \leq f \leq 1\Hz$.

It is useful to distinguish between what we will call merging binaries, systems close to coalescence that will merge during the LISA mission lifetime or a few years later, and slowly-chirping binaries still in the deep inspiral phase, of which we observe only a small snapshot in Fourier domain as they do not sweep to the end of the frequency band. The massive black hole binaries (MBH) that will be observed by LISA~\cite{LISA17} fall within the first category, with their merger in band, while the proposed population of stellar-origin black hole binaries (SOBH)~\cite{Sesana16}, with masses comparable to the LIGO/Virgo detections, will comprise both merging binaries, i.e. exiting the LISA band towards larger frequencies during observations, and slowly-chirping binaries hundreds or thousands of years away from merger.

We turn first to the slowly-chirping binaries. For these systems, we will focus on $\epsilon_{\Psi 2}$, which will be the most important error measure. Fig.~\ref{fig:lisafomPsiMcDeltat} shows contour levels for the value of our simple analytical estimate for $\epsilon_{\Psi 2}$ at the beginning of the observations, as a function of the chirp mass and time to merger, for both the orbital and the constellation response. The limit $\epsilon_{\Psi 2} = 1$ is used to single out areas where the perturbative treatment of Sec.~\ref{sec:formalism} is expected to break down, although the precise location of this boundary will vary, depending on the accuracy level required and on the orientation angles. We also single out the high-mass region where the start of the signal is set by its entry into the sensitive band at $f_{\rm min}$, making the error measure independent of the time to merger. We find that, for binaries in the LIGO/Virgo mass range $10^{1}-10^{2} \Msol$, we reach the limit $\epsilon_{\Psi 2}=1$ first for the orbital response, for time-to-merger values within the expected observed range~\cite{Sesana16}. We will introduce in Sec~\ref{subsec:comblisa} an alternative approach to deal with those signals. For completeness, we also show in Fig.~\ref{fig:lisafomPsiMcDeltat} the result for intermediate and massive black hole systems, although we expect to observe merging systems for this mass range.

We now turn to the case of merging binaries. For these we compute the three error measures $\epsilon_{\Psi 2}$, $\epsilon_{A1}$ and $\epsilon_{d}$. To go beyond the crude analytical estimates built from~\eqref{eq:timescalesNfstart} and~\eqref{eq:estimatederivorb}-\eqref{eq:estimatederivconst}, we use numerical derivatives for the timescales~\eqref{eq:defTf}-\eqref{eq:defTA} and for the derivatives in~\eqref{eq:deffom}. We summarize the results in Fig.~\ref{fig:fomLISA}, both for the orbital delay and the constellation response, considering equal-mass non-spinning systems with total masses of $M=10^{7} \Msol$, $10^{4} \Msol$ and $10^{2} \Msol$ with a starting frequency corresponding to $\Delta t =10 \yr$ of observation. Since individual signals can show significant variation depending on the orientation parameters, for the full computation with numerical derivatives we show a geometric average of $\epsilon$ over the sky position, inclination and polarization, together with $\pm 1\sigma$ geometric standard deviation. Here and in the following, for completeness we show results on the frequency band from $10^{-5}\Hz$ to $1\Hz$, while the signals are expected to have little signal-to-noise ratio outside of the band from $10^{-4}\Hz$ to $10^{-1}\Hz$, which is sometimes taken as a reference~\cite{LISA17}.

For $\epsilon_{\Psi 2}$, as expected from~\eqref{eq:TfN} we find a steep rise towards lower frequencies, which shows that considering merging binaries with a limited time to coalescence is crucial here. We note differences in behaviour between the orbital response, where for a fixed mass ratio the initial (10yr before merger) value of $\epsilon_{\Psi 2}$ grows towards lower masses, and the constellation response, where the initial $\epsilon_{\Psi 2}$ grows towards higher masses. Combining both parts of the response, for mergers with $10^2<M<10^7$, the error measure remains below $\epsilon_{\Psi 2} \sim 0.1$, where the corrections are still well manageable by the perturbative formalism (as we will show in Sec.~\ref{subsec:errorsLISA}).

For $\epsilon_{A1}$, we find that the structure in the waveform close to merger does have a noticable but limited impact on the error measure. For the orbital response, the error measure remains small for all masses. For the constellation response, we find that $\epsilon_{A1}$ can grow up to $\sim 0.01$ at the lower end of the frequency band for high masses.

The $\epsilon_{d}$ is signal-independent, and follows a quite different behaviour for the orbital and constellation responses. For the orbital response, $\epsilon_{d}^{0}$ grows for higher frequencies, reaching $\sim 0.1$ at around $1\Hz$. For the constellation response, $\epsilon_{d}^{L}$ is mainly important at lower frequencies, reaching $\sim 0.01$ at around $10^{-5}\Hz$.

Overall, we find that for merging binaries the perturbative approach presented in Sec.~\ref{sec:formalism} will be applicable on the whole LISA frequency band. The most relevant higher-order corrections are expected to be the phase corrections close to the starting frequency of the signal, and the delay correction at high frequencies for the orbital response. For slowly chirping binaries, we have identified regions in the parameter space where the perturbative formalism of Sec.~\ref{sec:formalism} will break down, requiring an alternative approach presented below in Sec.~\ref{subsec:comblisa}.


\subsection{Errors in the Fourier-domain response for merging binaries}
\label{subsec:errorsLISA}

Having at hand estimates for the relevance of higher-order corrections in the LISA response, we assess the performance of the leading-order treatment and we demonstrate that including higher-order corrections consistently reduces the reconstruction error of the Fourier-domain transfer function. We will consider two equal-mass and non-spinning systems for this illustration. The first will be a high-mass system with $M=10^{7} \Msol$, entering the LISA band at $f=10^{-5}\Hz$ and inspiralling for $\Delta t = 3.8 \yr$ before merging. The second will be a light system with $M=10^{2} \Msol$, close to the mass range of SOBHs~\cite{Sesana16}, starting at $\Delta t = 10\yr$ before merger and exiting the LISA band at $f=1\Hz$. We will discuss SOBH systems farther away from merger in the next section. We will also focus on the orbital response and on the single-arm constellation response $y_{132}$, as TDI combinations will appear (within the approximations listed in Sec.~\ref{subsec:modelLISA}) as linear combinations of these basic observables. The orientation angles chosen for these examples are $[\lambda,\beta,\iota,\psi] = [\pi/4,\pi/3,\pi/3,0]$ for the ecliptic longitude, latitude, inclination and polarization respectively.

On one hand, we perform a numerical inverse Fourier transform (IFFT) of the Fourier-domain PhenomD waveform, process the signal through the time-domain response~\eqref{eq:defresponse0} or~\eqref{eq:defresponseL}, and perform a numerical Fourier transform (FFT). This gives us the target Fourier-domain waveform, considered to be exact but for possible numerical artifacts\footnote{Numerical Fourier transforms include most notably oscillations induced by the necessary tapering of the signal. In the low-mass case, in practice we stitch together two frequency bands with different sampling rates, which leaves some visible residual in Figs.~\ref{fig:LISAerrorM1e2orb} and~\ref{fig:LISAerrorM1e2const}.}. On the other hand, we process the Fourier-domain signal through the response summarized in Sec.~\ref{subsec:executivesummary}, including various higher-order corrections. We then compare the output of the two procedures.

\begin{figure}
  \centering
  \includegraphics[width=.98\linewidth]{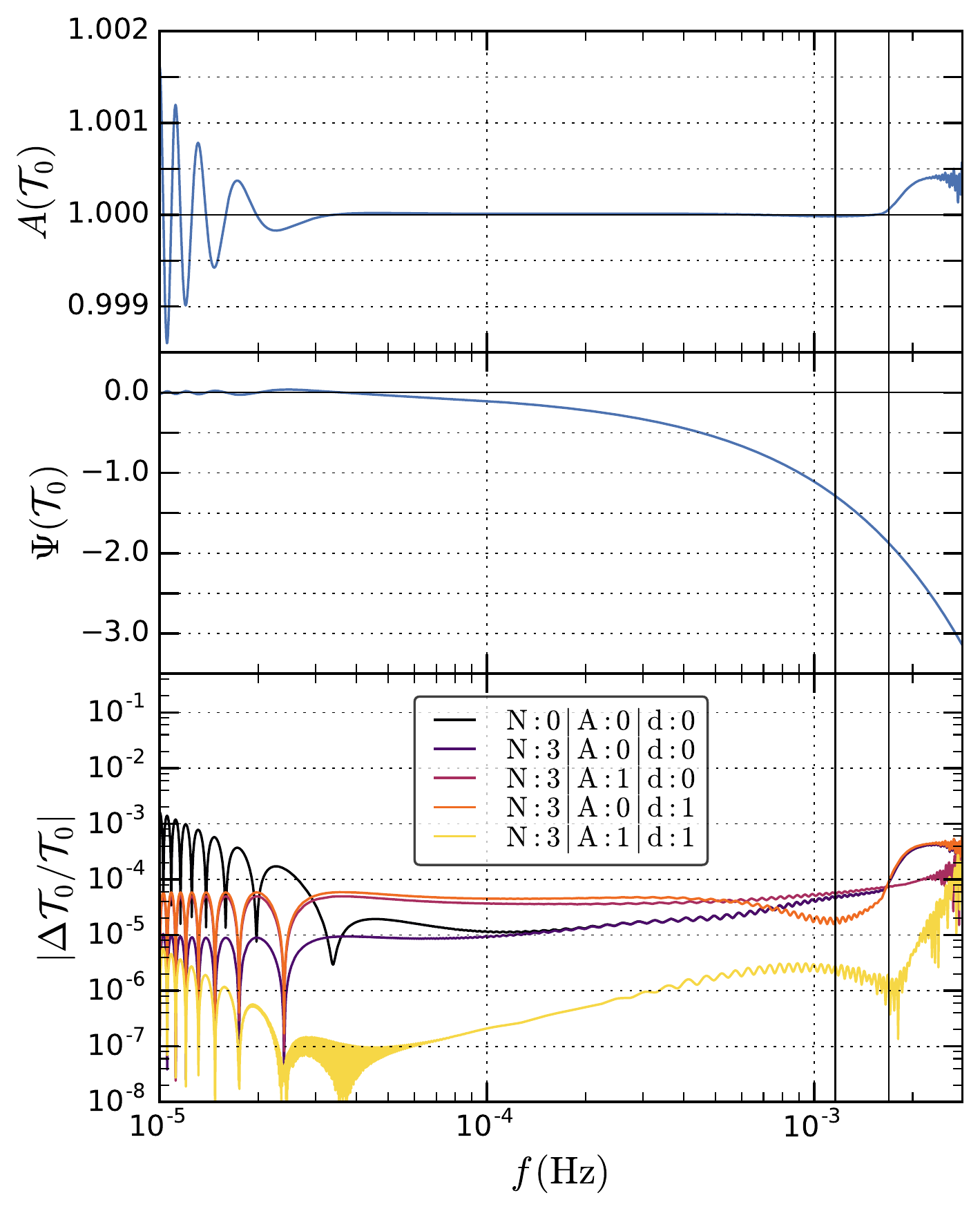}
  \caption{Transfer function and residual for the orbital response~\eqref{eq:defresponse0}, for an equal-mass system with $M=10^{7} \Msol$. The first and second panels show the amplitude and phase of the transfer function. The bottom panel shows the relative modulus of the residual of the perturbative treatment~\eqref{eq:summaryNAd}, with the color indicating the order of the approximation, compared to a numerical FFT. The starting frequency is set by the LISA sensitivity band at $10^{-5}\mathrm{Hz}$, corresponding to $3.8\mathrm{yr}$. The thick and thin vertical lines indicate the merger and ringdown frequencies.}
  \label{fig:LISAerrorM1e7orb}
\end{figure}

\begin{figure}
  \centering
  \includegraphics[width=.98\linewidth]{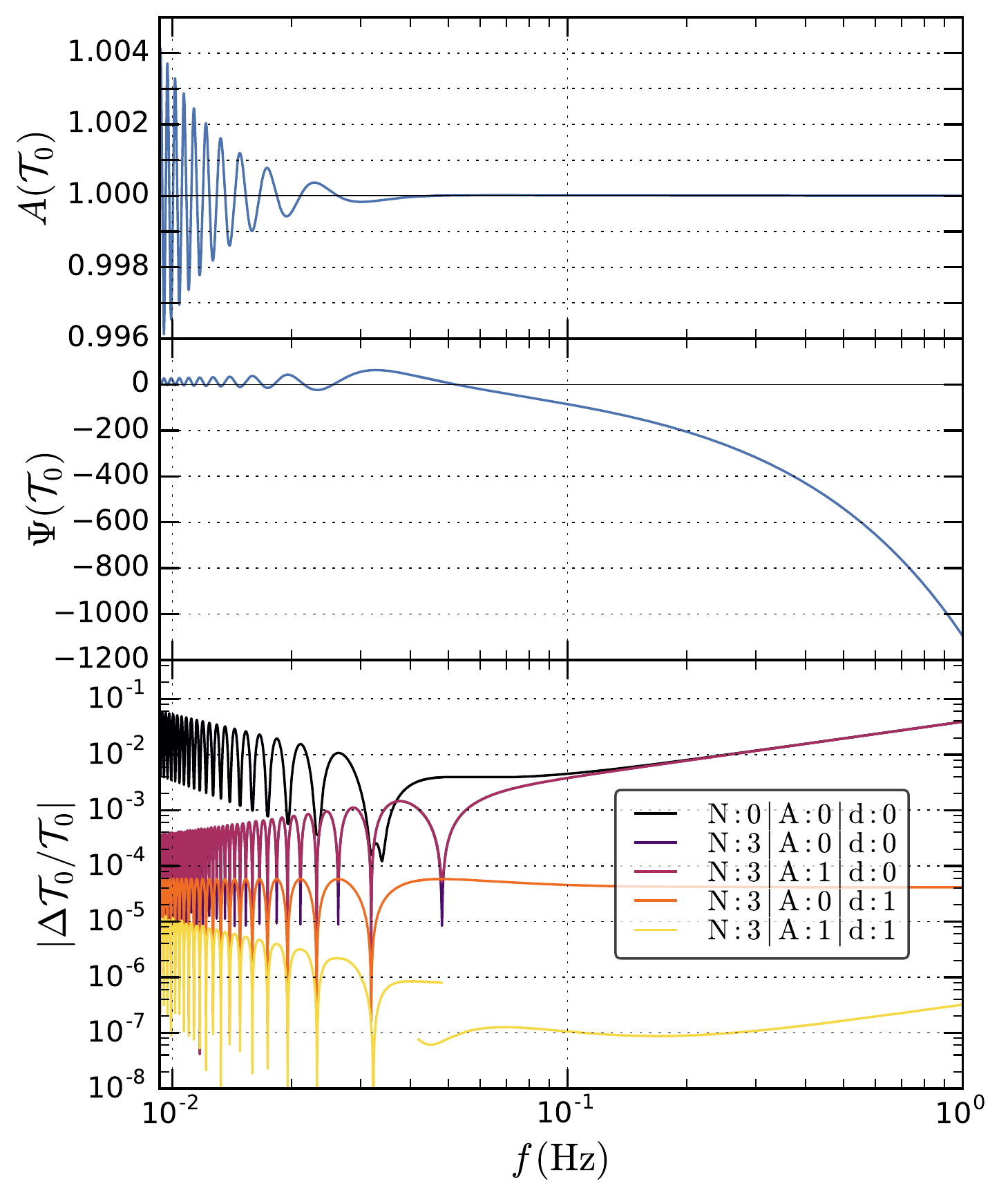}
  \caption{Transfer function and residual for the orbital response~\eqref{eq:defresponse0}, for an equal-mass system with $M=10^{2} \Msol$. The panels are as in Fig.~\ref{fig:LISAerrorM1e7orb}. The starting frequency corresponds to 10 years before merger. The ending frequency is set by the LISA sensitivity band at $1\mathrm{Hz}$, and the merger is out of band. The jump in the residuals for $\{N:3|A:1|d:1\}$ is due to the fact that we split the band in two, with two different sampling rates, for the FFT; it shows that, at this level, the details of the conditioning for the FFT start to cause numerical errors in the reference transfer function.}
  \label{fig:LISAerrorM1e2orb}
\end{figure}

The results for the orbital response~\eqref{eq:defresponse0} are shown in Figs.~\ref{fig:LISAerrorM1e7orb} and~\ref{fig:LISAerrorM1e2orb} in amplitude and phase form. In both cases, in accordance with~\eqref{eq:transfer0local}, the transfer function is essentially a phase, and contains many more cycles in the low-mass case in the higher frequency band. The leading-order treatment is accurate at $10^{-3}$ in the high-mass case, while in the low-mass case errors reach $10\%$ at the start of observations and a few percent at high frequencies. In both cases, including higher-order corrections does reduce the reconstruction errors, with the distinctive feature that including only one of the amplitude and delay corrections can make the error actually worse, while including both does improve the accuracy down to $10^{-5}$ or better.

\begin{figure}
  \centering
  \includegraphics[width=.98\linewidth]{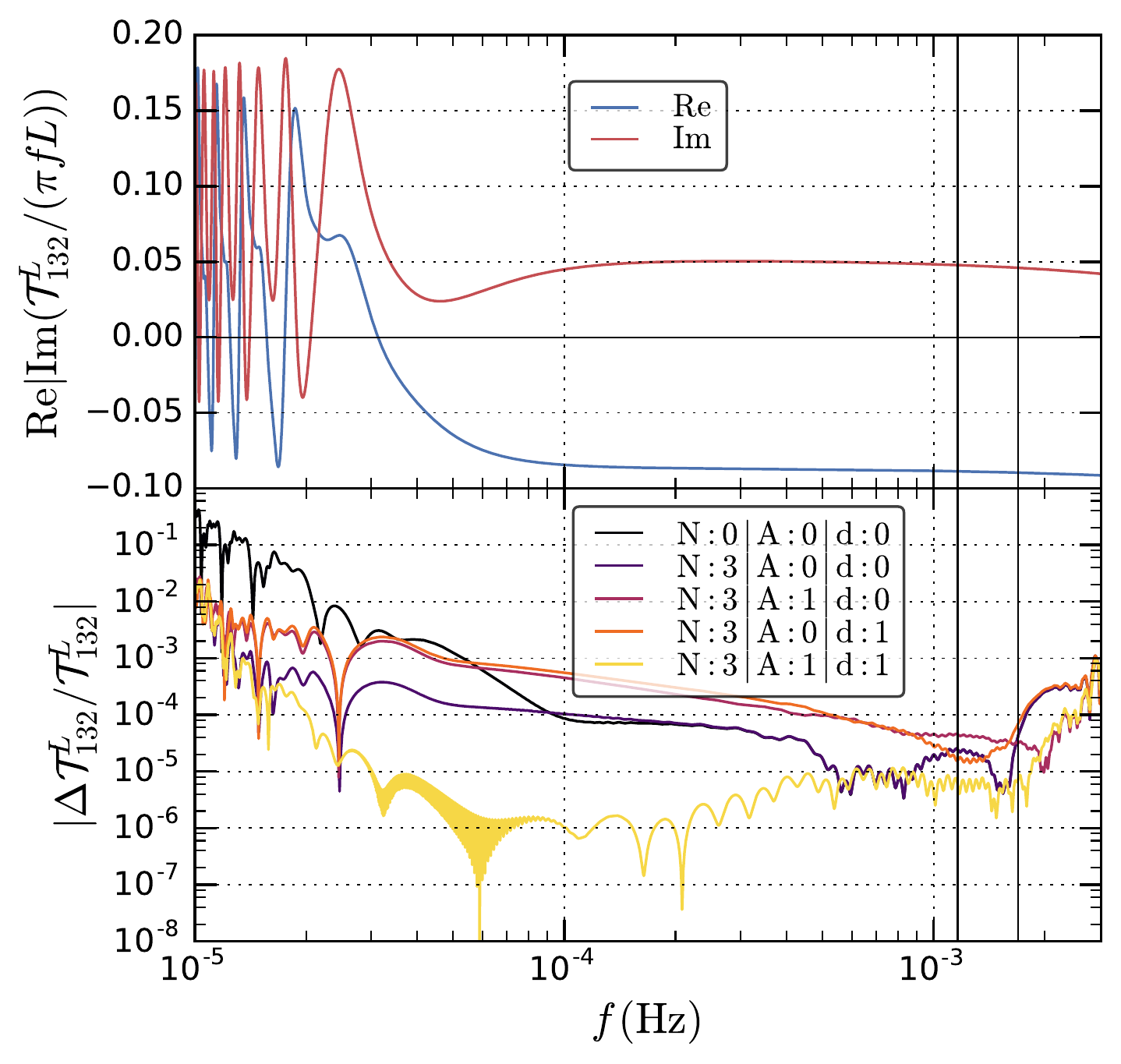}
  \caption{Transfer function and residual for the constellation response $y_{132}$~\eqref{eq:defresponseL}, for an equal-mass system with $M=10^{7} \Msol$. The top panel shows the real and imaginary parts of the transfer function, rescaled by the overall scaling $\pi f L$ of the response at low frequencies, while the bottom panel shows the relative residuals compared to an FFT. The frequencies are the same as in Fig.~\ref{fig:LISAerrorM1e7orb}.}
  \label{fig:LISAerrorM1e7const}
\end{figure}

\begin{figure}
  \centering
  \includegraphics[width=.98\linewidth]{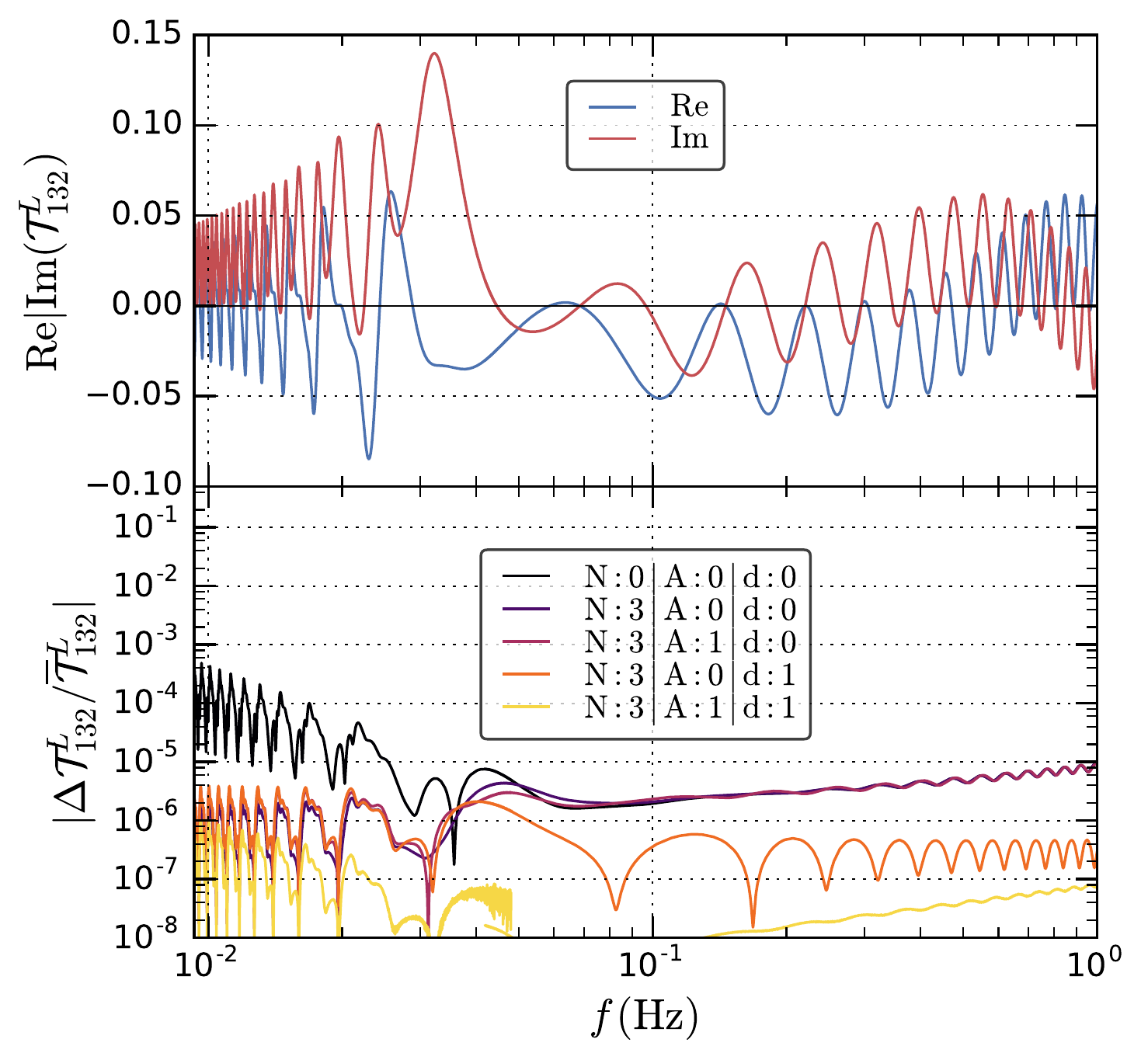}
  \caption{Transfer function and residual for the constellation response $y_{132}$~\eqref{eq:defresponseL}, for an equal-mass system with $M=10^{2} \Msol$. The top panel shows the real and imaginary parts of the transfer function, without the rescaling of Fig.~\ref{fig:LISAerrorM1e7const}. The bottom panels shows the residuals normalized by~\eqref{eq:transferLenvelope} to avoid zero-crossings. The frequencies are the same as in Fig.~\ref{fig:LISAerrorM1e2orb}, and the discontinuity in the residuals for $\{N:3|A:1|d:1\}$ is again due to the conditioning of the FFT.}
  \label{fig:LISAerrorM1e2const}
\end{figure}

For the constellation response~\eqref{eq:defresponseL}, we show the results in Figs.~\ref{fig:LISAerrorM1e7const} and~\ref{fig:LISAerrorM1e2const}, displaying the real and imaginary part of the transfer function. In the high-mass case, we rescale the response by the overall scaling $\pi f L$ in~\eqref{eq:GslrL}. The leading-order treatment reaches an inaccuracy of $\sim 10\%$ at the lowest frequencies of the high-mass case, while it is better than $10^{-3}$ for the low-mass case. Similarly to the orbital response, including all higher-corrections reduces the errors to better than $10^{-5}$, with the only exception of the high-mass case showing errors rising at the low-frequency end and beyond the ringdown frequency.

\subsection{Slowly chirping binaries and direct approach for periodic modulations and delays}\label{subsec:comblisa}

\begin{figure}
  \centering
  \includegraphics[width=.98\linewidth]{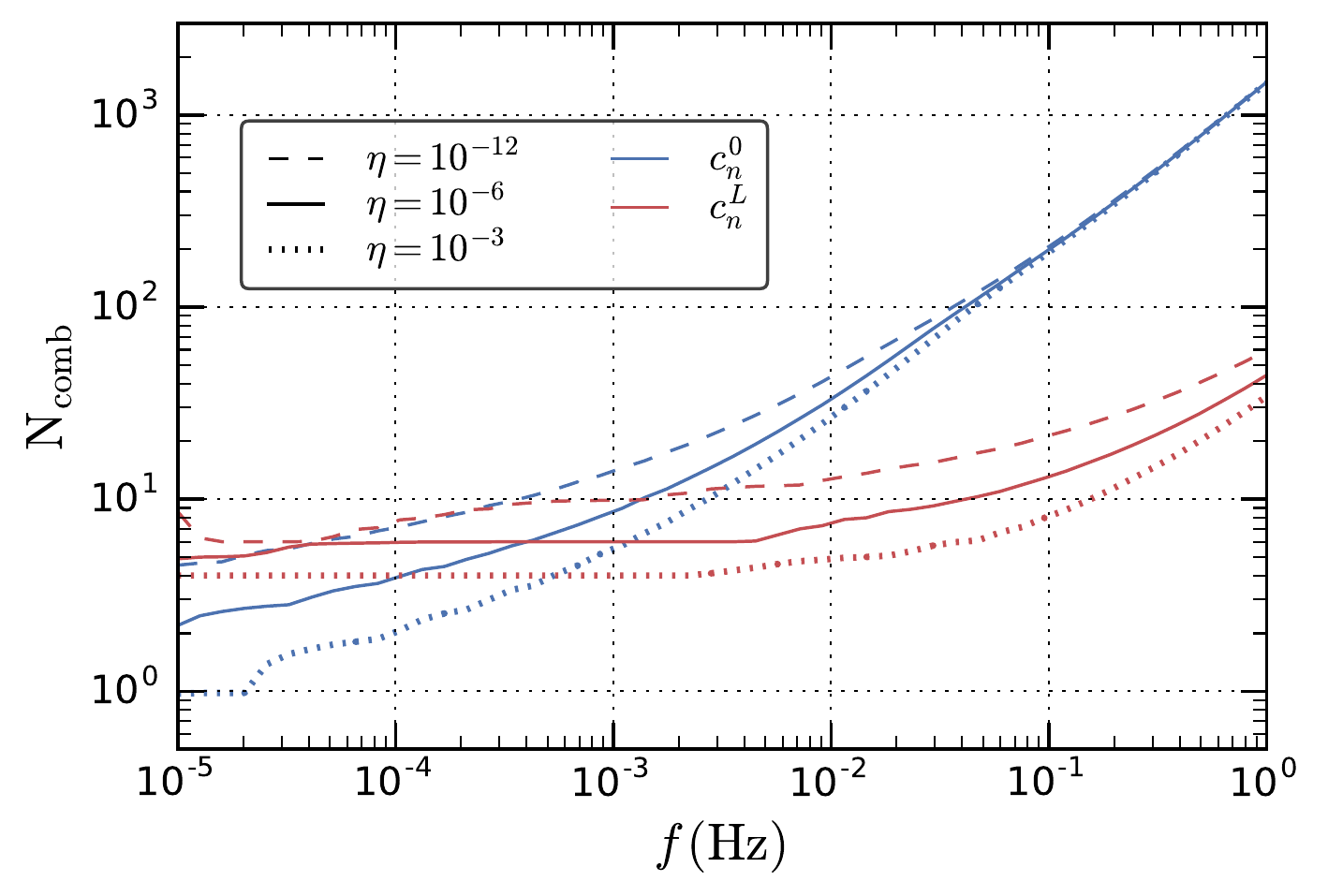}
  \caption{Extent of the Fourier-domain comb entering the convolution~\eqref{eq:transferdiscreteconvolution}, determined from the criterion~\eqref{eq:criteriontruncationcomb} with truncation levels $\eta=10^{-12}$, $10^{-6}$, $10^{-3}$. Blue corresponds to the orbital delay comb coefficients $c_{n}^{0}$ given by~\eqref{eq:cn0}, and red to the constellation comb $c_{n}^{L}$ for the response~\eqref{eq:GslrL}. The result is shown after averaging over 100 random orientation angles. Most of the slowly-chirping systems considered here will have $f \lesssim 10^{-2}\Hz$.}
  \label{fig:lisacombextent}
\end{figure}

\begin{figure}
  \centering
  \includegraphics[width=.98\linewidth]{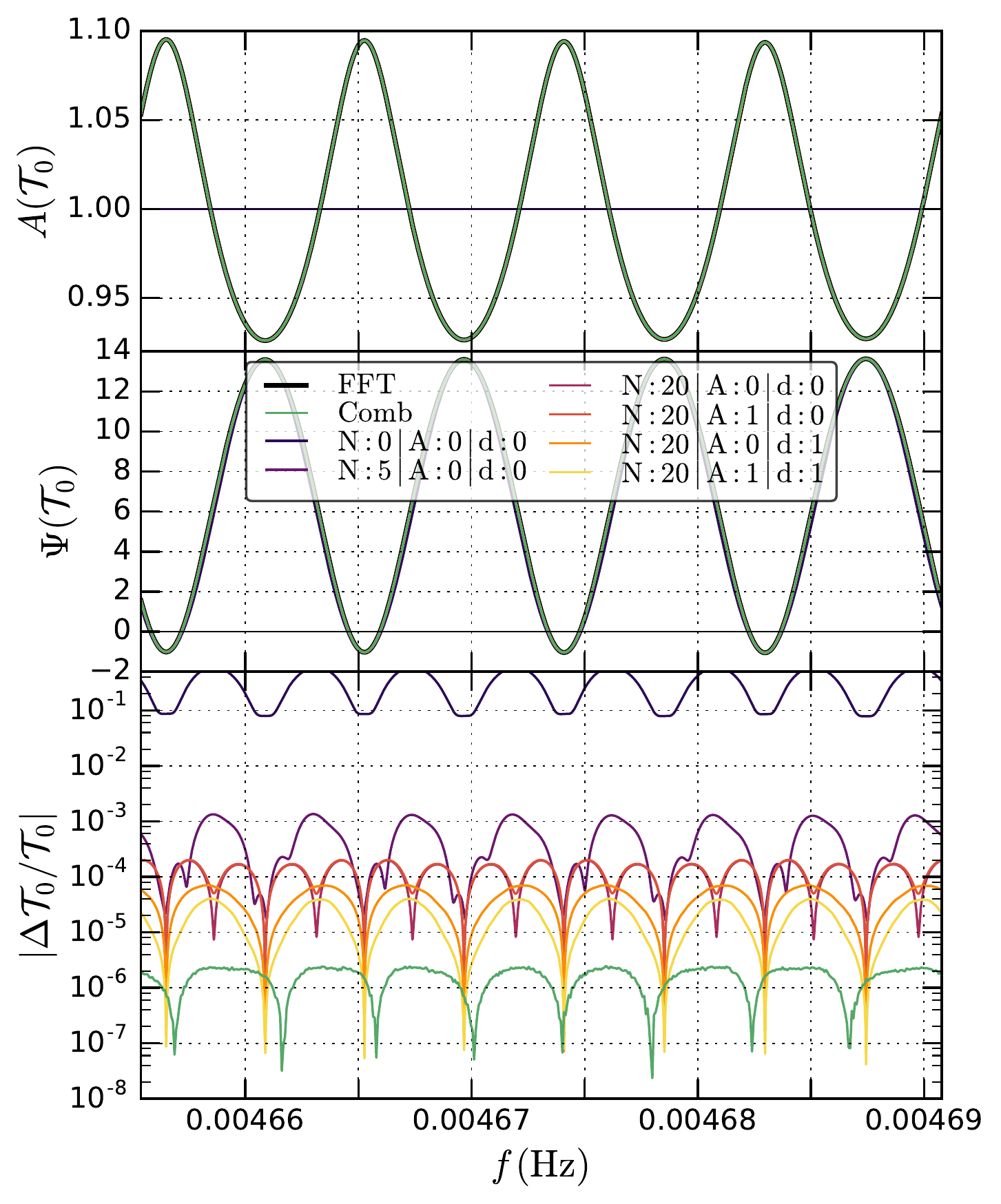}
  \caption{Transfer function and residual for the orbital delay $d_{0}$~\eqref{eq:delay0}, for an SOBH with $\epsilon_{\Psi 2}\sim 0.6$ for the orbital response. The frequency band corresponds to 4 years of observation for this system with $M=50\Msol$, $\Delta t = 200\yr$ away from merger. The top panels show the transfer function amplitude and phase at various orders of approximation in~\eqref{eq:summaryNAd}, with the numerical result of the FFT in black and the comb result~\eqref{eq:transferdiscreteconvolution} in green. The bottom panel shows the relative residuals with respect ot the FFT.}
  \label{fig:LISAerrorSOBHepsPsi1d0}
\end{figure}

\begin{figure}
  \centering
  \includegraphics[width=.98\linewidth]{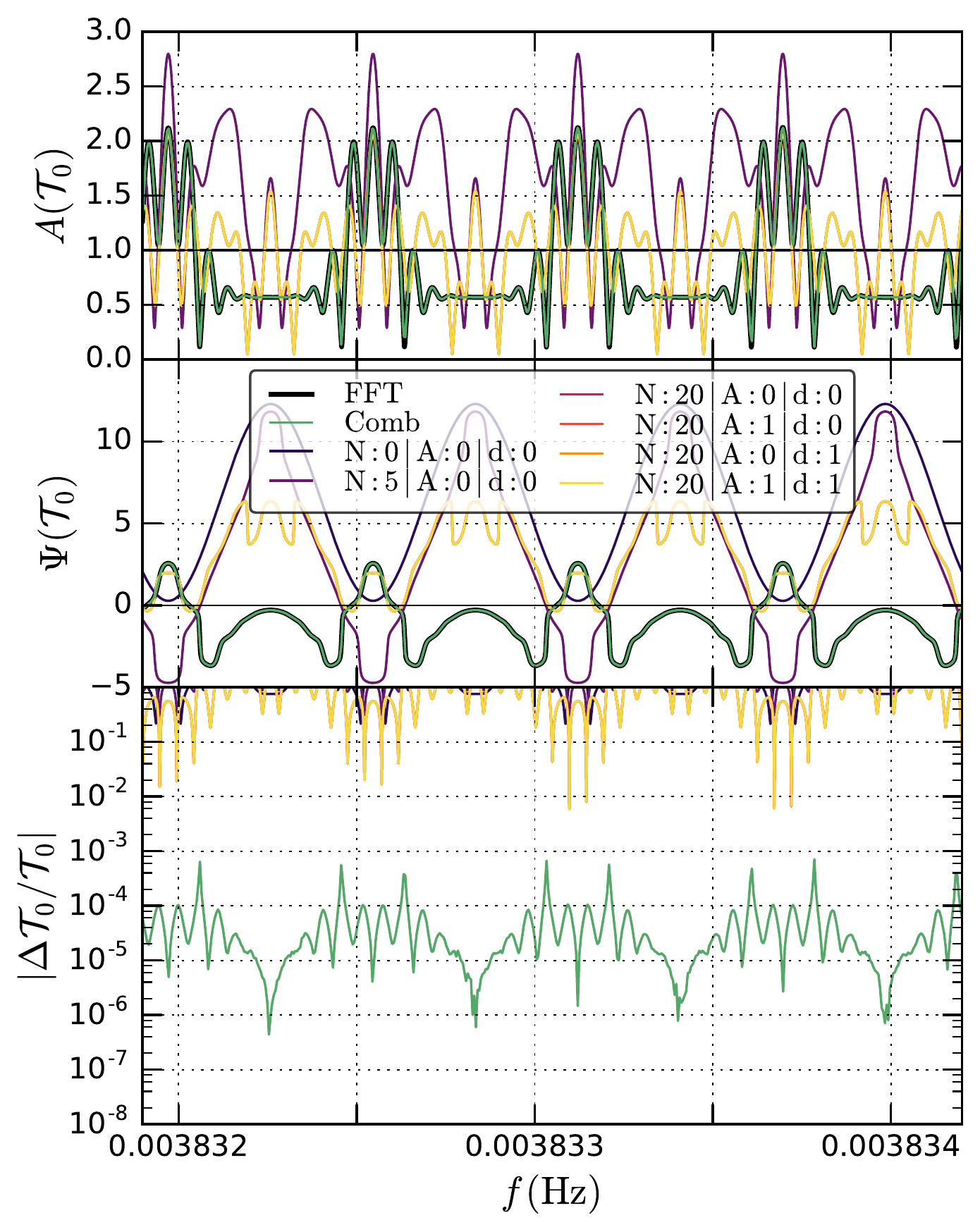}
  \caption{Same as Fig.~\ref{fig:LISAerrorSOBHepsPsi1d0}, but for an SOBH with $\epsilon_{\Psi 2}\sim 4.5$ for the orbital response. The frequency band corresponds to 4 years of observation for this system with $M=15\Msol$, $\Delta t = 1500\yr$ away from merger. The perturbative treatment breaks down in this case, with errors of order 1.}
  \label{fig:LISAerrorSOBHepsPsi10d0}
\end{figure}

\begin{figure}
  \centering
  \includegraphics[width=.98\linewidth]{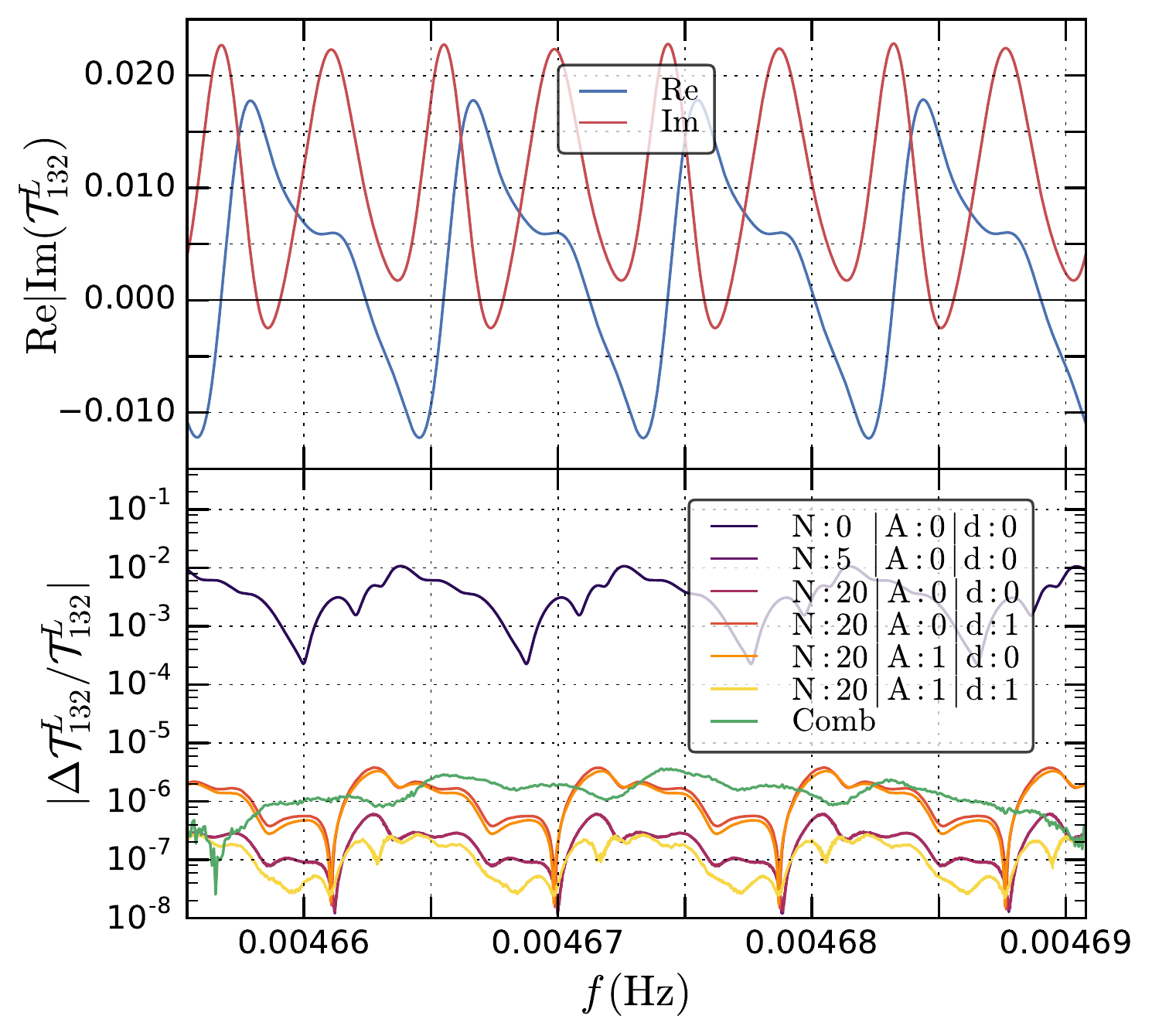}
  \caption{Transfer function and residual for the constellation response $y_{132}$~\eqref{eq:GslrL}, for the same SOBH as in Fig.~\ref{fig:LISAerrorSOBHepsPsi1d0}. Here $\epsilon_{\Psi 2}\sim 0.01$ for the constellation part of the response. The top panel shows the real and imaginary part of the transfer function.}
  \label{fig:LISAerrorSOBHepsPsi1const}
\end{figure}

\begin{figure}
  \centering
  \includegraphics[width=.98\linewidth]{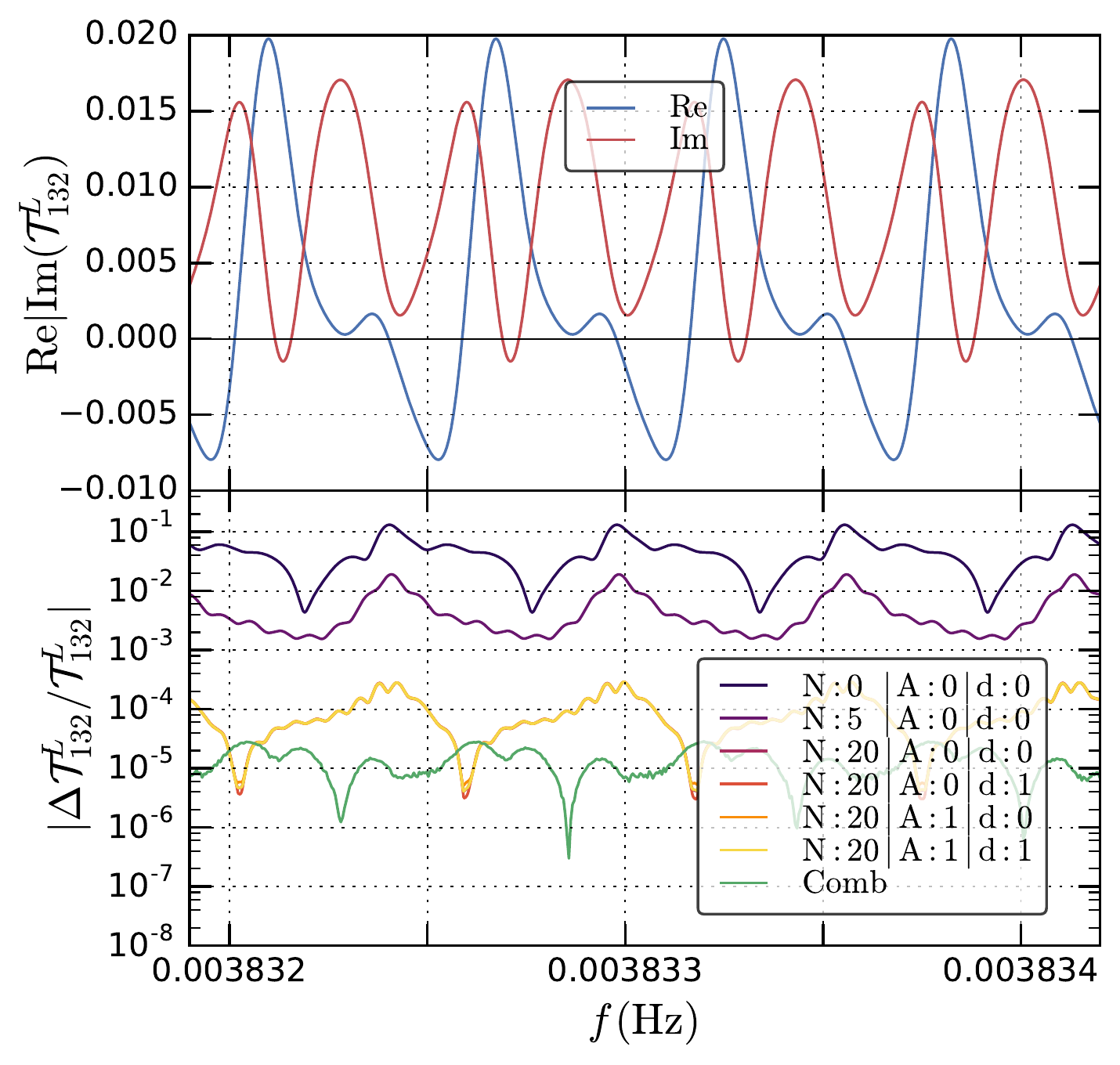}
  \caption{Transfer function and residual for the constellation response $y_{132}$~\eqref{eq:GslrL}, for the same SOBH as in Fig.~\ref{fig:LISAerrorSOBHepsPsi10d0}. Here $\epsilon_{\Psi 2}\sim 0.1$ for the constellation part of the response. The top panel shows the real and imaginary part of the transfer function.}
  \label{fig:LISAerrorSOBHepsPsi10const}
\end{figure}

As explained in Sec.~\ref{subsec:lisafom}, slowly-chirping binaries can be problematic for the perturbative formalism of Sec.~\ref{sec:formalism}. Indeed, when far enough from merger, SOBH systems~\cite{Sesana16} can have error estimates $\epsilon_{\Psi 2}$ reaching and exceeding $1$, as shown in Fig.~\ref{fig:lisafomPsiMcDeltat}. Here we investigate these sources in more details, and propose an alternative treatment for their instrument response.

First, we should mention that these signals share similarities with the galactic binaries that will provide numerous quasi-monochromatic signals in the LISA band~\cite{Nelemans+01, LISA17}. They are far from merger, slowly chirping, and span only a narrow frequency band over the course of the LISA mission lifetime, although being less monochromatic that galactic binaries. For the galactic binaries, an accurate and efficient numerical treatment of the response has been proposed and widely used in applications~\cite{CL07}. This treatment is referred to as the fast-slow decomposition, or heterodyning approach. If the gravitational wave signal extends only on the narrow frequency band $f \in [f_{*}, (1+\eta) f_{*}]$ with $\eta \ll 1$, then scaling out a carrier frequency of the signal by multiplying by $e^{-2 i \pi f_{*} t}$ will eliminate most of the time variability, allowing to process the signal through the time-domain response and to take a FFT with a Nyquist frequency shifted from $(1+\eta) f_{*}$ to $\eta f_{*}$, i.e. with a much smaller number of samples. This multiplication is simply equivalent to a shift in frequency domain, which can be restored after the numerical FFT has been computed. The efficiency of this approach is contingent to the smallness of $\eta$. Galactic binaries are extremely close to monochromatic~\cite{Nelemans+01}, and this quantity can be as low as $10^{-6}-10^{-7}$. For SOBH systems, however, $\eta$ will take a continuous set of values from roughly $10^{-4}$ to $5$. Although the systems for which $\eta$ is the largest are also the easiest to treat with the perturbative formalism of Sec~\ref{sec:formalism}, we propose here yet a third method, based on a discrete Fourier comb, making sure we can cover the intermediate ground of slowly-chirping systems with a large $\eta$. We leave for the future a more detailed study of SOBH systems as a population, and the investigation of the precise boundaries and overlap areas of the three methods (heterodyning, Fourier comb, perturbative) as well as their respective computational costs.

The Fourier comb approach we propose here exploits the fact that, in the LISA case, the modulations and delays entering~\eqref{eq:yslr} are periodic, with a period of one year and a frequency $f_{0} = \Omega_{0}/2\pi = 1/\mathrm{yr} \simeq 3.2\times 10^{-8}\mathrm{Hz}$. For any given frequency $f$, $G(f,t)$ is periodic in time, so that~\eqref{eq:defG} becomes a discrete Fourier series
\be\label{eq:Gdiscretefourier}
	G(f,t) = \sum_{n \in \mathbb{Z}} c_{n}(f) e^{-in\Omega_{0}t} \,,
\ee
with frequency-dependent discrete Fourier coefficients (that we will also call comb coefficients) given by the integrals (we recall our Fourier convention~\eqref{eq:defFT})
\be\label{eq:defcn}
	c_{n}(f) = \frac{\Omega_{0}}{2\pi} \int_{0}^{\frac{2\pi}{\Omega_{0}}} \ud t \, e^{i n \Omega_{0} t} G(f,t) \,.
\ee
For slowly-chirping systems, the orbital-delay part of the response gives a transfer function that has significant structure, by contrast with fast-chirping systems where it reduces essentially to an extra phase contribution. Thus, in practice this approach is to be applied to $G(f,t)$ representing the full response, orbital delay and constellation modulation.

For illustration purposes, however, we will keep the orbital and constellation response separated in the following. In the case of the orbital delay, with $G_{0}(f,t)$ given by~\eqref{eq:G0} and~\eqref{eq:delay0}, the particularly simple expression of the delay gives an analytic expression for the coefficients $c_{n}^{0}$ in terms of Bessel functions of the first kind, as
\be\label{eq:cn0}
	c_{n}^{0} = i^{n} e^{i n \lambda} J_{n} \left[ -2 \pi f R \cos \beta \right]\,.
\ee
By contrast, for the constellation response (or for the full response), to our knowledge the coefficients $c_{n}^{L}$ do not admit such a simple close-form expression, and they must be computed numerically using the integrand given by~\eqref{eq:GslrL}. Truncating~\eqref{eq:Gdiscretefourier} to a finite order $N$, this computation reduces to an FFT and the $c_{n}$ coefficients are given by~\eqref{eq:ykDFT} and~\eqref{eq:ckyk}.

Inserting~\eqref{eq:Gdiscretefourier} into the convolution~\eqref{eq:FDkernel} leads to the following generalized discrete convolution:
\be\label{eq:transferdiscreteconvolution}
	\tilde{s}(f) = \sum_{n \in \mathbb{Z}} c_{n}(f - n f_{0}) \tilde{h} (f - n f_{0}) \,.
\ee
Thus, computing the Fourier-domain response now requires to convolve the signal with a discrete comb with frequency-dependent coefficients $c_{n}$. In practice, this sum is to be truncated at $|n|\leq N_{\rm comb}$, for some finite order $N_{\rm comb}$ determining the accuracy of the approximation. To assess the expected truncation error, we use a simple criterion based on the $L^{1}$-norm of the $\{c_{n}\}$ sequence. For a given target truncation error $\eta$, we define the truncation $N_{\rm comb}(f, \eta)$ as the smallest integer such that
\be\label{eq:criteriontruncationcomb}
	\sum_{|n| > N(f, \eta)} |c_{n}(f)| < \eta \,.
\ee
The truncation order $N_{\rm comb}(f,\eta)$ is frequency-dependent, and also depends on the orientation angles. For $c_{n}^{0}$, the expression~\eqref{eq:cn0} can provide an asymptotic bound on the width of the comb. For large values of $n$ we have indeed the equivalent $| J_{n}[z] | \sim (e z/2n)^{n}/\sqrt{2\pi n} \; \left[ n\rightarrow + \infty \right]$ (see (10.19) in~\cite{DLMF}), so that a conservative estimate for the truncation order (almost independent of $\eta$) is given by $N_{\rm comb} \geq  |e \pi f R \cos \beta|$. Fig.~\ref{fig:lisacombextent} shows numerical computations for $N_{\rm comb}(f,\eta)$, averaged over orientation angles, for both the constellation and the orbital response and for the truncation levels  $\eta=10^{-12}$, $10^{-6}$, $10^{-3}$. We see that, especially at high frequencies, the orbital response requires more coefficients than the constellation response due to the longer baseline of the delay. The majority of systems for which we wish to apply the comb method will have frequencies $\lesssim 10^{-2}\Hz$, where for the orbital response $N_{\rm comb} \simeq 40$.

An important point is that, since the signals from slowly-chirping binaries extend on a narrow frequency band, for both responses the frequency-dependence of $c_{n}^{0}(f)$, $c_{n}^{L}(f)$ will be very mild, so that the coefficients can be computed at two or three frequencies and interpolated in-between. The computational cost of this approach is then set by the convolution~\eqref{eq:transferdiscreteconvolution} that must be evaluated at as many frequencies as necessary to interpolate the transfer functions as functions of frequency (for examples, see Figs.~\ref{fig:LISAerrorSOBHepsPsi1d0}, \ref{fig:LISAerrorSOBHepsPsi10d0}, \ref{fig:LISAerrorSOBHepsPsi1const} and~\ref{fig:LISAerrorSOBHepsPsi10const}). In general, this Fourier comb approach will be more expensive than the perturbative response of Sec.~\ref{subsec:executivesummary}. We leave for future work a proper assessment of the computational performance of this method compared to the others.

We now illustrate this approach by considering two examples of equal-mass SOBH systems. The first has $M = 50\Msol$ and is $\Delta t = 200\yr$ away from merger, and for the orbital response $\epsilon_{\Psi 2} \simeq 0.6$. The second has $M = 15\Msol$ and is $\Delta t = 1500\yr$ away from merger, and for the orbital response $\epsilon_{\Psi 2} \simeq 4.5$. The phase error measures for the constellation response are smaller, $\epsilon_{\Psi_{2}} \simeq 0.1$ and $0.01$ respectively. In keeping with the previous section, in our presentation we separate the orbital and the constellation response, keeping in mind that in practice the full response is to be handled in one step.

The transfer functions for the orbital response and residual errors are shown for the two systems in Figs.~\ref{fig:LISAerrorSOBHepsPsi1d0} and~\ref{fig:LISAerrorSOBHepsPsi10d0}. We see that, contrarily to Sec.~\ref{subsec:errorsLISA}, the transfer function starts to develop more structure that just a phase contribution. The case $M=50\Msol$ with $\epsilon_{\Psi 2} \simeq 0.6$ shows that the leading-order treatment leads to relative errors of order $10\%$, while increasing $N$ to 5 or 20 brings the errors back to $\sim 10^{-4}$, with a marginal improvement from further amplitude and delay corrections. The case $M=50\Msol$ with $\epsilon_{\Psi 2} \simeq 4.5$, by contrast, is clearly outside the range of applicability of the perturbative formalism, as all orders of approximation give errors of order $100\%$. In both cases, the comb treatment performs well, at better than $10^{-4}$.

For the constellation response, the transfer functions and residual errors are shown for the two systems in Figs.~\ref{fig:LISAerrorSOBHepsPsi1const} and~\ref{fig:LISAerrorSOBHepsPsi10const}. Here, with $\epsilon_{\Psi 2} \sim 0.01$ and $0.1$, both systems are within reach of the perturbative formalism, although the case  $\epsilon_{\Psi 2} \sim 0.1$ shows errors of $10\%$ at leading order and requires a rather large $N=20$ to reach errors below $10^{-3}$. The comb treatment yields again errors below $10^{-4}$ in both cases.

Thus, we have shown that some slowly-chirping systems will be out of reach of the perturbative treatment of Sec.~\ref{subsec:executivesummary}, and we have demonstrated that the Fourier comb approach presented here could be applied to these systems with a good accuracy.


\section{Application to waveforms from precessing binaries}
\label{sec:precession}

In this Section, we illustrate the application of the formalism described in~\ref{sec:formalism} to signals from precessing binaries, with modulations induced by the precession. We investigate how the separation of the relevant timescales evolves in the late inspiral and after the merger occurs. We will also give in App.~\ref{app:precpreviousapproaches} a short overview of previous approaches to this problem of the Fourier-domain precession, and how they relate to our formalism.


\subsection{Precession and frame decomposition}
\label{subsec:precdef}

As we already discussed in Sec.~\ref{subsec:modulationPrec}, in the presence of spin components that are not aligned with the orbital angular momentum, an inspiraling binary system will undergo precession of its orbital plane~\cite{Apostolatos+94, Kidder95}. This is a crucial effect to be taken into account in the modelling of such signals.

As proposed by several authors~\cite{BCV03b, BCPTV05, Schmidt+10, OShaughnessy+11, Boyle+11}, if one performs the mode decomposition of the waveform~\eqref{eq:defmodes} not in a fixed inertial frame but rather in a time-dependent, rotating frame that follows the plane of the orbit, it is possible to restore much of the structure of a non-precessing waveform. In particular, one recovers qualitatively the hierarchy of mode amplitudes that prevails for non-precessing systems, with the modes $h_{22}$ and $h_{2,-2}$ being dominant, and with each harmonic mode characterized by a chirp with smoothly evolving amplitude and phase. In the following, we will identify the $z$-axis of the co-precessing frame as the dominant eigenvector of the matrix representing the action of the angular momentum operator acting on the waveform modes, as proposed in~\cite{OShaughnessy+11} (see App.~\ref{app:wigner} for more details).

We define $(\alpha, \beta, \gamma)$ as the Euler angles of the active rotation from the inertial frame to the precessing frame, in the $(z,y,z)$ convention. The first two angles, $\alpha$ and $\beta$, are the two spherical angles tracking the direction of the radiation axis, which during the inspiral follows essentially the normal to the orbital plane. The last angle $\gamma$ parametrizes the remaining freedom of rotation around this radiation axis. To fix this third degree of freedom, we use the minimal rotation condition~\cite{Boyle+11}, enforcing the absence of rotation of the precessing frame around the radiation axis. In terms of Euler angles, this condition translates to
\be\label{eq:gammadot}
	\dot{\gamma} = -\dot{\alpha}\cos \beta \,.
\ee
A natural choice for the $z$ axis of the inertial frame is the direction of $\bm{J}$, the total angular momentum, which is almost constant\footnote{Except in the cases of transitional precession~\cite{Apostolatos+94}, where for high mass ratios and large antialigned spins the orbital angular momentum and the spin can almost cancel each other.}. Quite generically, and in particular in the case of simple precession~\cite{Apostolatos+94, Kidder95} but also in the model we will take for the post-merger precession, the radiation axis precesses on a cone around the direction of $\bm{J}$, with $\alpha$ increasing, while $\beta$, the opening angle of the precession cone, is slowly varying.

The modes in the inertial frame $h_{\ell m}^{\rm I}$ and the modes in the precessing frame $h_{\ell m}^{\rm P}$ are then related by~\cite{Goldberg+67}
\begin{subequations}
\label{eq:wignerrot}
\begin{align}
	h_{\ell m}^{\rm I} = \sum\limits_{m=-\ell}^{\ell} \calD^{\ell *}_{mm'} (\alpha,\beta,\gamma) h_{\ell m'}^{\rm P} \,, \\
	h_{\ell m}^{\rm P} = \sum\limits_{m=-\ell}^{\ell} \calD^{\ell }_{m'm} (\alpha,\beta,\gamma) h_{\ell m'}^{\rm I} \,.
\end{align}
\end{subequations}
Notice that there is no mixing of the modes with different values of $\ell$. Here the coefficients $\calD^{\ell}_{mm'}$ are given by Wigner D-matrices~\cite{Wigner59} as
\be\label{eq:defWignerD}
	\calD^{\ell}_{mm'} (\alpha, \beta, \gamma) = e^{im \alpha} d^{\ell}_{mm'}(\beta) e^{im' \gamma}\,.
\ee
Here, the real-valued Wigner d-matrix $d^{\ell}_{mm'}(\beta)$ takes the form of a polynomial in $\cos (\beta/2)$, $\sin (\beta/2)$, and acts as an amplitude for the modulation function. We refer to App.~\ref{app:wigner} for explicit expressions.

In the following, our objective will be to compute mode-by-mode transfer functions $\calT^{\ell}_{mm'}$, defined as
\be\label{eq:defprectransfer}
	\mathrm{FT} \left[ \calD^{\ell *}_{mm'} (\alpha,\beta,\gamma) h_{\ell m'}^{\rm P} \right] (f) \equiv \calT^{\ell}_{mm'}(f) \tilde{h}_{\ell m'}^{\rm P} (f)
\ee
such that the complete Fourier-domain inertial-frame waveform will be given as the sum of these individual mode contributions,
\be\label{eq:defprechIsum}
	\tilde{h}_{\ell m}^{\rm I} (f) = \sum\limits_{m'=-\ell}^{\ell} \calT^{\ell}_{mm'}(f) \tilde{h}_{\ell m'}^{\rm P} (f) \,.
\ee
In the notation that we used in Sec.~\ref{sec:formalism}, in the absence delays, we wish to compute the convolutions
\be\label{eq:precconvolution}
	\tilde{s} (f) = \int df' \; \tilde{F}(f') \tilde{h} (f-f') \,,
\ee
with $\tilde{s}(f)$ being one mode contribution to $\tilde{h}_{\ell m}^{\rm I} (f)$ in~\eqref{eq:defprechIsum}, $\tilde{h}$ being one of the P-frame modes $\tilde{h}_{\ell m'}^{\rm P} (f)$, and the modulation $F(t)$ being one of the time-dependent Wigner matrices $\calD^{\ell *}_{mm'}$.

An important qualitative observation is that the opening angle $\beta$ is typically small. Large misalignments between $\bm{J}$ and $\bm{\ell}$ can only be reached with large spins and large mass ratios. When $\beta$ is small, from the explicit expression~\eqref{eq:defWignerdapp}, one can see that $d^{\ell}_{mm'}(\beta)$ is greatly suppressed in the limit $\beta \rightarrow 0$ when increasing $|m-m'|$. Intuitively, this means that, in the limit of a small misalignment between frames, the rotation produces mainly mode contributions with the same mode number. Additionally, when $\beta$ is constant or slowly varying, \eqref{eq:gammadot} gives $\gamma \simeq - \alpha \cos\beta$, and we have in that case
\be\label{eq:wignerphasesimpleprec}
	\calD^{\ell *}_{mm'} (\alpha, \beta, \gamma) \propto e^{i(m' \cos\beta - m) \alpha} \,,
\ee
with $\cos\beta \simeq 1$ for small $\beta$. This shows that the modulation for $m=m'$ has a suppressed phase, while increasing $|m-m'|$ increases the magnitude of the modulation phase. Thus, modulation functions for distant mode contributions (for instance from the $h^{\rm P}_{22}$ mode all the way to the $h^{\rm I}_{2,-2}$ mode) have larger phase evolutions, but smaller amplitudes.

The relations~\eqref{eq:wignerrot} do not mix different values of $\ell$. Furthermore, rotations leave invariant the combined square amplitude $\calA_{\ell}$ for each $\ell$, as well as the total square amplitude:
\be\label{eq:defsumamplitude}
	\calA^{2} = \sum\limits_{\ell \geq 2}\calA_{\ell}^{2} = \sum\limits_{\ell \geq 2}\sum\limits_{m=-\ell}^{\ell} |h_{\ell m}|^{2} \,.
\ee
One can therefore use these amplitudes (in our case, limited to $\ell = 2$) to define a frame-independent peak amplitude of the precessing waveform.

The dominant features of the the precessing-frame waveform are approximately reflection symmetric about the orbital plane (exactly valid for non-precessing systems, see~\eqref{eq:symmetryhlminusm}), implying
\be\label{eq:approxsymmetryhlminusm}
	h^{\rm P}_{\ell,-m} \simeq (-1)^{\ell} h^{\rm P*}_{\ell,m} \,.
\ee
Similarly, in the Fourier domain, one can neglect either the negative or positive frequency band in the Fourier transform of precessing-frame modes (as in~\eqref{eq:zeronegativef})
\begin{align}\label{eq:approxzeronegativef}
	\tilde{h}_{\ell m}^{\rm P} (f) &\simeq 0 \text{ for } f<0, \; m>0 \nn\,,\\
	\tilde{h}_{\ell m}^{\rm P} (f) &\simeq 0 \text{ for } f>0, \; m<0 \,,
\end{align}
and neglect altogether the $m=0$ modes, $\tilde{h}_{\ell 0}^{\rm P} (f) \simeq 0$.

When using the approximations~\eqref{eq:approxsymmetryhlminusm}-\eqref{eq:approxzeronegativef}, one can derive a symmetry relation in the transfer functions themselves. From the explicit expression of the Wigner matrices, we have indeed (see App.~\ref{app:wigner})
\be
	\calD^{\ell *}_{-m,-m'} = (-1)^{m+m'}\calD^{\ell}_{mm'} \,.
\ee
Since for a function $g$ we have in general for its conjugate
\be
	\mathrm{FT}[g^{*}](-f) = \tilde{g}(f)^{*} \,,
\ee
we can write, using~\eqref{eq:approxsymmetryhlminusm},
\be
	\mathrm{FT}[\calD^{\ell *}_{mm'} h_{\ell m'}^{\rm P}](-f)^{*} = (-1)^{\ell+m+m'} \mathrm{FT}[\calD^{\ell *}_{-m,-m'} h_{\ell, -m'}^{\rm P}](f) \,,
\ee
or, for transfer functions,
\be
	\calT^{\ell}_{mm'}(f) = (-1)^{m+m'} \calT^{\ell *}_{-m,-m'}(-f) \,.
\ee
This means that such a model is required to cover only the positive frequency band and the values $m'>0$, since
\begin{align}\label{eq:hIlmposnegfreq}
	\tilde{h}^{\rm I}_{\ell m}(f) &= \sum_{m'>0} \calT^{\ell}_{mm'}(f) \tilde{h}^{\rm P}_{\ell, m'}(f) \text{ for } f>0 \,, \nn\\
	\tilde{h}^{\rm I}_{\ell m}(f) &= \sum_{m'>0} (-1)^{\ell+m+m'} \calT^{\ell *}_{-m,m'}(-f) \tilde{h}^{\rm P *}_{\ell, m'}(-f) \text{ for } f<0 \,.
\end{align}

Further simplifications occur when including only the dominant harmonics $h^{\rm P}_{22}$, $h^{\rm P}_{2,-2}$ in the precessing-frame waveform (as is done for instance in PhenomP~\cite{Hannam+13}, as well as in our toy model~\ref{subsec:precmodel}), we have in this case for $f<0$
\be
	\calT^{2}_{m,-2}(f) = (-1)^{m} \calT^{2 *}_{-m,2}(-f) \,,
\ee
which for both $f>0$ and $f<0$ translates into
\be\label{eq:symmetryhIfor22only}
	\tilde{h}^{\rm I}_{2 m}(f) = (-1)^{m} \tilde{h}^{\rm I *}_{2,-m}(-f) \,.
\ee
When reconstructing the polarizations $h_{+},h_{\times}$ according to~\eqref{eq:hpcfrommodes}, we have in this case
\be
	\tilde{h}_{+,\times} (f) = \sum_{m = -2}^{2} L^{+,\times}_{2 m} \tilde{h}^{\rm I}_{2 m}(f) \,,
\ee
where we defined
\begin{align}
	L^{+}_{\ell m} &\equiv \frac{1}{2} \left( {}_{-2}Y_{\ell m} + (-1)^{m} {}_{-2}Y_{\ell, -m}^{*} \right) \,, \nn\\
	L^{\times}_{\ell m} &\equiv \frac{i}{2} \left( {}_{-2}Y_{\ell m} - (-1)^{m} {}_{-2}Y_{\ell, -m}^{*} \right) \,.
\end{align}

Finally, we note that with the restriction to $\ell = 2$ and $m' = 2$, in the approximate phase given in Eq.~\eqref{eq:wignerphasesimpleprec}, the coefficient of $\alpha$ is close to $0$ for small $\beta$ for $m=2$, and increasingly positive when going down from $m=1$ to $m=-2$. In our Fourier convention~\eqref{eq:defFT}, the Fourier transforms $\tilde{F}$ thus have most of their support on negative frequencies. This means that the support of the convolution integral~\eqref{eq:precconvolution} extends to the right side for $\tilde{h}$, i.e. $f-f' > f$. This point will become important when considering the high-frequency part of the signal in Sec.~\ref{subsec:trigopoly}. Note, however, that this statement might not be true anymore when including more modes in our model, in particular modes $\ell \neq m$.

\subsection{Simplified model for precessing IMR waveforms}
\label{subsec:precmodel}

We wish to apply the formalism presented in Sec.~\ref{sec:formalism} to IMR waveforms of precessing binaries, exploring the separation of the timescales involved through the inspiral and merger, and assessing the relevance of the higher-order corrections summarized in~\ref{subsec:executivesummary}. To this end, we will use a simplified model for the precession, allowing a number of simplifications and idealizations.

We want to be able to investigate generic-spin signals with a long inspiral phase, beyond the range covered by numerical relativity simulations, and we want to include the merger and ringdown phase. We will use the following three ingredients for this simplified model:
\begin{itemize}
	\item PhenomD~\cite{Husa+15, Khan+15} for the IMR Fourier-domain waveform in the precessing frame;
	\item SEOBNRv3~\cite{Pan+13, BTB16} for the Euler angles during the inspiral;
	\item an effective extension of the Euler angles post-merger based on~\cite{OShaughnessy+12} (see~\eqref{eq:OmegaframeQNM} below).
\end{itemize}

For the precessing-frame waveform, we make all the simplifying assumptions described above in Sec.~\ref{subsec:precdef}: we approximate it as a non-precessing waveform, further enforcing the approximations~\eqref{eq:approxsymmetryhlminusm} and~\eqref{eq:approxzeronegativef}. Limiting our analysis to the dominant harmonics $h^{\rm P}_{22}$ and $h^{\rm P}_{2,-2}$, we will use the PhenomD waveform model~\cite{Husa+15, Khan+15}, an aligned-spin Fourier-domain model publicly available in the LIGO Algorithm Library (LAL) that covers the inpiral, merger and ringdown for binaries with generic spin magnitude\footnote{See~\cite{London+17} for a recent extension of the model to include higher harmonics.}. The amplitude and phase of the waveform are produced as piecewise analytical functions of the frequency. Using a Fourier-domain approximant that is smooth by construction avoids the Gibbs oscillations induced by the tapering of of finite-length time-domain waveforms, and will allow us to easily take Fourier-domain derivatives of the waveform\footnote{Note that the PhenomD amplitude and phase are smooth on three separate frequency bands (inspiral, intermediate, and ringdown) with junction conditions that are only of class $C^{1}$. To avoid spurious discontinuities, we introduce decaying corrective terms of the form $(f-f_{\rm join})^{2} e^{-\lambda(f-f_{\rm join})^{2}}$ on each side of $f_{\rm join}$, the junction frequency between any two given bands, resulting in functions of class $C^{2}$.}.

For the frame trajectory in the inspiral phase, we use an SEOBNRv3 waveform~\cite{Pan+13, BTB16}, that incorporates all degrees of freedom of both spins. As the inspiral of the SEOBNRv3 waveform is constructed by using the instantaneous orbital plane of the dynamics as the precessing frame, the frame extraction procedure of~\cite{OShaughnessy+11} simply returns the normal to the orbital plane as the radiation axis. The frame thus presents oscillatory features of small amplitude at the orbital timescale, due to the nutation of the orbital plane, which we smooth out using a Gaussian filtering with width based on the orbital phase as in~\cite{Blackman+17a}.

\begin{figure}
  \centering
  \includegraphics[width=.98\linewidth]{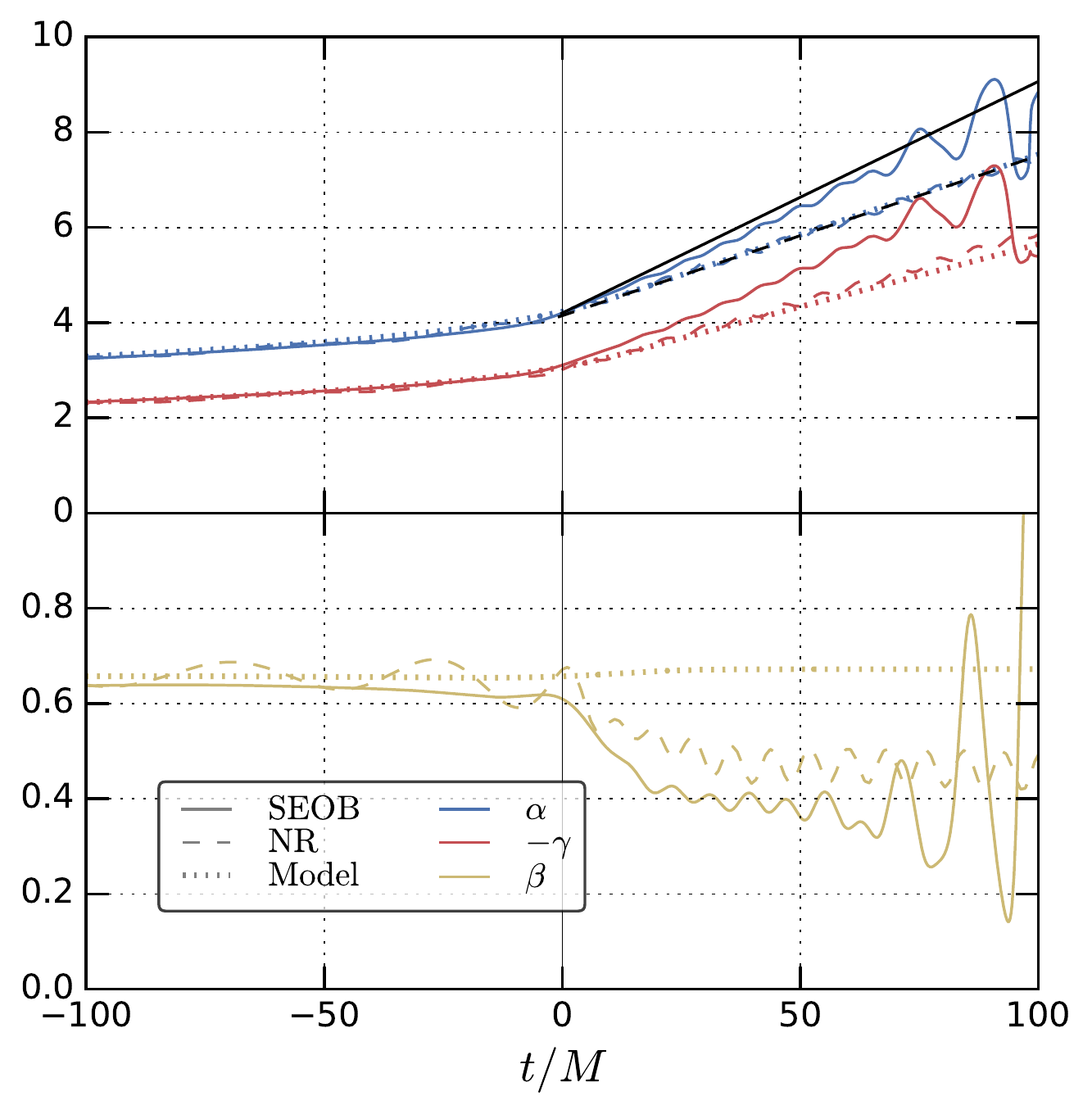}
  \caption{Evolution of Euler angles $(\alpha, \gamma)$ (top panel) and $\beta$ (bottom panel) near merger for the example SpEC waveform SXS:BBH:0058~\cite{SpEC, SXScatalog, Mroue+12, Mroue+13}. Full, dashed and dotted lines represent the NR, SEOB and extended SEOB waveforms respectively. The vertical line at $t=0$ indicates the time of merger. In the left panel the full and dashed black lines indicate the asymptotic behaviour~\eqref{eq:OmegaframeQNM} for the rotation of the frame around the direction of the final $\bm{J}$. The difference between NR and SEOB there is due to the value of the final spin, with $\chi_{f}^{\rm NR} = 0.54$ and $\chi_{f}^{\rm SEOB} = 0.42$, which leads to a different $\Omega_{\rm frame}$. In the right panel, $\beta$ is asymptotically constant in our toy model.}
  \label{fig:precmodel}
\end{figure}

An extension is required in specifying a model for the precession post-merger. After the merger occurs, the intuitive interpretation of the precessing frame as roughly following the orbital plane of the binary is lost. However, the prescription of~\cite{OShaughnessy+11} to extract the radiation axis is based entirely on the waveform itself, and can be applied to the ringdown as well. For $\ell=2$, analyzing numerical relativity waveforms, Ref.~\cite{OShaughnessy+12} described a qualitative model where the radiation axis essentially keeps precessing around the final angular momentum $\bm{J}$, but transitions to a faster precession rate, with an angular velocity determined by the quasinormal mode (QNM) frequencies of the remnant black hole as
\be\label{eq:OmegaframeQNM}
	\Omega_{\rm frame} = \omega_{220}^{\rm QNM} - \omega_{210}^{\rm QNM} \,.
\ee
This implies a significant acceleration of the precession post-merger with respect to the inspiral (see Fig.~\ref{fig:precmodel}). For the opening angle of the precessing cone $\beta$, Ref.~\cite{OShaughnessy+12} proposes an exponential decay driven similarly by the difference between damping frequencies. We found by inspection of various precessing waveforms currently available in the SXS catalog (see e.g.~\cite{SXScatalog, Mroue+12, Mroue+13}) that this picture is at least qualitatively correct for $\ell=2$, with $\beta$ appearing to transition to a lower value rather than decaying all the way to zero. By contrast, we found that the current ringdown prescription in the SEOBNRv3 model can cause a qualitatively different behaviour, with oscillations in $\beta$.

In our model, after reaching the time of peak amplitude as defined in~\eqref{eq:defsumamplitude}, we will use the prescription~\eqref{eq:OmegaframeQNM} for the post-merger frame precession and simply keep $\beta$ constant,
\begin{subequations}\label{eq:eulerQNM}
\begin{align}
	\alpha_{\rm post-merger}(t) &= \alpha(t_{\rm peak}) + (\omega_{220}^{\rm QNM} - \omega_{210}^{\rm QNM})(t-t_{\rm peak}) \,, \\
	\beta_{\rm post-merger}(t) &= \beta(t_{\rm peak}) \,, \\
	\gamma_{\rm post-merger}(t) &= \gamma(t_{\rm peak}) - (\alpha(t) - \alpha(t_{\rm peak}) )\cos \beta \,,
\end{align}
\end{subequations}
where we use the minimal-rotation condition~\eqref{eq:gammadot} for a constant $\beta$. To compute the QNM frequencies as a function of the final spin $\chi_{f}$, we use fits constructed in Ref.~\cite{Berti+05}, illustrated in Fig.~\ref{fig:QNM}. The final spin is taken to be the same as the one computed internally to the SEOBNRv3 code~\cite{Pan+13, BTB16}, where the final-spin fit formula of Ref.~\cite{BR09}, built for spin-aligned systems, is applied to the spin components projected on the orbital angular momentum $\bm{L}$ at merger.

Fig.~\ref{fig:precmodel} presents a comparison of the frame trajectory at merger for the SEOBNRv3, NR waveforms and for our model~\eqref{eq:eulerQNM}. The example catalog waveform is SXS:BBH:0058, generated with SpEC~\cite{SXScatalog, SpEC, Mroue+12, Mroue+13}, with a mass ratio of $q=5$, and a single in-plane spin of $\chi_{1} = 0.5$. The NR and SEOB post-merger frames disagree mainly due to a different value of the final spin $\chi_{f}$\footnote{To compute $\chi_{f}$, SEOBNRv3 uses fitting formulas from~\cite{BR09}. Using updated formulas incorporating additional NR data (see e.g.~\cite{HBR16}) would reduce or remove this disagreement.}, which leads to a different $\Omega_{\rm frame}$. Given their different value of $\chi_{f}$, however, both agree with the qualitative description~\eqref{eq:OmegaframeQNM}. The oscillations developing after $t\sim 50 M$ occur in a regime where the overall waveform amplitude has decayed to small values. Our simple model~\eqref{eq:eulerQNM} follows well the SEOBNRv3 behaviour for $\alpha$ and $\gamma$, but departs more for $\beta$ as we do not model its variation at merger.

In this paper, we will need this model only as an example for demonstrating our frequency-domain precession response, and we only ask that it represents the qualitative features of a precessing waveform. In that context, we would like to underline a number of caveats and limitations of our simplified model. First, it is clear from Fig.~\ref{fig:precmodel} that the acceleration of the precession post-merger is going to be the most challenging feature for the separation of timescales at the basis of Sec.~\ref{sec:formalism}. While we checked the qualitative soundness of the decomposition outlined in Sec.~\ref{subsec:precdef} and of the post-merger precession~\eqref{eq:OmegaframeQNM} in some of the available numerical relativity waveforms~\cite{SXScatalog}, it is important to stress that there is no guarantee that another prescription for the precessing frame would not yield a better separation of timescales. Secondly, approximating the precessing-frame waveform by a spin-aligned one comes with known limitations, ignoring for instance mode asymmetries departing from~\eqref{eq:approxsymmetryhlminusm} as shown in~\cite{Boyle+14}. Thirdly, more exploration of the parameter space would be needed, as the QNM frequencies entering~\eqref{eq:OmegaframeQNM} vary rapidly for large $\chi_{f}$. Finally, we ensure the smoothness both of the Fourier-domain precessing-frame waveform and of the time-domain modulation, allowing us to take the derivatives required by the formalism of Sec.~\ref{sec:formalism}, which is an idealization.

Leaving aside the question of a more in-depth investigation of the best representation of precessing waveforms in their post-merger phase, we will proceed to investigating the separation of timescales in three chosen examples. The parameters of these example cases are summarized in Table~\ref{tab:precparams}. We consider a mass ratio of $q=4$, close-to-maximal spins of $\chi_{1} = \chi_{2} = 0.95$, and vary both spins misalignmnent angles to be $\pi/6$ (case labeled $++$, close to aligned spins), $\pi/2$ ($\perp\perp$, spins in the plane) and $5\pi/6$ ($--$, close to anti-aligned spins). Table~\ref{tab:precparams} also shows the remnant spins that are internally computed in both the SEOBNRv3 and the PhenomD/PhenomP codes, as well as the QNM frequencies and the resulting frame precession frequency $\Omega_{\rm frame}$. Note that the case $--$ shows a larger variation in the direction of $\bm{J}$, and the spin of the remnant is almost 0, due to the fact that the spin is large and antialigned with a large mass ratio. This example therefore serves the purpose of lying at the edge of validity of the picture of standard precession along a cone around an almost-fixed direction, and the standard hierarchy between mode contributions is not valid in this case.

\begin{table}[t]
\begin{ruledtabular}\caption{Parameters of of the three example cases that we use to illustrate our formalism. The three cases differ by their spin alignment angles $\theta_{A} = \bm{\chi}_{A} \cdot \hat{\bm{L}}_{i}$, the spins being almost aligned, in the orbital plane and almost anti-aligned. The table gives also the final spins magnitudes (for SEOB as well as PhenomD for comparison), angles between the initial and final angular momenta $\theta(\bm{J}_{i}, \bm{J}_{f})$, as well as QNM frequencies and frame rotation velocity used for the post-merger extension.}\label{tab:precparams}
\begin{tabular}{ccccccc}\label{tab:precexamples}
	$f_{\rm min}$ & $ M_{\rm min} $ & $q$ & $\chi_{1}$ & $\chi_{2}$ & $ \phi_{1} $ & $ \phi_{2} $ \\
	\hline
	$20\mathrm{Hz}$ & $20\Msol$ & $ 4.0 $ & $ 0.95 $ & $ 0.95 $ & $0$ & $\pi/2$ \\
	\hline\hline
	Case && ++ && $\perp\perp$ && $--$ \\
	\hline
	$\theta_{1}$ && $\pi/6$ && $\pi/2$ && $5\pi/6$ \\
	$\theta_{2}$ && $\pi/6$ && $\pi/2$ && $5\pi/6$ \\
	\hline
	$\chi_{f}$ && $0.89$ && $0.50$ && $0.02$ \\
	$\chi_{f}^{\rm Ph}$ && $0.90$ && $0.47$ && $0.002$ \\
	$\theta(\bm{J}_{i}, \bm{J}_{f})$ && $0.02$ && $0.04$ && $0.27$ \\
	$M \omega_{220}^{\rm QNM}$ && $0.66$ && $0.47$ && $0.377$ \\
	$M \omega_{210}^{\rm QNM}$ && $0.51$ && $0.42$ && $0.375$ \\
	$M \Omega_{\rm frame}$ && $0.15$ && $0.04$ && $0.002$ \\
\end{tabular}
\end{ruledtabular}
\end{table}


\subsection{Estimates for the separation of timescales}
\label{subsec:sizecorrPrec}

\begin{figure}
  \centering
  \includegraphics[width=.98\linewidth]{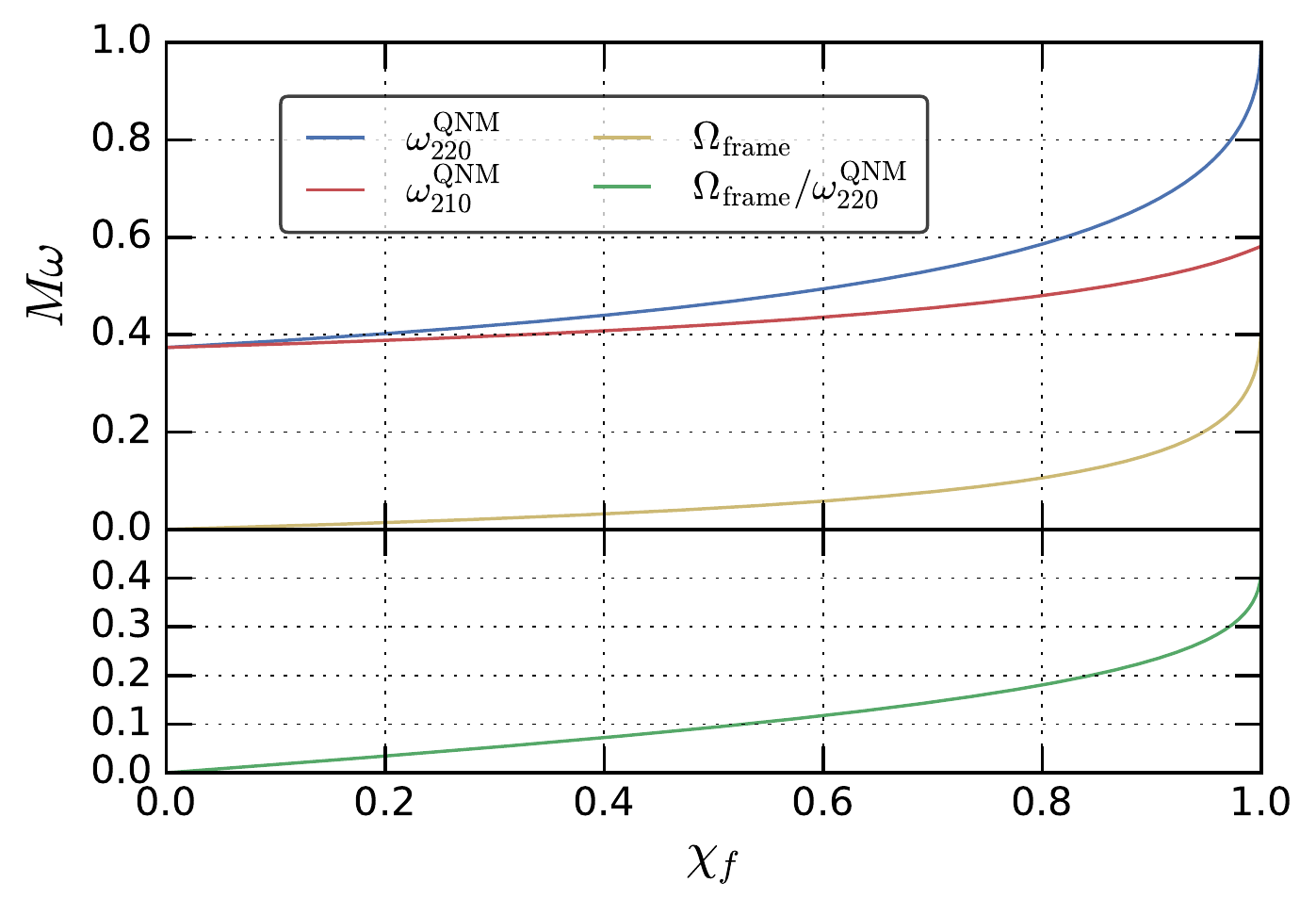}
  \caption{Quasi-normal mode frequencies for the 0th overtone of the modes $22$ and $21$, as a function of the dimensionless spin of the final black hole $\chi_{f}$, as well as their difference $\Omega_{\rm frame}$ (see~\eqref{eq:OmegaframeQNM}). The lower panel shows the  ratio $\Omega_{\rm frame} / \omega^{\rm QNM}_{220}$ characteristic of the separation of timescales between the phase of the modulation~\eqref{eq:wignerphasesimpleprec} and the P-frame waveform phase.}
  \label{fig:QNM}
\end{figure}

We now turn to the separation of timescales and to the magnitude of the higher-order corrections derived in Sec.~\ref{sec:formalism} in the case of precessing binaries, as can be estimated by the quantities $\epsilon_{\Psi 2}$, $\epsilon_{A1}$ and $\epsilon_{A 2}$ introduced in~\eqref{eq:deffom}.

In the inspiral phase, we can obtain a qualitative picture of the timescales involved by using well-known leading-order post-Newtonian results. To simplify things further, we will consider a single-spin system, and use orbit-averaging to ignore nutation features in the normal to the orbital plane. In this configuration, both the orbital angular momentum $\bm{L}$ and the spin vector $\bm{S}$ undergo simple precession on a cone around the total angular momentum $\bm{J}$ (see e.g.~\cite{Apostolatos+94, Kidder95}), with an opening angle of the cone and a precession velocity $\Omega_{\rm prec}$ that vary only on the radiation reaction timescale. In the general case, the presence of two spins complicates the evolution of the system, but the picture of a precession cone for $\bm{L}$ remains approximately valid (see~\cite{Kesden+14} for a classification of generic precession trajectories).

We will use the following notation: for $m_{1}$, $m_{2}$ the masses of the two bodies, we set $M=m_{1}+m_{2}$, $\nu=m_{1}m_{2}/M^{2}$, $\delta = (m_{1}-m_{2})/M$, and $|\bm{S}_{A}|=Gm_{A}^{2} \chi_{A}$ for $A=1,2$, with $\chi$ the dimensionless spin between $0$ and $1$ for Kerr black holes. We define $\bm{\ell}$ as the unit vector normal to the orbital plane. We take the convention $m_{1} \geq m_{2}$ and assume that only the more massive object has a spin, $\bm{S}_{2} = 0$. As in Sec.~\ref{subsec:SPA}, we will use the notation $v = (G M \omega/c^{3})^{1/3}$ with $\omega = \dot{\varphi}$ the orbital frequency, which translates to $v=(G \pi M f/c^{3})^{1/3}$ for the 22 mode when the SPA applies. At the Newtonian order, the orbital angular momentum is $\bm{L} = L_{N} \bm{\ell}$, with
\be\label{eq:defLN}
	L_{N} = \frac{G M^{2} \nu}{c v} \,.
\ee
Since we neglect radiation reaction, $\bm{J} = \bm{L} + \bm{S}_{1}/c$ is treated as a constant that we use to set the z-axis so that $\bm{J} = J \bm{e}_{z}$, and we decompose the spin in its aligned and perpendicular components as $\bm{S}_{1} = S_{1}^{z} \bm{e}_{z} + \bm{S}_{1}^{\perp}$. At leading order, the precession equations read
\begin{align}
	\dot{\bm{S}}_{1} &= \bm{\Omega}_{1} \times \bm{S}_{1} \,, \nn\\
	\dot{\bm{\ell}} &= - \frac{1}{c L_{N}}  \dot{\bm{S}}_{1}\,,
\end{align}
with the spin precession velocity~\cite{Kidder95}
\be
	\bm{\Omega}_{1} = \Omega_{1} \bm{\ell} = \frac{c^{3}}{G M} \left( \frac{3}{4} + \frac{\nu}{2} - \frac{3\delta}{4} \right) v^{5} \bm{\ell} \,.
\ee
which is formally a 1PN quantity. Considering only the leading PN order, we ignore effects quadratic in the spin that would enter here at 1.5PN. Decomposing the vectors in their in-plane component and projection on $z$, and using $\delta^{2} = 1-4\nu$ to make explicit the overall scaling in $\nu$, we obtain for the frame precession velocity~\cite{Kidder95}
\be
	\dot{\alpha} \equiv \Omega_{\rm prec} = \frac{c^{4}}{G^{2} M^{3}} \frac{7+\delta}{2(1+\delta)} v^{6} J\,.
\ee
Separating the factors as
\be\label{eq:defLambdaxi}
	\Lambda \equiv \frac{7+\delta}{4(1+\delta)} \,, \quad	\xi \equiv 1 + \frac{v S_{1}^{z}}{G M^{2} \nu} \,,
\ee
we see that $\Lambda$ is a mass ratio-dependent factor chosen to be always of order 1, varying from $7/4$ for equal masses to $1$ in the test-mass limit, while the factor $\xi$ contains the contribution of the aligned component of the spin to the precession rate. With this notation,
\be\label{eq:Omegaprec}
	\Omega_{\rm prec} = \frac{2c^{3} \nu}{G M} \Lambda \xi v^{5} \,.
\ee
If precession effects are in general larger for larger mass ratios, as the precession cones widens, the overall scaling of $\nu$ in~\eqref{eq:Omegaprec} above shows that, as long as the orbital angular momentum still dominates the spin in $\xi$, increasing the mass ratio yields a slower precession rate. For high mass ratios and spins, the correction to $\xi$ in~\eqref{eq:defLambdaxi} starts to become important. Here, $\Omega_{\rm prec}$ evolves only on the radiation-reaction timescale through its dependence in $v$; this is a consequence of our simple-precession assumption, as $\Omega_{\rm prec}$ varies on the precession timescale in the generic case (see e.g.~\cite{Chatziioannou+17}).

We now turn to the Euler angles and modulation functions $\calD^{\ell *}_{mm'}(\alpha, \beta, \gamma)$, given explicitly in~\eqref{eq:defWignerD}. As explained in~\ref{subsec:precdef}, in the case of simple precession, the opening angle of the precession cone $\beta$ is essentially constant, so that the minimal rotation condition~\eqref{eq:gammadot} gives $\gamma  = -\alpha \cos \beta$ (up to a constant), and the only variable part of the precession modulation functions in~\eqref{eq:wignerrot} are the phases, according to~\eqref{eq:wignerphasesimpleprec}. The constant rate of rotation around $\bm{J}$ translates into $\dot{\alpha} = \Omega_{\rm prec}$. From the closure relation $\bm{J} = \bm{L} + \bm{S}_{1}/c$, we have
\be\label{eq:betaconst}
	\cos \beta = \sqrt{1 - \left( \frac{vS_{1}^{\perp}}{G M^{2}\nu} \right)^{2}}\,.
\ee
For a given mode contribution $\calD^{\ell *}_{mm'}$, \eqref{eq:wignerphasesimpleprec} shows that we will have factors of $(m' \cos\beta - m)$ when taking derivatives.

We have now everything we need to compute the error estimates~\eqref{eq:deffom} for the transfer function $\calT^{\ell}_{mm'}$~\eqref{eq:defprectransfer} with $m' \neq 0$. Given our restrictive  assumptions of orbit-averaged leading-order PN and single-spin simple precession, the result will only be a crude order-of-magnitude estimate. Using the Fourier-domain leading-order timescales generalized for modes $h^{\rm P}_{\ell m'}$ in the text below~\eqref{eq:timescalesN} yields:
\begin{subequations}\label{eq:precfomPN}
\begin{align}
	\epsilon_{\Psi 2} &= \left(\frac{2}{m'}\right)^{\frac{2}{3}} \frac{5\nu}{96} \Lambda^{2} \xi^{2} (m' \cos\beta - m)^{2} v^{-1} \,, \\
	\epsilon_{A 1} &= \left(\frac{2}{m'}\right)^{\frac{5}{3}} \frac{(7 - 2\kappa_{\ell m'})\nu}{6} \Lambda \xi |m' \cos\beta - m| v^{2} \,, \\
	\epsilon_{A 2} &= \left(\frac{2}{m'}\right)^{\frac{10}{3}} \frac{(7 - 2\kappa_{\ell m'}) (13 - 2\kappa_{\ell m'}) \nu^{2}}{72} \nn\\
	& \quad\quad\quad \cdot \Lambda^{2} \xi^{2} (m' \cos\beta - m)^{2} v^{4} \,,
\end{align}
\end{subequations}
where $v = (GM \pi f/c^{3})^{1/3}$. It is worth noting that $\epsilon_{\Psi 2}$ has an overall frequency scaling of $v^{-1}$, formally at $-0.5$PN order, which means that the relative size of this correction grows towards smaller frequencies, away from merger. Remember however that the quantities $\epsilon$ are meant to fractional errors, and the opening angle of the precession cone, giving the overall normalization for precession effects in the waveform, also goes to 0 as $v$ in that limit. The amplitude error estimates $\epsilon_{A1}$, $\epsilon_{A2}$, by contrast, have the more usual behaviour of PN corrections growing towards merger. The geometric factors $(m' \cos\beta - m)$ show that mode contributions with a larger $|m-m'|$ are harder to model, again in a relative sense, but those contributions are however suppressed in amplitude (see discussion below~\eqref{eq:defprechIsum}).

The overall $\nu$ scaling indicates a better separation of timescales when increasing the mass ratio away from equal mass. These expressions also show that, since the total angular momentum appears as a factor in $\Omega_{\rm prec}$, higher-order corrections will be larger for roughly spin-aligned systems than for anti-aligned spins, an effect that becomes significant for large mass ratios. Note that the regime where $\xi$ gets close to 0 corresponds to the transitional precession range~\cite{Apostolatos+94}, with the spin of the primary compensating the orbital angular momentum, and our analysis based on simple precession is not valid anymore.

We can somewhat complement this picture for the post-merger precession by considering the accelerated frame rotation rate described by the model~\eqref{eq:OmegaframeQNM} and illustrated in Fig.~\ref{fig:precmodel}. We show in Fig.~\ref{fig:QNM} the dependency of the QNM frequencies $\omega_{220}^{\rm QNM}$, $\omega_{210}^{\rm QNM}$ with the final spin of the remnant black hole $\chi_{f}$, together with the ratio $\Omega_{\rm frame} / \omega^{\rm QNM}_{220}$. This ratio is characteristic of the separation of timescales, in the ringdown regime, between the phase of the modulation~\eqref{eq:wignerphasesimpleprec} and the P-frame waveform phase, and increases monotonically with $\chi_{f}$, reaching $\sim 0.4$ for $\chi_{f} \rightarrow 1$. We cannot translate readily the time-domain separation of timescales in the ringdown regime to the Fourier-domain error measures defined in~\eqref{eq:deffom}, as the time-to-frequency correspondence~\eqref{eq:deftf} does not reach times beyond the merger, as shown by Fig.~\ref{fig:tf}. However, a faster frame rotation will yield a more extended Fourier transform of the modulation~\eqref{eq:defG}, and will be more challenging to accomodate with the formalism of Sec.~\ref{sec:formalism}.

To go beyond the above order-of-magnitude picture, we now present a numerical computation of the error estimates~\eqref{eq:deffom} for our post-merger extended precession model presented in Sec.~\ref{subsec:precmodel}, for the three examples summarized in Table.~\ref{tab:precexamples}. Here and in the following we consider only the $h^{\rm I}_{2 m}$ mode contributions induced by $h^{\rm P}_{22}$ for positive frequencies, knowing that the ones induced by $h^{\rm P}_{2,-2}$ for negative frequencies can be deduced using~\eqref{eq:hIlmposnegfreq}. We use analytic derivatives of the PhenomD phase and amplitude for the Fourier-domain based timescales $\Tf$, $T_{A1}$ and $T_{A2}$, and numerical derivatives for the time-domain modulation.

\begin{figure*}
  \centering
  \includegraphics[width=.98\linewidth]{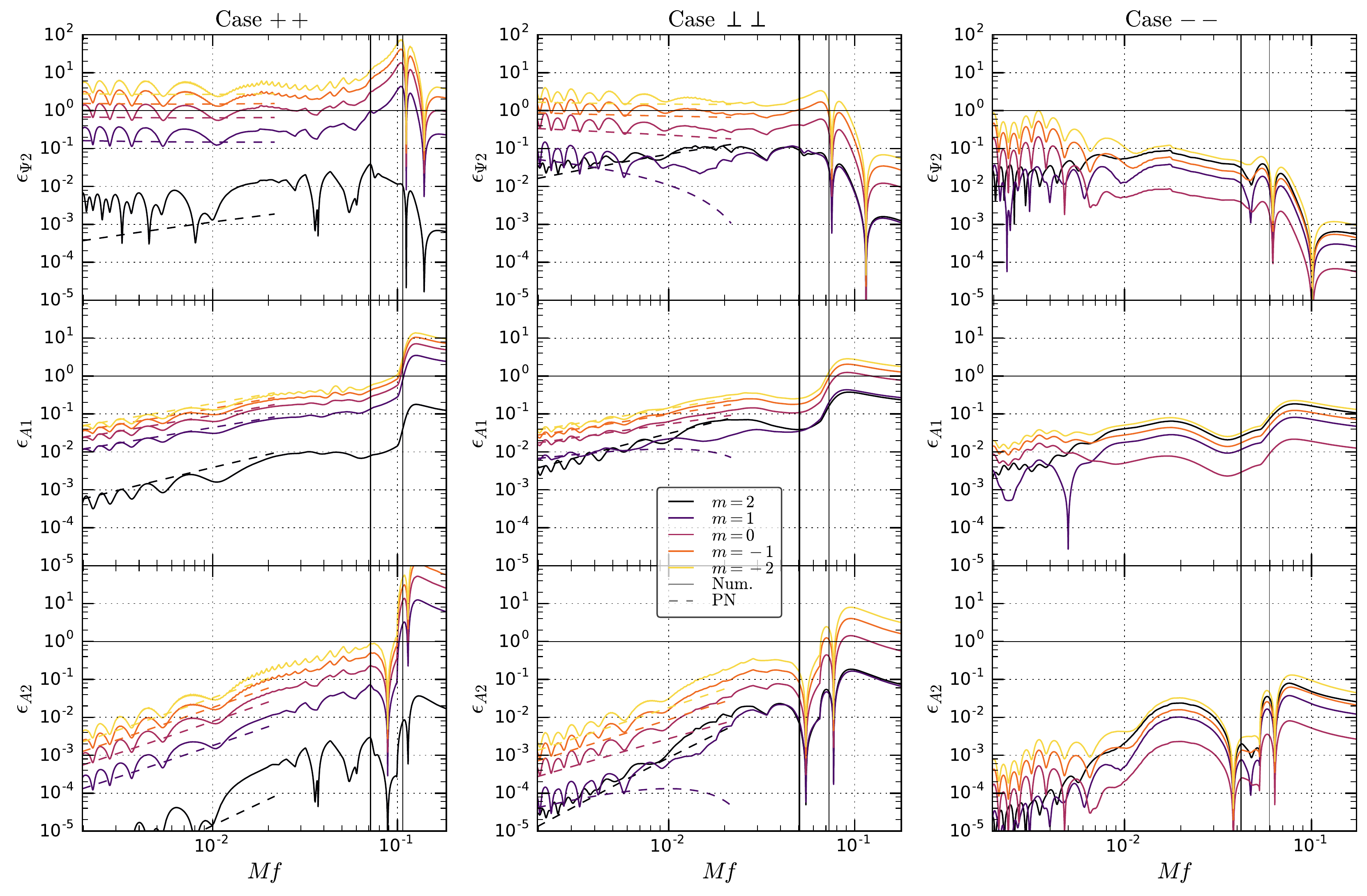}
  \caption{Error estimates of the approximation as defined in~\eqref{eq:deffom} for the precession modulations, for the three cases $++$, $\perp\perp$ and $--$ listed in Table~\ref{tab:precexamples}. The thick and thin vertical lines show, respectively, the merger frequency and the asymptotic ringdown frequency. The colors correspond to different values of $m$ in $\calT^{2}_{m 2}$ as defined in~\eqref{eq:defprectransfer}. The full lines are computed with numerical derivatives for the full waveform and modulation, while in the cases $++$ and $\perp\perp$ the dashed lines show the leading-order PN estimates~\eqref{eq:precfomPN} for single-spin simple precession. We show the three error estimates $\epsilon_{\Psi 2}$, $\epsilon_{A1}$, $\epsilon_{A2}$ defined in~\eqref{eq:deffom}, and the range $\epsilon \gtrsim 1$ indicates a breakdown of the formalism of Sec.~\ref{sec:formalism}.}
  \label{fig:fomprec}
\end{figure*}

The results are shown in Fig.~\ref{fig:fomprec}, using the merger frequency and the ringdown frequency, shown by the vertical lines, to give an idea of the separation between the inspiral and post-merger phases. The leading-order PN prediction for single-spin simple precession~\eqref{eq:precfomPN} is overlayed for the cases $++$ and $\perp\perp$, but not for the case $--$ as it departs significantly from simple precession. We can take $\epsilon \sim 1$ as an order-of-magnitude indication of the breakdown of a perturbative treatment. An important point about Fig.~\ref{fig:fomprec} is that the error estimates are relative for each mode, thus higher $|m-m'|$ modes, which are found to be the hardest to model precisely, can be very subdominant in the final waveform.

The magnitude of the error estimates for $++$ and $\perp\perp$ is roughly in agreement with the PN-inspired computation~\eqref{eq:precfomPN} above for the inspiral part of the signal, with $\epsilon_{\Psi 2}$ showing a negative slope $v^{-1}=(Mf)^{-1/3}$, and with a hierarchy between modes due to the factors of $|m' \cos \beta - m|$. The PN estimates are missing oscillations on the precession timescale due to the double-spin precession. The case $--$ departs from the simple precession picture, and~\eqref{eq:precfomPN} does not apply. Before merger, we see that amplitude corrections are within the perturbative regime, contrarily to $\epsilon_{\Psi 2}$ that can exceed $1$ for subdominant modes in the $++$ and $\perp\perp$ cases. The amplitude corrections are found to be in the perturbative regime for all cases during the inspiral, however both the $++$ and $\perp\perp$ cases show a sharp increase of $\epsilon_{A1}$, $\epsilon_{A2}$ post-merger. The case $--$, with its small remnant spin and mild post-merger frame rotation, shows no such increase.

Overall, the conclusion to be drawn of Fig.~\ref{fig:fomprec} is that for the $++$ and $\perp\perp$ cases we expect the perturbative approach to be applicable only during the inspiral, with a possible breakdown for the post-merger phase, especially for the case $++$ with its fast post-merger frame rotation. The higher $|m-m'|$ modes are expected to be more challenging for the perturbative formalism, including during the inspiral. To make this picture quantitative, we need a full comparison of the signals processed at different orders of approximation against a numerical Fourier transform, which will be presented in Sec.~\ref{subsec:precerror} below.

\subsection{Direct convolution approach for the merger-ringdown phase}
\label{subsec:trigopoly}

\begin{figure*}
  \centering
  \includegraphics[width=.98\linewidth]{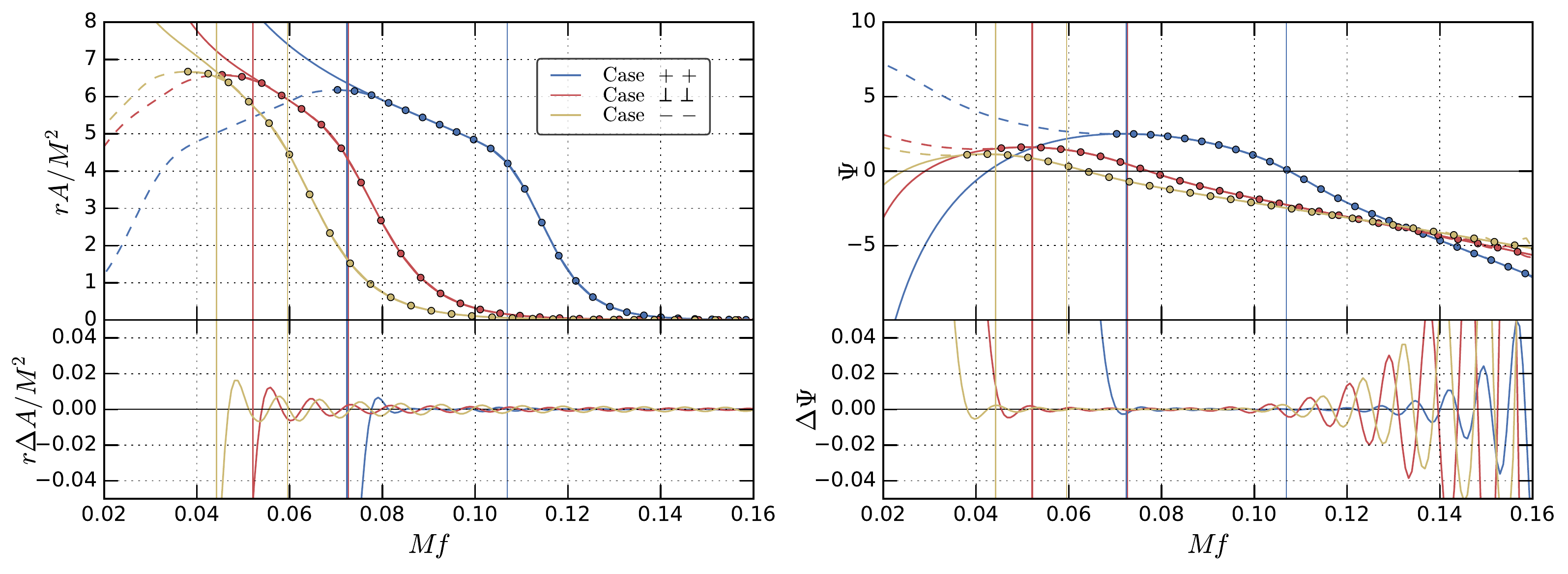}
  \caption{Amplitude (left panel) and phase (right panel), compared to its trigonometric polynomial representation for the three cases listed in Table~\ref{tab:precexamples}. The thick and thin vertical lines indicate the frequency of the merger and ringdown (the QNM frequency) respectively. The continuous line shows the target signal $\tilde{h}(f)$, while the dashed line shows the artificially symmetrized trigonometric polynomial~\eqref{eq:hsymtrigo}. The dots show the discrete samples entering~\eqref{eq:ykDFTprec}. The lower panels show the amplitude and phase residuals of~\eqref{eq:hsymtrigo} compared to the original $\tilde{h}(f)$.}
  \label{fig:trigopoly}
\end{figure*}

As shown by Fig.~\ref{fig:fomprec}, the faster evolution of the modulation functions in the post-merger phase and the resulting weaker separation of timescales can be expected to be challenging for the perturbative formalism layed out in Sec.~\ref{sec:formalism}. Motivated by this forecasted shortcoming of the Taylor-like expansion approach, here we investigate an alternative way of handling the merger-ringdown part of the signal.

In the correspondence between the time and Fourier domain, sharp features in the time domain map to extended features in the Fourier domain, and vice versa. The merger and ringdown part of the Fourier-domain signal extends over a wide range of frequencies, while it corresponds to a short interval of times, much shorter than the inspiral part. A possible approach would be to separate the time-domain waveform between the inspiral and merger phase, build a Fourier-domain model for the inspiral part only, to which the formalism of Sec.~\ref{sec:formalism} could be applied, while the merger and ringdown part could be handled by a direct FFT, which would be a cheap operation on a limited range in time. Although this approach should always be applicable, here we will use an alternative method, simpler to implement as it allows us to keep the setting of a Fourier-domain precessing-frame waveform combined with a time-domain precession modulation.

As argued in Sec.~\ref{subsec:precdef}, the support of the convolution~\eqref{eq:precconvolution} will be mainly one-sided towards the high-frequency part of $\tilde{h}(f)$, which is featureless and slowly varying as a function of $f$. Taking advantage of this, we will adopt a one-sided trigonometric polynomial representation for $\tilde{h}(f)$. For frequencies high enough that the support of the convolution~\eqref{eq:precconvolution} does not extend beyond the range covered by this trigonometric representation, the result will be obtained directly as an FFT/IFFT with a limited number of samples. The limitation to the high-frequency range is crucial here: the Fourier-domain amplitude and phase diverge as $f^{-7/6}$ and $f^{-5/3}$ respectively for $f \rightarrow 0$, and our procedure would require a much finer sampling to cover part of the inspiral, going back to being equivalent to an IFFT of the full signal if we were to cover all frequencies.

We consider the high-frequency part of the signal above some frequency $f_{0}$, above which the signal has limited amplitude and phase evolution, up to some maximal frequency $f_{\rm max}$ where the Fourier-domain amplitude of the signal has decayed to a negligible level. In practice, we define $f_{\rm knee}$ as the peak of $f^{2}A(f)$, representing the onset of the decay in amplitude, and roughly corresponding to $\omega_{22}^{\rm QNM}/\pi$. We set $f_{0}\equiv 2/3f_{\rm knee}$, which is close in practice to the merger frequency, and $f_{\rm max}$ is chosen so that the amplitude is $10^{-4}$ of the amplitude at $f_{\rm knee}$. We also eliminate a constant and a linear term in the phase by choosing another frequency central to the high-frequency range we want to represent, which we take to be $f_{p} = f_{\rm knee}^{2/3} f_{\rm max}^{1/3}$ with corresponding time $t_{p}\equiv \tf(f_{p})$. Note that the method should only be weakly sensitive to the precise choice of $f_{p}$ and $t_{p}$.

Instead of tapering the signal to $0$ below $f_{0}$, to limit the deviation from the original signal we take advantage of the one-sidedness and only flatten the amplitude\footnote{In practice, this is implemented as the discrete integral of a cosine window function on just the first two samples. We also taper to 0 the last three samples before $f_{\rm max}$, and $0$-pad by a factor of 2.}, and artificially symmetrize the signal. To ensure continuity, this artificial symmetrization to a fictitious range $f\in [f_{0} - (f_{\rm max} - f_{0}), f_{0}]$ is done by imposing symmetric amplitudes and phases about $f_{0}$. Defining $\Delta f \equiv 2 (f_{\rm max} - f_{0})$, we write
\begin{widetext}
\be\label{eq:defhsym}
	\tilde{h}_{\rm sym}(f) =
	\begin{cases}
		\exp\left[i \Psi(f_{0}) - 2i\pi (f-f_{0}) t_{p} \right] \tilde{h}(f) \,,  &\text{ for } f \in [f_{0}, f_{0} + \Delta f /2] \\
	\tilde{h}_{\rm sym}(2 f_{0} - f)^{*} \,,  &\text{ for } f \in [f_{0} - \Delta f/2, f_{0}]
	\end{cases}
\ee
\end{widetext}

Next, we build a a trigonometric polynomial representation of $\tilde{h}_{\rm sym}$, a construction intimately related to the FFT, that we recall in App.~\ref{app:notation}. Over the frequency range $f\in [f_{0}, 2f_{\rm max} - f_{0}]$, we can approximate
\be\label{eq:hsymtrigo}
	\tilde{h}(f) \simeq \tilde{h}_{\rm sym } (f) \simeq \sum\limits_{k=-M}^{+M} (-1)^{k} c_{k} e^{2i\pi k \frac{f-f_{0}}{\Delta f}} \,,
\ee
where the factor $(-1)^{k}$ comes from the fact that $f_{0}$ is here at the center of the interval. The coefficients $c_{k}$ are built following the rules~\eqref{eq:ckyk} from the IFFT coefficients
\be\label{eq:ykDFTprec}
	y_{k} = \frac{1}{N} \sum\limits_{j=0}^{N-1} \tilde{h}_{\rm sym}\left( f_{0} + \frac{2j - N}{N} \Delta f \right) \omega^{jk} \,.
\ee
Taking the point of view of an interpolation problem, Fig~\ref{fig:trigopoly} shows the accuracy of this representation of the high-frequency part of the signal, by comparing the orginal $\tilde{h}(f)$ to its  trigonometric-polynomial representation~\eqref{eq:hsymtrigo}, for the three cases listed in Table~\ref{tab:precexamples}. We see that we can achieve a good agreement already for 32 samples (128 counting the symmetrization and 0-padding). Errors in the phase $\Psi$ grow towards high frequencies because they are essentially errors in a relative sense, and amplitudes are decaying in this region.

The crucial point in this approach is that the convolution integral in~\eqref{eq:precconvolution} has support mainly on $f'<0$, as discussed in Sec.~\ref{subsec:precdef}, which means that when we try to compute the transfer function $\calT(f)$ at a given frequency $f$, we only need the trigonometric-polynomial representation~\eqref{eq:hsymtrigo} to be accurate for frequencies $>f$ in Fig.~\ref{fig:trigopoly}. Thus, the trigonometric representation of $\tilde{h}(f)$ can be used to compute the convolution almost all the way down to $f_{0}$, effects of the tapering aside. However, this statement is tied to our restriction to the $22$-mode, and extending the method to a precessing-frame waveform that includes more modes $h^{\rm P}_{\ell m}$ will require care.

When inserting this representation~\eqref{eq:defhsym} and~\eqref{eq:hsymtrigo} into~\eqref{eq:precconvolution}, we obtain
\begin{align}\label{eq:resultdirectconvol}
	\tilde{s}(f) &= e^{-i \Psi(f_{0})} e^{2i\pi (f-f_{0}) t_{p}} \nn\\
	& \qquad \cdot\sum\limits_{k=-M}^{+M} (-1)^{k} c_{k} e^{2i\pi k \frac{f-f_{0}}{\Delta f}} F(t_{p} + k\delta t) \,,
\end{align}
where we defined the time sampling $\delta t \equiv 1/\Delta f$. We see that we are left with an FFT-like expression to compute from $N+1$ time samples of the modulation function $F$, centered around $t_{p}$. We see that, apart from the conditioning described above with the artificial symmetrization, this is analogous to an FFT of the product of the modulation with the time-domain signal obtained through an IFFT.

In terms of computational performance, the implementation of this approach is expected to have a reasonable cost. In the following, we will use $M=64$ samples ($32$ useful samples before 0-padding, $N=128$ samples in total for the artificially symmetrized signal). The computation of~\eqref{eq:ykDFTprec} and~\eqref{eq:resultdirectconvol} amounts to two FFT/IFFT operations, and is done only once for a given waveform and modulation function. The number of samples is of the same order of magnitude as the one required to represent the Fourier-domain waveform with an interpolating cubic spline for its amplitude and phase (a few hundreds for the full frequency band, see e.g~\cite{Puerrer14}), thus the cost should be comparable to the Taylor-like approach presented in Sec.~\ref{subsec:executivesummary}.


\subsection{Error control for the Fourier-domain precession modulation}
\label{subsec:precerror}

We now assess the accuracy of the transfer function computation, applying the formalism of Secs.~\ref{subsec:executivesummary} and~\ref{subsec:trigopoly} to the three examples listed in Table~\ref{tab:precexamples}, and comparing the result to a reference numerical computation.

To obtain the latter, we first have to perform an IFFT of our Fourier-domain P-frame waveform, to obtain $h^{\rm P}_{22}$ as a function of time. To mitigate the effect of the necessary tapering of the waveform when computing this numerical inverse Fourier transform, we apply a Planck-window tapering on the range $f\in 16-20 \Hz$ for a total mass of $M=20 \Msol$. In parallel, we generate an SEOB waveform with the appropriate length in time. Both are aligned to peak at $t=0$, we build the post-merger extended modulation functions as explained in Sec.~\ref{subsec:precmodel}, compute the inertial-frame modes $h^{\rm I}_{2m}$ following~\eqref{eq:wignerrot} before computing their Fourier-domain counterparts $\tilde{h}^{\rm I}_{2m}$ with an FFT. The Fourier-domain transfer functions are then computed using~\eqref{eq:defprectransfer}.

\begin{figure*}
  \centering
  \includegraphics[width=.98\linewidth]{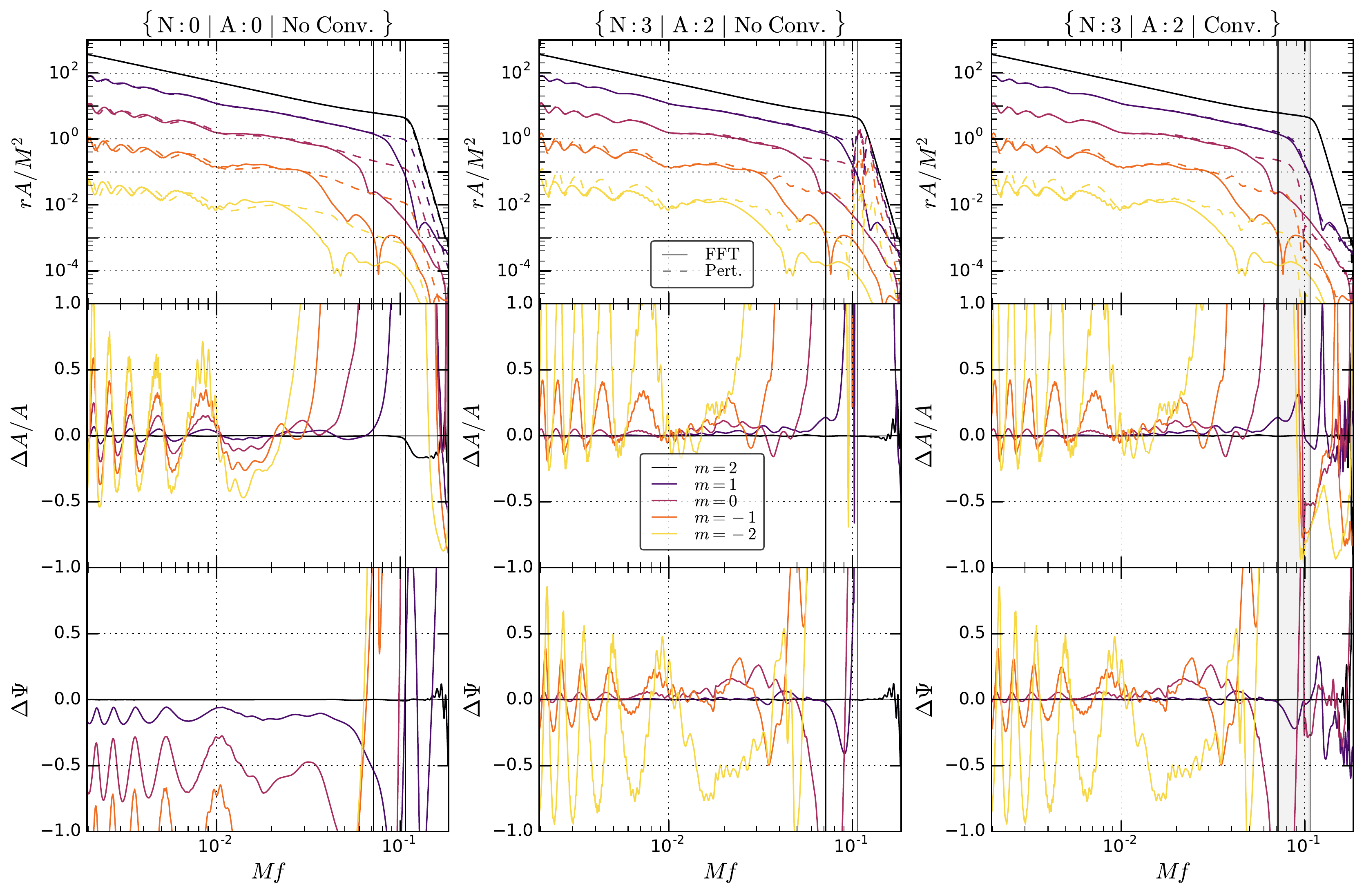}
  \caption{Fourier-domain amplitudes of the modes $\tilde{h}^{\rm I}_{\ell m}$ (top panels), reconstruction errors in the transfer fonctions $\calT^{2}_{m2}$ for the amplitudes (middle panels) and for the phases (bottom panels),  for the configuration $++$ of Table~\ref{tab:precexamples}, with spins almost aligned. The thick and thin vertical lines show the merger and asymptotic ringdown frequencies respectively. For the solid curves in the top panel, transfer functions were computed from an FFT of the full time-domain precessing signal, while the dashed curves show the results of the Fourier-domain reconstruction techniques presented here. The columns show the orders of approximation $\{N:0 | A:0 | \text{No Conv.}\}$, $\{N:3 | A:2 | \text{No Conv.}\}$, $\{N:3 | A:2 | \text{Conv.}\}$. In the right column, the region of transition between the perturbative treatment of Sec.~\ref{subsec:executivesummary} and the high-frequency treatment of Sec.~\ref{subsec:trigopoly} is shaded in grey. Note that the errors shown here are relative to a given mode, not to the leading mode, and the errors are largest in this sense for the most subdominant modes; the top panels illustrate the mode hierarchy.}
  \label{fig:precerrors++}
\end{figure*}

\begin{figure*}
  \centering
  \includegraphics[width=.98\linewidth]{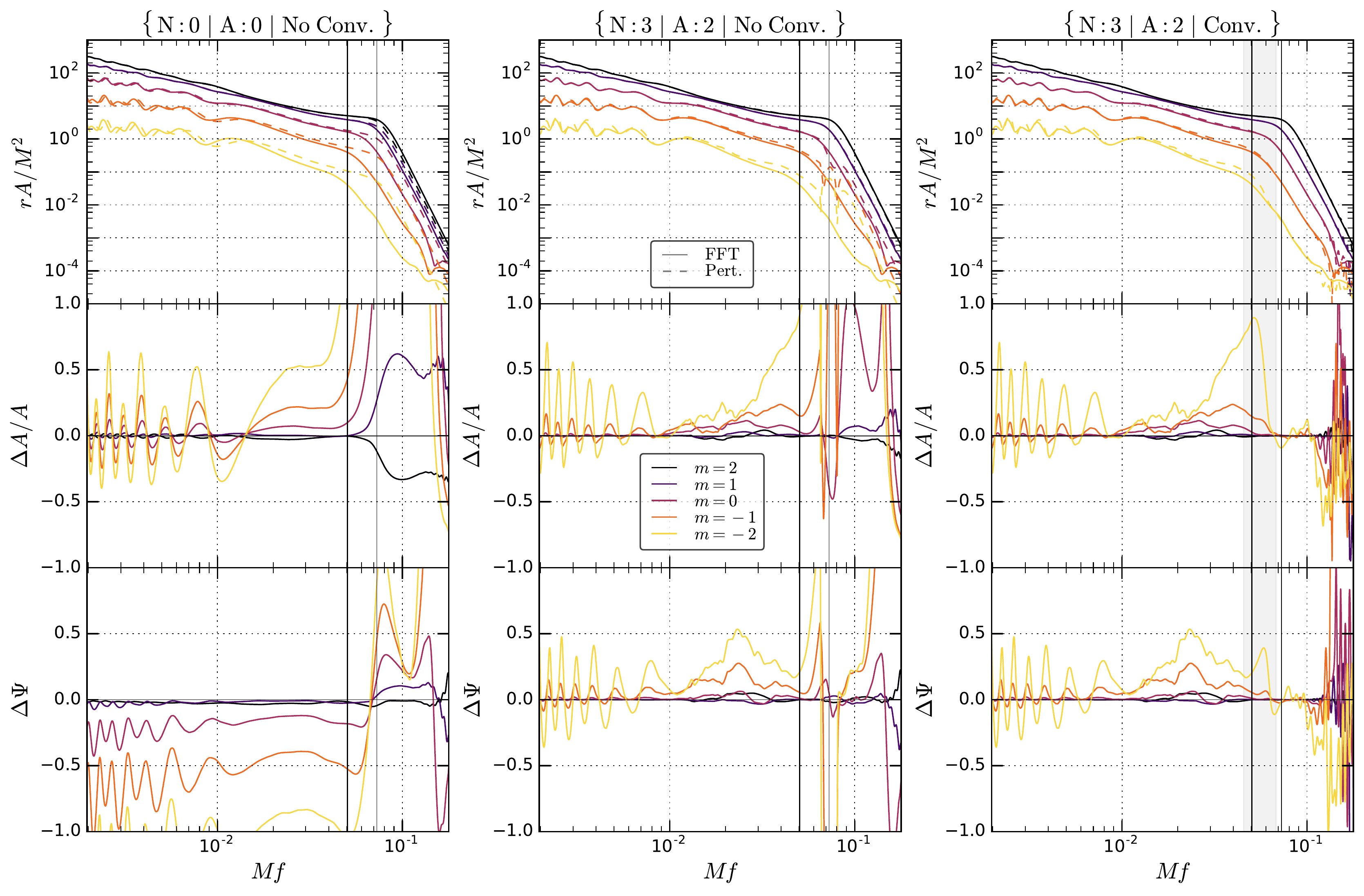}
  \caption{Same as Fig.~\ref{fig:precerrors++}, but for the configuration $\perp\perp$ of Table~\ref{tab:precexamples}, with both spins in the orbital plane.}
  \label{fig:precerrorspp}
\end{figure*}

\begin{figure*}
  \centering
  \includegraphics[width=.98\linewidth]{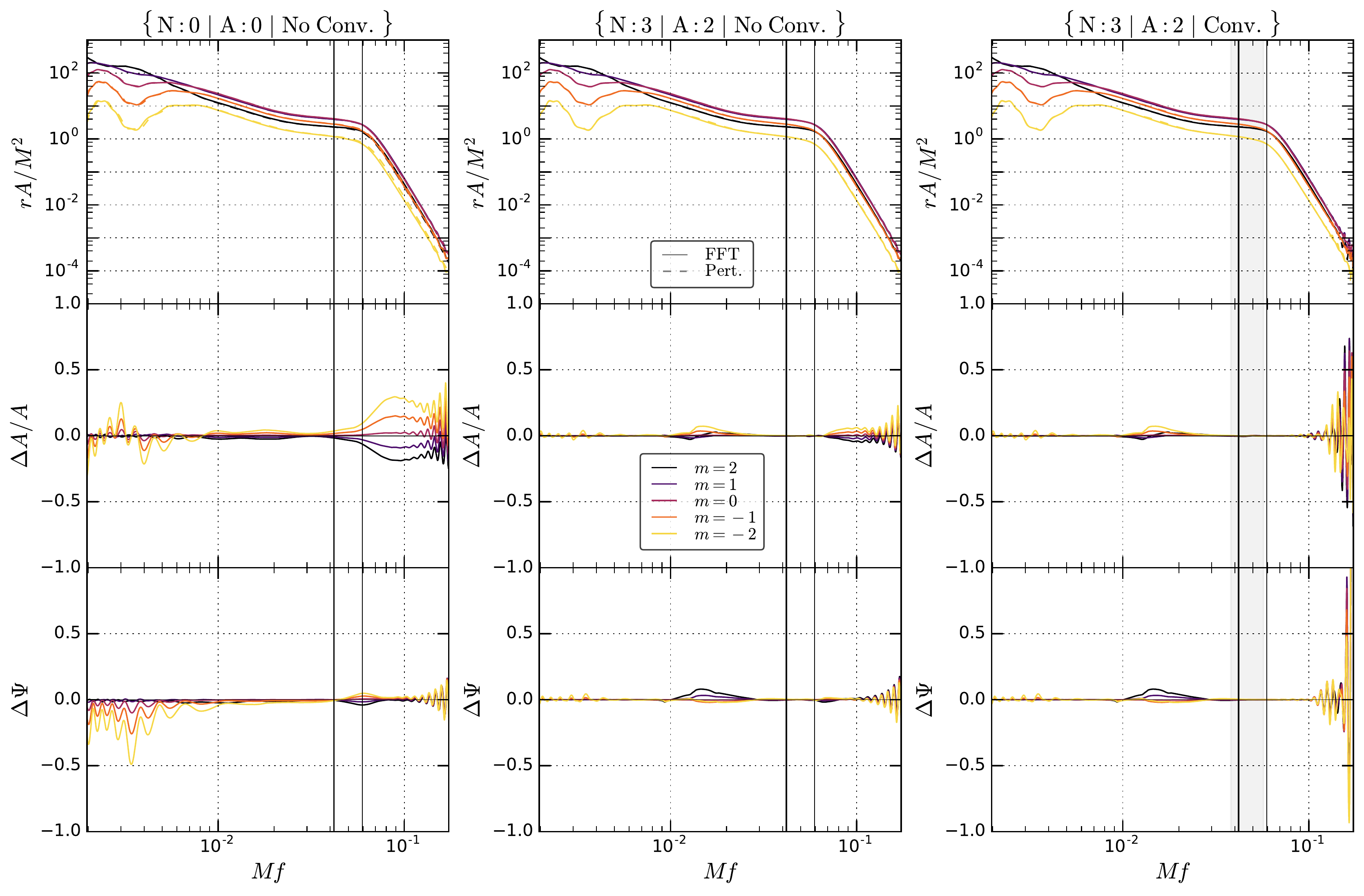}
  \caption{Same as Fig.~\ref{fig:precerrors++}, but for the configuration $--$ of Table~\ref{tab:precexamples}, with almost anti-aligned spins.}
  \label{fig:precerrors--}
\end{figure*}

The figures Fig.~\ref{fig:precerrors++}, Fig.~\ref{fig:precerrorspp} and Fig.~\ref{fig:precerrors--} show the result for the cases $++$, $\perp\perp$ and $--$ of Table~\ref{tab:precexamples} at three successive approximations, following the notation of~\eqref{eq:summaryNA}: $\{N:0 | A:0 | \mathrm{No \; Conv.}\}$, which is simply the leading-order transfer function~\eqref{eq:transferlocal}, ignoring all the corrections; $\{N:3 | A:2 | \mathrm{No \; Conv.}\}$, which incorporates both the phase corrections~\eqref{eq:stencilfresnel}, using a stencil size $N=3$, and the amplitude corrections up to second order in~\eqref{eq:summaryNA}; and $\{N:3 | A:2 | \text{Conv.}\}$, which is the same as the previous setting for the inspiral but uses the convolution formalism of Sec.~\ref{subsec:trigopoly} to cover the high frequency range, with a smooth transition in the shaded range. The panels show the Fourier-domain amplitude for each of the modes, both exact and from the reconstruction, the fractional errors in amplitude, and the errors in phase. The errors here are relative to each mode, not to the dominant mode. Thus, the amplitude plots importantly allow to visualize the mode hierarchy, giving an idea of the impact of relative errors in the subdominant modes on the full waveform.

The $++$ case, shown in Fig.~\ref{fig:precerrors++}, is the most challenging. As expected from the analysis of Sec.~\ref{subsec:sizecorrPrec}, the presence of strong aligned spins and a large spin of the remnant degrades the separation of timescales, and the reconstruction shows large relative errors, at least in subdominant mode contributions. Applying the higher-order corrections of Sec.~\ref{subsec:executivesummary} does improve the accuracy in the inspiral, but the most difficult modes $m=-1$ and $m=-2$ still show large errors. We also find that for this case higher-order corrections do not reduce the errors in the merger-ringdown region, consistently with the breakdown of the perturbative formalism indicated by Fig.~\ref{fig:fomprec}. Using the convolution treatment improves the main modes $m=2$, $m=1$, but not the most challenging modes $m=-1$ and $m=-2$, which can be seen to depart from the perturbative treatment before the range covered by the convolution and for which the convolution~\eqref{eq:precconvolution} extends to very high frequencies where our trigonometric polynomial representation of the signal is not accurate (see Fig.~\ref{fig:trigopoly}). Note however that, as will be shown below by unfaithfulness computations, these large fractional errors for the subdominant modes do not affect much the full waveform.

The $\perp\perp$ case is shown in Fig.~\ref{fig:precerrorspp}. Errors are increasing to larger $|m-m'|$, higher-order corrections improve the reconstruction but are unsufficient for the merger-ringdown, and the convolution works for all modes on the frequency range where it is applied. The errors at the very high end of the frequency band occur when the overall amplitudes are low, and are therefore unimportant.

In the case $--$, shown in Fig.~\ref{fig:precerrors--}, the mode hierarchy is not respected, as the frame trajectory does not quite follow the picture of simple precession, and the estimates~\eqref{eq:precfomPN} do not apply. The precession velocity is much milder, both in the inspiral and in the merger-ringdown range, thanks to a low remnant spin, and the separation of timescales is better, as shown in Fig.~\ref{fig:fomprec}. This leads to smaller errors, and, apart from some amplitude errors in the merger-ringdown region, we find that in this case even the leading-order treatment gives good results.

In order to illustrate what these errors really mean for the analysis of GW signals, we also compute the unfaithfulness (or mismatch) between various orders of approximation and the exact, numerical result. This unfaithfulness figure, although giving a simplified view of waveform inaccuracies, is commonly used to quantify disagreements between template families and to compare them to numerical relativity waveforms. To define the unfaithfulness, different prescriptions are possible, taking into account or not the detector orientation and optimizing over different sets of parameters.

Here, we will use directly the wave polarizations $h_{+}$, $h_{\times}$, optimizing over time, phase, and polarization angle. For real or complex signals $a$, $b$, one introduces the usual Hermitian noise-weighted scalar product~\cite{CF94}
\be\label{eq:defoverlap}
	\left( a | b \right) = 2 \int \ud f \frac{\tilde{a}_{1}(f) \tilde{b}_{2}(f)^{*}}{S_{n}(f)} \,,
\ee
where the integral extends over all frequencies (in practice, two intervals $[-f_{\rm max}, -f_{\rm min}]$ and $[f_{\rm min}, f_{\rm max}]$). In the above, $S_{n}(f)$ the noise power spectral density, for which we use the aLIGO ZDHP noise curve~\cite{LIGOProspects13}. We use the notation $h \equiv h_{+} - i h_{\times}$ for the complex strain, and write $h[\delta t, \delta \Phi, \delta \psi]$ to represent a signal where we shifted the time of coalescence, the phase of the line of sight to the observer and the polarization angle. The dependencies in time and polarization are given by~\eqref{eq:shifttime} and $h[\delta \psi] = e^{2 i \delta \psi} h$, while the phase $\delta \Phi$ affects differently each waveform mode in~\eqref{eq:defmodes}. In the norm $(h|h)$, only the dependence on $\delta\Phi$ remains. For two signals $h_{1}$ and $h_{2}$, the mismatch between them is then defined as
\be\label{eq:defMM}
	\mathrm{MM} \equiv \mathrm{Min}_{\delta t, \delta \varphi, \delta \psi} \left( 1 - \frac{ \mathrm{Re} \left[ (h_{1}[\delta t, \delta \Phi, \delta \psi] | h_{2}) \right]}{\sqrt{(h_{1}|h_{1})[\delta\Phi]}\sqrt{(h_{2}|h_{2})}} \right)
\ee
where an optimization over a shift in time, phase and polarization is performed. This definition is the same as in~\cite{Blackman+17a} (see App.~D there), except that we added the optimization over the phase. The target unfaithfulness for waveform models depends on the application. For ground-based detectors like advanced LIGO and Virgo, it is often set to 1\% or 3\%, while detectors of the next generations, including LISA, will require a lower unfaithfulness due to their higher signal-to-noise ratios.

\begin{figure*}
  \centering
  \includegraphics[width=.98\linewidth]{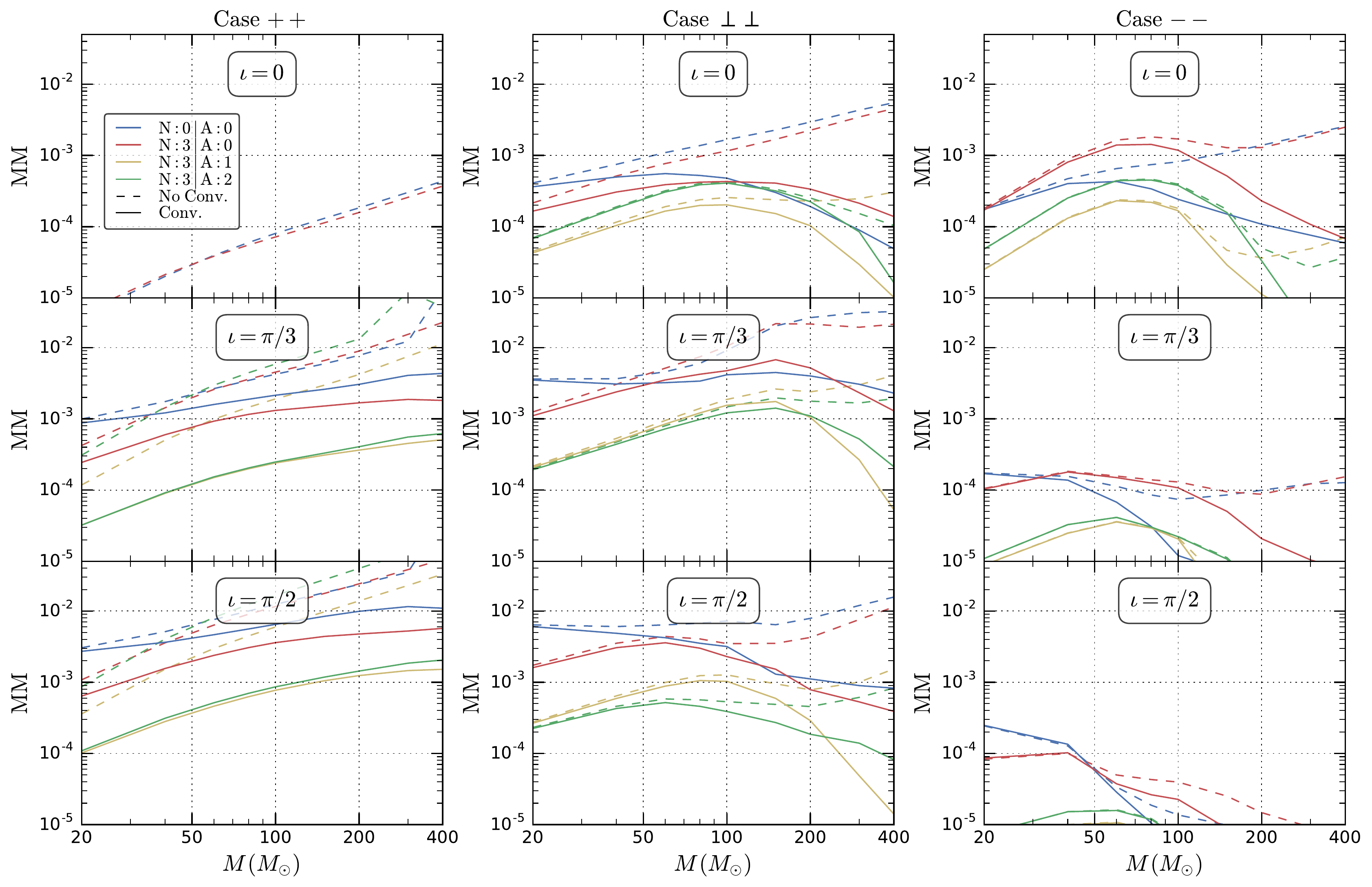}
  \caption{Unfaithfulness at various orders of approximation as compared with an FFT of the full time-domain precessing signal, for the three precessing systems of Table~\ref{tab:precexamples}. The color indicates the order of approximation of the perturbative treatment of Sec.~\ref{subsec:executivesummary}. Dashed lines are obtained applying the treatment of Sec.~\ref{subsec:executivesummary} to the whole frequency band, while full lines are obtained using Sec.~\ref{subsec:trigopoly} for the high frequency range. The computation is done using~\eqref{eq:defMM} with $f_{\rm min} = 20\mathrm{Hz}$, for masses between 20 and 400 $\Msol$, and with the aLIGO ZDHP noise curve~\cite{LIGOProspects13}. The inclination is defined here as the angle between the line of sight and the final $\bm{J}$.}
  \label{fig:precunfaithfulness}
\end{figure*}

In this unfaithfulness computation, we take our illustrative model of Sec.~\ref{subsec:precmodel} as the reference waveform, thus ignoring all limitations of our model for precession and focusing only on the passage from time-domain to Fourier-domain for a given waveform. We show in Fig.~\ref{fig:precunfaithfulness} the unfaithfulness obtained for three inclination angles between the line-of-sight and the direction of the final angular momentum $\bm{J}$, $0$ (face-on), $\pi/3$ and $\pi/2$ (edge-on). Again, we use different approximation levels in~\eqref{eq:summaryNA}, indicated by the color, including or not the treatment of Sec.~\ref{subsec:trigopoly} for the high frequencies as indicated by a continuous or dashed line. The total mass ranges from $20\Msol$ (where the inspiral dominates) to $400\Msol$ (where the merger-ringdown phase dominates).

As discussed in App.~\ref{app:precpreviousapproaches}, the approximation $\{N:0 | A:0 | \text{No Conv.}\}$ corresponds to the leading-order formula~\eqref{eq:transferlocal}, which is close to the treatment of PhenomP~\cite{Hannam+13}, with the difference that this model uses effective Fourier-domain Euler angles instead of a time-domain modulation through merger and ringdown. The order of approximation $\{N:3 | A:0\}$ is equivalent to formalism of Ref.~\cite{KCY14} with a stencil order of $N=3$, except that the latter formalism is based on an SPA representation of the P-frame waveform and limited to the inspiral, while our Fourier-domain approach is an attempt at covering the whole frequency band.

Fig.~\ref{fig:precunfaithfulness} shows that, tor the $++$ case, the face-on case shows good agreement even at leading order, as the effects of precession on the full waveform are suppressed for a zero inclination. By contrast, with $\pi/3$ or $\pi/2$ inclination the mismatch can reach $10^{-2}$ for high masses above $M \simeq 100 \Msol$, even with the perturbative corrections. Using the convolution treatment of the merger and ringdown limits the mismatch, and including both phase (going from blue to red) and amplitude corrections (going from red to yellow and green) makes a clear improvement. Similarly, for the case $\perp\perp$, both the perturbative corrections and the convolution at high frequencies reduce the unfaithfulness. In both cases, one can see that, as expected, the high-frequency convolution becomes unimportant at low masses, where the inspiral dominates. The $--$ case does not show the same dependency on inclination as the other ones, as we measure the inclination angle with respect to the final direction of $\bm{J}$, and this case is close to transitional precession and departs from the picture of simple precession on a cone. As expected from the error estimates and the transfer function errors shown in Figs.~\ref{fig:fomprec} and~\ref{fig:precerrors--}, the milder precession velocity means that the mismatches are small in this case already for the leading-order treatment. We note, however, that in all cases increasing the order of the amplitude corrections from $A:1$ to $A:2$ in~\eqref{eq:summaryNA} does not yield an improvement, and gives even slightly worse errors in some cases. This points to a limitation of our analysis. Higher-order time derivatives are harder to extract numerically, and this might be an indication that, even in our smooth model of Sec.~\ref{subsec:precmodel}, we should limit the formalism to first derivatives.

Overall, we found that, although the configurations listed in Table~\ref{tab:precexamples} are strongly spinning ($\chi = 0.95$) with a quite high mass ratio ($q=4$), which both enhance precession effects, the unfaithfulness due to the leading-order Fourier-domain treatment of the precession remains mainly below 1\%, except at larger masses. This indicates that this treatment (used in particular in PhenomP~\cite{Hannam+13}, see the discussion of App.~\ref{app:precpreviousapproaches}) should be good enough for current LIGO applications. All types of corrections we investigated, either perturbative in phase and amplitude from Sec.~\ref{subsec:executivesummary}, or generated with the convolution treatment of Sec.~\ref{subsec:trigopoly}, consistently lower the unfaithfulness below the 1\% level. One exception are second-order amplitude corrections, which we found to make little or no difference. Using all the tools at our hands, we are able to keep the unfaithfulness of our three example waveforms below $2.10^{-3}$ for all masses and inclinations. We leave for future work the investigation of the full parameter space of spinning binaries, and of higher harmonics $\ell \neq 2$.

\section{Summary and Conclusions}
\label{sec:discussion}

We presented a formalism to process gravitational-wave signals described in the Fourier domain through a time-domain modulation and delay of the type created by the motion and response of a LISA-type detector and by the precessing motion of spinning compact binaries. The natural leading-order of this perturbative formalism simply amounts to building a time-to-frequency correspondence~\eqref{eq:deftf}, generalizing the SPA correspondence~\eqref{eq:deftfSPA} through merger and ringdown.

This formalism is based on the separation of timescales between the modulation and the signal. By a Fourier-domain expansion of the integrand of convolution integrals, corrections beyond the leading order are straightforwardly generated. The most important are quadratic phase corrections, for which the formalism yields a natural extension of the results of~\cite{KCY14} to IMR signals, thanks to the generalization of the radiation reaction timescale with a Fourier-domain expression~\eqref{eq:defTf}. Other corrections include amplitude and delay terms. We provided order-of-magnitude error estimates~\eqref{eq:deffom} allowing a simple diagnosis of the relevance of these various corrections, forecasting where the perturbative formalism will fail due to an insufficient separation of timescales.

We applied our approach to the case of non-precessing comparable-mass binary systems seen by LISA, investigating examples of supermassive black hole mergers as well as inspirals of stellar-mass black holes. The population of stellar-mass binary black hole signals seen in LISA will include chirping systems, merging within years of the LISA observations, as well as slowly-chirping systems hundreds or thousands of years away from merger. We found that our treatment is well applicable for both high-mass and chirping low-mass systems. The leading-order treatment performs quite well, with at most a localized $\sim10\%$ in amplitude or phase error, while the corrections we derived can reduce the errors in the transfer functions to much lower levels if needed. Quadratic phase corrections at the start of the waveform and delay corrections at the high end of the frequency band are the most relevant. On the other hand, the morphology of the slowest-chirping stellar-mass signals is close to quasi-monochromatic galactic binaries, and they cannot be handled by this perturbative formalism. We developed a separate method, using discrete Fourier coefficients, to compute the instrument response in this case.

To explore the application of the formalism to precessing binaries, we built a toy model for the precession of a two-spins system including a post-merger precessing-frame trajectory. For the high spin, moderate mass ratio examples we explored, we found that subdominant precessing mode contributions can be challenging for the formalism and have large fractional errors, although they are suppressed in amplitude in the complete waveform. For the inspiral, the most relevant corrections are the quadratic in phase ones that were already derived in~\cite{KCY14}. Our formalism extends to the merger and ringdown, where our model for the post-merger precession gives a weaker separation of timescales. We developed an alternative method to handle the high-frequency, late-time part of the signal, with a direct convolution of the Fourier-domain signal. We showed that although the leading-order treatment is acceptable for our examples with an unfaithfulness at or below 1\% for typical LIGO masses, including in the inspiral all the corrections we derived, as well as using the convolution treatment for the merger, both help lowering the unfaithfulness down to the $10^{-4}$-$10^{-3}$ range.

Our investigation of the precession provides tools that will be potentially useful for the next generations of Fourier-domain waveform models for IMR precessing systems. Our treatment of the Fourier-domain LISA instrumental response will allow us to leverage modern, fast Fourier-domain codes producing IMR waveforms to accelerate computationally intensive data analysis applications, most notably Bayesian source parameter estimation on synthetic data, that have become urgent with the selection of LISA as a mission by ESA. The LISA response described in Sec.~\ref{sec:LISA} has been implemented in the LISA Data Challenge~\cite{LDC} software suite.

Our results suggest a few straightforward extensions. First, we only considered the leading harmonic of non-spinning waveforms in the LISA response. The same approach is also well-suited to treating higher harmonics and aligned spin signals, but precession will leave a definite imprint on the Fourier-domain signal. One option in this case would be to handle the precession and response modulation together, as in~\cite{Klein+15}. However, we present in App.~\ref{app:precLISA} an argument showing that our formalism should apply straightforwardly to waveforms with simple precession. We leave the general case for future study.

Another natural application of our formalism developed for the LISA response is to third-generation ground-based detectors. For detectors like the Einstein Telescope, signals are sufficiently long for the rotation of the Earth to matter, and their arms are long enough that the long wavelength approximation is not valid at high frequencies and that the response becomes frequency-dependent~\cite{Essick+17}. A preliminary order-of-magnitude analysis shows that the leading-order treatment should be applicable, but more investigation is needed.

Finally, while we restricted ourselves to comparable-mass binary systems, Extreme Mass Ratio Inspirals (EMRI) will also be very important signals for LISA. Those signals generally include not only strong precession but also large eccentricity. An effective frequency-domain LISA response may be possible by applying our treatment separately to each of the numerous individual harmonics produced (see e.g.~\cite{Chua+17}).


\vspace{4.5mm}

\hspace{0.85in}
{\bf Acknowledgments}

\vspace{3.5mm}

We are indebted to Stanislav Babak, Alejandro Boh\'{e}, Alessandra Buonanno, Tito Dal Canton, Scott Field, Philip Graff, Sebastian Khan, Antoine Klein, Serguei Ossokine, Michael P\"{u}rrer for useful discussions and comments. We used the software GWFrames~\cite{Boyle13, Boyle+14} in our study. This work was supported by a NASA Astrophysics Theory Program grant 11-ATP-046.


\appendix

\section{Notation and conventions}
\label{app:notation}

The convention we will be using for the Fourier transform of a signal $h(t)$ and its inverse is
\begin{subequations}
\label{eq:defFT}
\begin{align}
	\tilde{h}(f) &= \mathrm{FT}[h](f) =  \int \ud t \, e^{+2i\pi f t} h(t) \,, \\
	h(t) &= \mathrm{IFT}[\tilde{h}](t) =  \int \ud f \, e^{-2i\pi f t} \tilde{h}(f) \,.
\end{align}
\end{subequations}
Notice that this sign convention is not the most frequently used in the literature. We chose it to ensure that, with the conventions of~\cite{BlanchetLiving}, spin-weighted spherical modes $h_{\ell m}$ with $m>0$ will have support mostly for positive frequencies. One can revert to the more usual convention by taking $f\rightarrow -f$.

The effect of a shift in time of the time-domain signal translates into a linear phase contribution added to the Fourier-domain signal. For $h_{\Delta t}(t) \equiv h(t+\Delta t)$ with $\Delta t$ a constant, in our convention we have
\be\label{eq:shifttime}
	\tilde{h}_{\Delta t} (f) = e^{-2i\pi f \Delta t} \tilde{h}(f) \,.
\ee

A useful representation of the gravitational waveform is given by its decomposition in spin-weighted spherical harmonics. The waveform emitted in the direction $(\Theta, \Phi)$, with its two polarizations $h_{+},h_{\times}$, can be decomposed as a superposition of modes as~\cite{Thorne80}
\be\label{eq:defmodes}
	h_{+} - i h_{\times} = \sum\limits_{\ell \geq 2} \sum\limits_{m=-\ell}^{\ell} {}_{-2}Y_{\ell m}(\Theta,\Phi) h_{\ell m} \,,
\ee
where the ${}_{-2}Y_{\ell m}(\Theta,\Phi)$ are spin-weighted spherical harmonics~\cite{Goldberg+67}. Conversely, the individual polarizations are obtained from the individual modes as
\begin{subequations}
\begin{align}
	h_{+} = \frac{1}{2} \sum\limits_{\ell, m} \left[ {}_{-2}Y_{\ell m}h_{\ell m} + {}_{-2}Y_{\ell m}^{*} h_{\ell m}^{*} \right] \,,\\
	h_{\times} = \frac{i}{2} \sum\limits_{\ell, m} \left[ {}_{-2}Y_{\ell m}h_{\ell m} - {}_{-2}Y_{\ell m}^{*} h_{\ell m}^{*} \right] \,.
\end{align}
\end{subequations}

For a non-precessing system, with a fixed orbital plane, the individual modes have the additional symmetry property
\be\label{eq:symmetryhlminusm}
	h_{\ell, -m} = (-1)^{\ell} h_{\ell m}^{*} \,.
\ee
Since we will work in the Fourier domain, it will be useful to write the contributions of the individual modes to the Fourier transforms of the polarizations as
\begin{subequations}\label{eq:hpcfrommodes}
\begin{align}
	\tilde{h}_{+}(f) &= \frac{1}{2} \sum\limits_{\ell \geq 2} \sum\limits_{m=-\ell}^{\ell} \left[ {}_{-2}Y_{\ell m} \tilde{h}_{\ell m}(f) + {}_{-2}Y_{\ell m}^{*} \tilde{h}_{\ell m}(-f)^{*} \right] \,, \\
	\tilde{h}_{\times}(f) &= \frac{i}{2} \sum\limits_{\ell \geq 2} \sum\limits_{m=-\ell}^{\ell} \left[ {}_{-2}Y_{\ell m} \tilde{h}_{\ell m}(f) - {}_{-2}Y_{\ell m}^{*} \tilde{h}_{\ell m}(-f)^{*} \right] \,,
\end{align}
\end{subequations}
where we used $\widetilde{h_{\ell m}^{*}}(f) = \tilde{h}_{\ell m}(-f)^{*}$. An additional approximation often used in waveform models consists in considering that the Fourier transforms $\tilde{h}_{\ell m}$ have support only on one side of the spectrum, either for positive of for negative frequencies depending on the sign of $m$. This holds in particular within the stationary phase approximation (see Sec.~\ref{subsec:SPA}). With our sign convention~\eqref{eq:defFT}, this approximation reads
\begin{align}\label{eq:zeronegativef}
	\tilde{h}_{\ell m} (f) &\simeq 0 \text{ for } f<0, \; m>0 \nn\,,\\
	\tilde{h}_{\ell m} (f) &\simeq 0 \text{ for } f>0, \; m<0 \,.
\end{align}
When both and~\eqref{eq:zeronegativef} apply, \eqref{eq:hpcfrommodes} becomes simpler. For $f>0$, we have then
\be
	\tilde{h}_{+,\times} (f) = \sum_{\ell \geq 2} \sum_{m = 1}^{\ell} K^{+,\times}_{\ell m} \tilde{h}_{\ell m}(f) \,,
\ee
where we set
\begin{align}
	K^{+}_{\ell m} &\equiv \frac{1}{2} \left( {}_{-2}Y_{\ell m} + (-1)^{\ell} {}_{-2}Y_{\ell, -m}^{*} \right) \,, \nn\\
	K^{\times}_{\ell m} &\equiv \frac{i}{2} \left( {}_{-2}Y_{\ell m} - (-1)^{\ell} {}_{-2}Y_{\ell, -m}^{*} \right) \,.
\end{align}
The range $f<0$ can be obtained readily, since $h_{+},h_{\times}$ are real quantities, from $\tilde{h}_{+,\times} (-f) = \tilde{h}_{+,\times}(f)^{*}$.

Dropping the mode indices $\ell$, $m$, we will decompose a given mode\footnote{Not to be confused with the commonly used notation $h = h_{+} - i h_{\times}$ for the complex strain.} $\tilde{h}_{\ell m} = \tilde{h}$ into a Fourier-domain amplitude $A$ and a phase $\Psi$ according to
\be\label{eq:defAPsi}
	\tilde{h}(f) \equiv A(f) e^{-i\Psi(f)} \,.
\ee

Throughout this paper we will refer to Discrete Fourier Transform (DFT) with the acronyms FFT/IFFT for the Fast Fourier Transform and its inverse. In our convention~\eqref{eq:defFT}, the link between the DFT and the trigonometric polynomial representation of a function goes as follows. For a periodic function $F(x)$ defined on $x\in [x_{0}, x_{0} + \Delta x]$, and represented by $N$ samples $x_{j} = \ov{x} + j \Delta x/N$, $j=0,\dots,N-1$, with $N$ large enough to satisfy, at least approximately, the Nyquist criterion, we can build a trigonometric interpolant $P(x)$ as
\be
	P(x) = \sum\limits_{k=-M}^{+M} c_{k} e^{2i\pi k \frac{x-x_{0}}{\Delta x}} \,,
\ee
that will satisfy the system $P(x_{j}) = F(x_{j})$ for $j=0,\dots, N-1$. Here we set $M=N/2$, assuming N is even. This trigonometric polynomial representation is equivalent to truncating the formal Fourier series, representing the full signal, to a finite order $M$. The coefficients $c_{k}$, in the full series, are defined as
\be\label{eq:defcnApp}
	c_{n}(F) = \frac{1}{\Delta x} \int_{0}^{\Delta x} \ud x \, e^{\frac{2 i \pi n x }{ \Delta x}} F(x) \,.
\ee
In both interpretations, either the truncated approximation of the Fourier series or  the trigonometric interpolation formulation, these coefficients are related to the coefficients of the IFFT. If we set $\omega \equiv e^{2i\pi/N}$ and define
\be\label{eq:ykDFT}
	y_{k} = \frac{1}{N} \sum\limits_{j=0}^{N-1} F(x_{j}) \omega^{jk} \,,
\ee
which is the expression of the IFFT in our sign convention~\eqref{eq:defFT}, the coefficients $c_{k}$ are given by
\begin{align}\label{eq:ckyk}
	c_{k} &= y_{k} \text{ for } k=0,\dots, M-1 \,, \nn\\
	c_{k} &= y_{k+N} \text{ for } k=-M+1,\dots, -1 \,, \nn\\
	c_{M} &= c_{-M} = \frac{y_{M}}{2} \,,
\end{align}
where the condition $c_{M} = c_{-M}$ is an arbitrary condition enforced to match the number of degrees of freedom. In practice, a good representation of the Fourier series of the signal is achieved when the truncation order $M$ is sufficient so that the coefficients $c_{|n|\geq M}$ become negligibly small.


\section{Wigner matrices and precessing frame}
\label{app:wigner}

In this Appendix, we summarize our conventions for the Wigner matrices and give a brief description of the construction of a precessing-frame directly from the waveform.

If the $h_{\ell m}^{\rm I}$ are the spin-weighted spherical harmonics~\eqref{eq:defmodes} of the waveform in a fixed inertial frame (I), and if the $h_{\ell m}^{\rm P}$ are the modes of the waveform in a time-dependent precessing frame (P) constructed from the inertial frame by an active rotation with Euler angles $(\alpha, \beta, \gamma)$ (in the convention $(z,y,z)$), then the modes are related by
\begin{subequations}
\label{eq:wignerrotApp}
\begin{align}
	h_{\ell m}^{\rm I} = \sum\limits_{m=-\ell}^{\ell} \calD^{\ell *}_{mm'} (\alpha,\beta,\gamma) h_{\ell m'}^{\rm P} \,, \\
	h_{\ell m}^{\rm P} = \sum\limits_{m=-\ell}^{\ell} \calD^{\ell }_{m'm} (\alpha,\beta,\gamma) h_{\ell m'}^{\rm I} \,.
\end{align}
\end{subequations}
Here we introduced Wigner $\calD$-matrices
\be\label{eq:defWignerDapp}
	\calD^{\ell}_{mm'} (\alpha, \beta, \gamma) = e^{im \alpha} d^{\ell}_{mm'}(\beta) e^{im' \gamma}\,,
\ee
with the real-valued Wigner $d$-matrix reading
\begin{widetext}
\be\label{eq:defWignerdapp}
	d^{\ell}_{mm'}(\beta) = \sum\limits_{k=k_{\rm min}}^{k_{\rm max}} \frac{(-1)^{k}}{k!} \frac{\sqrt{(l+m)! (l-m)! (l+m')! (l-m')!}}{(l+m-k)! (l-m'-k)! (k-m+m')!} \left( \cos\frac{\beta}{2} \right)^{2\ell+m-m'-2k} \left( \sin\frac{\beta}{2} \right)^{2k-m+m'}\,,
\ee
\end{widetext}
where the boundaries of the sum, $k_{\rm min} = \mathrm{max}(0, m-m')$ and $k_{\rm max} = \mathrm{min}(\ell+m, \ell-m')$, can also be read by enforcing that the arguments of the factorials must be non-negative. Note that our convention differs from the convention of~\cite{ABFO09} by a transposition,
\be
	\calD^{\ell}_{m m'} (\alpha, \beta, \gamma) = D^{\ell \, \mathrm{ABFO}}_{m' m} (\alpha, \beta, \gamma) \,.
\ee
In Sec.~\ref{sec:precession}, we considered only $\ell = 2$ and we restricted ourselves to the contributions of the precessing-frame mode $h_{22}^{\rm P}$. In that case, the relevant explicit expression for the $d$-matrix are
\begin{subequations}
\begin{align}
	d^{2}_{22} (\beta) &= \cos^{4} \frac{\beta}{2}\,, \\
	d^{2}_{12} (\beta) &= 2 \cos^{3} \frac{\beta}{2} \sin \frac{\beta}{2}\,, \\
	d^{2}_{02} (\beta) &= \sqrt{6} \cos^{2} \frac{\beta}{2} \sin^{2} \frac{\beta}{2} \,, \\
	d^{2}_{-12} (\beta) &= 2 \cos \frac{\beta}{2} \sin^{3} \frac{\beta}{2} \,, \\
	d^{2}_{-22} (\beta) &= \sin^{4} \frac{\beta}{2} \,.
\end{align}
\end{subequations}

Different prescriptions have been proposed to construct a precessing frame. Intuitively, the frame follows the plane of the orbit as the spins force the latter to precess, so one could define the precessing frame from the trajectory during the inspiral. This is in fact what is done for SEOBNRv3~\cite{Pan+13}, where a non-precessing waveform is generated in the orbital plane before being rotated back to an inertial frame. Alternatively, the precessing frame can be directly read off the waveform, an approach that is less coordinate-dependent and that can readily be extended through merger where there is no notion of an orbital plane. Proposals in this sense include a maximization of the modes $22$ and $2,-2$~\cite{Schmidt+10}, and the construction of an angular velocity for the waveform~\cite{Boyle13}.

Here, we use the dominant eigenvector prescription of~\cite{OShaughnessy+11}, to which we refer (as well as~\cite{OOS12, Boyle13}) for more details. One defines the matrix
\be
	\langle \bm{L} \bm{L} \rangle_{ab} \equiv \frac{\sum_{\ell, m, m'} h^{*}_{\ell m'} \bra{\ell, m'} L_{a} L_{b} \ket{\ell, m} h_{\ell m} }{ \sum_{\ell, m, m'} h^{*}_{\ell m'} h_{\ell m} } \,,
\ee
where the $\ket{\ell, m}$ are a notation for the spin-weighted spherical harmonics ${}_{-2}Y_{\ell m}$, the indices $a,b$ are ordinary spatial incides, and $L_{a}$ stands for the angular-momentum operator, acting on the bras and kets $\ket{\ell, m}$ in the usual fashion. Defining $L_{\pm} = L_{x} \pm i L_{y}$,
\begin{subequations}
\begin{align}
	L_{+} \ket{\ell, m} &= \begin{cases}\sqrt{\ell (\ell+1) - m (m+1)} \ket{\ell, m+1} \;\text{if} \; m<\ell, \\ \;\; 0 \;\;\text{otherwise} \end{cases} \\
	L_{-} \ket{\ell, m} &= \begin{cases}\sqrt{\ell (\ell+1) - m (m-1)} \ket{\ell, m-1} \; \text{if} \; m>-\ell, \\ \;\; 0 \;\;\text{otherwise} \end{cases} \\
	L_{z} \ket{\ell, m} &= m \ket{\ell, m} \,.
\end{align}
\end{subequations}
Note that in the original proposal~\cite{OShaughnessy+11}, $\psi_{4}$ was used instead of the strain. The $z$-axis of the precessing frame is then chosen to be the dominant eigenvetor of the matrix $\langle \bm{L} \bm{L} \rangle_{ab}$. The construction of the precessing frame is then supplemented by the minimal rotation condition~\cite{Boyle+11}, see~\eqref{eq:gammadot}.

\section{Explicit expression for the stencil coefficients}
\label{app:stencil}

\begin{table}[t]
\begin{ruledtabular}\caption{Stencil coefficients $a_{N,k}^{\epsilon}$ entering the formula~\eqref{eq:stencilfresnel}, given here for $\epsilon=1$. The case $\epsilon=-1$ is obtained by complex conjugation.}\label{tab:stencil}
\begin{tabular}{c|cccccc}
	$N \backslash k$ & $0$ & $1$ & $2$ & $3$ & $4$ & $5$ \\
	\hline
	$0$ & $1$ & - & - & - & - & - \\
	$1$ & $1+i$ & $-i$ & - & - & - & - \\
	$2$ & $\frac{1+5i}{4}$ & $\frac{3-4i}{3}$ & $\frac{-3+i}{12}$ & - & - & - \\
	$3$ & $\frac{17 i-3}{18}$ & $\frac{13-7 i}{8}$ & $\frac{-5-i}{10}$ & $\frac{11 i+15}{360}$ & - & - \\
	$4$ & $\frac{185 i-69}{288}$ & $\frac{209-47 i}{120}$ & $\frac{-41 i-67}{120}$ & $\frac{251 i+147}{2520}$ & $\frac{-29 i-7}{3360}$ & \\
	$5$ & $\frac{1669 i-690}{3600}$ & $\frac{2393-135 i}{1440}$ & $\frac{-645 i-646}{1260}$ & $\frac{3295 i+831}{20160}$ & $\frac{26-345 i}{15120}$ & $\frac{429 i-115}{302400}$ \\
\end{tabular}
\end{ruledtabular}
\end{table}

In this Appendix, we give explicit expressions for the stencil coefficients entering~\eqref{eq:stencilfresnel}. In this work we used only low order stencils with $N\leq 20$, and one can trivially invert of the system~\eqref{eq:stencilsystem}, as this operation has to be done only once. One can also obtain closed-form expressions for these coefficients, thanks to the particular choice of samples at $\pm kT$ that gives to the system~\eqref{eq:stencilsystem} the form of a Vandermonde system. Defining the Vandermonde matrix $V(x_{0},\dots,x_{N})$ as $V_{ij} = (x_{j})^{i}$, setting $x_{j} = j^{2}$, and $b_{p} \equiv (-i\epsilon)^{p}(2p-1)!!$, the linear system~\eqref{eq:stencilsystem} becomes
\be
	b_{p} = \sum\limits_{k=0}^{N} V_{pk} a_{N,k}^{\epsilon} \,.
\ee
The expression of the inverse of a Vandermonde matrix in terms of symmetric polynomials then allows us to write the $a_{N,k}^{\epsilon}$, for every finite order $N$, as:
\begin{align}
	a_{N,k}^{\epsilon} &= \frac{1}{\prod\limits_{\substack{q=0 \\ q\neq k}}^{N} (q^{2}-k^{2})} \sum\limits_{p=0}^{N} (i\epsilon)^{p}(2p-1)!! \nn\\ & \quad \cdot \sum\limits_{\substack{ 0 \leq j_{1} < \dots < j_{N-p} \leq N \\ j_{1}, \dots, j_{N-p} \neq k}} j_{1}^{2}\dots j_{N-p}^{2}
\end{align}
The two cases $\epsilon\pm 1$ correspond simply to a complex conjugation of the coefficients $a_{N,k}^{\epsilon}$. Table~\ref{tab:stencil} gives the resulting complex rational values for these coefficients for $N\leq 5$.

\section{Precession in current Fourier-domain waveform models}
\label{app:precpreviousapproaches}

In this Appendix, we give a more detailed overview of the treatment of precession in existing waveform models that generate signals directly in the Fourier domain. Our objective is not to give an exhaustive account of the various existing models, but rather to highlight how their treatment relates to ours.

To describe previous approaches to the problem of producing precessing waveforms directly in the Fourier domain, avoiding the use of a time-domain generation followed by an FFT, it is useful to introduce schematically four different approximations (dropping the mode indices):
\begin{itemize}
	\item unstable SPA: $\calT(f) \tilde{h}^{\rm P}(f) = \mathrm{SPA}\left[ \calD^{*} h^{\rm P} \right](f)$,
	\item $0^{\text{th}}$-order SUA: $\calT(f) = \calD^{*}(\tfSPA) $,
	\item Fourier-domain $0^{\text{th}}$-order SUA: $\calT(f) = \calD^{*}(\omega = \pi f ) $,
	\item SUA: $\calT(f) = \frac{1}{2} \sum\limits_{k} a_{k} \calD^{*}(\tfSPA \pm k \Tf^{\rm SPA})$.
\end{itemize}

In the first option, one applies directly the SPA, as described in Sec.~\ref{subsec:SPA}, to the product of the modulation and the signal. This is known (see e.g.~\cite{KCY13}) to lead to possible pathologies, since the prefactor $1/\sqrt{\ddot{\phi} + \ddot{\Phi}_{\rm prec}}$ can blow up due the precessing contributions to the phase of the inertial-frame waveform.

The second option corresponds to simply multiplying the Fourier-domain signal by the modulation function evaluated at the time $\tfSPA$ as defined in~\eqref{eq:deftfSPA}. In the SUA formalism of Ref.~\cite{KCY14}, this treatment could be called the 0th order as it amounts to using a stencil reduced to a single point, i.e. using only one term with $a_{0,0} = 1$ in~\eqref{eq:stencilfresnel}. It is equivalent to the leading order of our formalism ($\{N:0 | A:0 | d:0 \}$ in the terminology of Sec.~\ref{subsec:executivesummary}), with the difference that we use a more general definition of $t_{f}$ (see~\eqref{eq:deftf}) that extends through the merger and ringdown.

While keeping implictly the same level of approximation, one can also use frequency-based expressions for the modulation, as in Refs.~\cite{LOS13, Hannam+13}. The precessing-frame evolution is then modelled using post-Newtonian expressions for the Euler angles $(\alpha^{\rm PN}, \beta^{\rm PN}, \gamma^{\rm PN})$ as functions of the orbital frequency $\omega$. Using the SPA-inspired correspondence~\eqref{eq:deftfSPA} between the orbital and Fourier-domain frequency, one then writes
\be\label{eq:precPhenomP}
	\calT^{\ell}_{mm'}(f) = \calD^{\ell *}_{mm'}(\alpha^{\rm PN}, \beta^{\rm PN}, \gamma^{\rm PN}) \left( \omega=\pi f\right) \,.
\ee
When the SPA is valid for the underlying signal $h^{\rm P}_{\ell m}$, the condition $\omega(t) = \pi f$ is equivalent by definition to $t=t^{\rm SPA}_{f}$. In~\cite{LOS13}, only the inspiral phase is modelled, and only single-spin configurations, for which the system undergoes simple precession~\cite{Apostolatos+94, Kidder95}. The PhenomP model~\cite{Hannam+13} maps the two spins to a single effective spin, and uses Euler angles computed in the limit of a small opening angle of the precession cone and at the spin-orbit level (see also~\cite{BBF11, MBBB13}).

In the PhenomP model~\cite{Hannam+13}, the frequency-based transfer function~\eqref{eq:precPhenomP} is used as an effective prescription covering the whole Fourier-domain frequency band, including the merger and ringdown. Note that this procedure amounts to using PN results outside of their range of validity, since in the merger-ringdown phase both the PN perturbative treatment and the SPA approximation break down. In particular, when computed as a function of time from the balance equation between emitted flux and orbital energy $\calF = -dE/dt$, the orbital $\omega^{\rm PN}(t)$ blows up and cannot be extended through merger. In practice, one finds however that the extended frequency-based expressions~\eqref{eq:precPhenomP} are mildly varying when extrapolating to higher frequencies, and this approach has been validated by comparisons to numerical relativity waveforms~\cite{Hannam+13}. Future versions of PhenomP~\cite{PhenomPv3InPrep} will move beyond the single-spin approximation by incorporating analytic solutions for the precession trajectory of double-spin systems~\cite{Chatziioannou+17}.

In the SUA formalism proposed in~\cite{KCY13, KCY14}, one extends the SPA to incorporate the leading-order correction during the inspiral. The transfer function is then computed by evaluating the modulation on a stencil of times centered around the time-of-frequency given by the SPA, arriving at the formula~\eqref{eq:stencilfresnel} with the times $t = \tfSPA$ and $T = \TfSPA$ as defined in~\eqref{eq:deftfSPA} and~\eqref{eq:TfSPA}. As described in Sec.~\ref{sec:formalism}, our formalism reduces to the SUA when keeping only the quadratic phase correction, but the SUA is a priori limited to the inspiral phase of the signal through the definitions of the times $t^{\rm SPA}_{f}$ and $T^{\rm SPA}_{f}$. For inspiral waveforms, Ref.~\cite{KCY14} showed that this treatment improves the accuracy with respect to the 0th-order approximation.

\section{Precessing waveform and the LISA Fourier-domain response}
\label{app:precLISA}

In this Appendix, we present an argument for the applicability of the treatment of Section~\ref{sec:formalism} to a Fourier-domain waveform that already contains precession effects, thus combining our investigations of Secs.~\ref{sec:LISA} and~\ref{sec:precession}. Let us first note that one possible approach would be to incorporate both the precession modulation and the LISA modulation and delay in a single $G(f,t)$ kernel. This is what is done notably in~\cite{Klein+15}, with inspiral-only precessing waveforms (and a low-frequency approximation for the response). In that approach, comparing the results of Secs.~\ref{sec:LISA} and Sec.~\ref{sec:precession}, it appears that the precession is more challenging to model than the LISA response -- thus the accuracy of the final result should depend mainly on the precession treatment. In the following, we take a different route and investigate wether we can directly process Fourier-domain precessing mode contributions of the form $\calT^{\ell}_{m m'}(f) \tilde{h}_{\ell m'}(f)$ through the LISA response. We will make simplifying assumptions, assuming simple precession and keeping to order-of-magnitude estimates only. We leave for future work a more thorough investigation of precessing waveforms for LISA beyond simple precession and for IMR waveforms.

For simplicity, we assume that the precession transfer function has been computed using the leading-order approximation, as including corrections should not change the timescales involved. Thus, we consider the signal
\be\label{eq:htildeasaprecmodecontrib}
	\tilde{h} (f) \equiv \calT^{\ell}_{m m'}(f) \tilde{h}_{\ell m'}(f) = \calD^{\ell *}_{m m'}(\alpha, \beta, \gamma)(\tf) \tilde{h}_{\ell m'}(f)
\ee
to be processed through the LISA response. If we assume simple precession, we can use the results of Sec.~\ref{subsec:sizecorrPrec} for the behaviour of $(\alpha, \beta, \gamma)$ as a function of $t$. Restricting ourselves to the inspiral part, we will also use the SPA expressions~\eqref{eq:tfSPA} and~\eqref{eq:TfSPA} for the times $\tf$ and $\Tf$. According to~\eqref{eq:wignerphasesimpleprec}, the mode contribution~\eqref{eq:htildeasaprecmodecontrib} above acquires a Fourier-domain phase $\Psi \rightarrow \Psi + \Phi_{\rm prec}$ with
\be\label{eq:additionalPhiprec}
	\Phi_{\rm prec} = -(m' \cos \beta(\tf) - m) \alpha(\tf) \,
\ee
with $\beta(t)$ varying on the radiation-reaction timescale and $\alpha(t)$ varying on the precession timescale with $\dot{\alpha} = \Omega_{\rm prec}$. The Fourier-domain amplitude becomes $A\rightarrow A \times d^{\ell}_{mm'} (\beta(\tf))$. If $\beta$ varies on the radiation-reaction timescale, this extra factor can be absorbed in a redefinition of the $a(\tf)$ amplitude factor in~\eqref{eq:ASPA}, so that it won't affect the separation of timescales.

Using the derivatives
\be
	\frac{\ud \tf}{\ud f} = 2\pi \Tf^{2}
\ee
and
\be
	\frac{\ud \Tf}{\ud f} = -2\pi \Tf^{5} \ddot{\omega}(\tf) = - 2\pi \frac{11}{12} \frac{\Tf}{\pi f} \,,
\ee
where we have used $\ddot{\omega} = 11/3 (\dot{\omega})^{2}/\omega$ following~\eqref{eq:omegaphiN}, the extra Fourier-domain phase contribution~\eqref{eq:additionalPhiprec} cause new contributions in~\eqref{eq:deftf} and~\eqref{eq:defTf} as
\begin{subequations}
\begin{align}
	-\frac{1}{2\pi} \frac{\ud \Phi_{\rm prec}}{\ud f} &\sim (m' \cos \beta - m) \Tf^{2} \Omega_{\rm prec} \,, \label{eq:additionaltf}\\
	-\frac{1}{4\pi^{2}} \frac{\ud^{2} \Phi_{\rm prec}}{\ud f^{2}} &\sim (m' \cos \beta - m) \Tf^{2} \left[ \frac{11}{6} \frac{\Omega_{\rm prec}}{\pi f} - \Tf^{2} \dot{\Omega}_{\rm prec} \right] \,, \label{eq:additionalTf}
\end{align}
\end{subequations}
where we ignored $\dot{\beta}$. The new term~\eqref{eq:additionaltf} means that, when processing the signal through the LISA response, we would attribute a different $\tf$ to each mode contribution.

The new term~\eqref{eq:additionalTf} will enter $\epsilon_{\Psi 2}$ as defined in~\eqref{eq:deffom}. Comparing to $\Tf^{2}$, and ignoring the prefactor depending on the mode number (of order a few), we have two terms in the bracket. The first term is $\Omega_{\rm prec} / \omega$, which is a quantity of order 1PN, and should remain well below 1. The second term is $\dot{\Omega}_{\rm prec} / 2\dot{\omega}$. Under the simple precession assumption, \eqref{eq:Omegaprec} gives $\dot{\Omega}_{\rm prec} \sim \dot{\omega}/\omega \Omega_{\rm prec}$, and this term suppressed like the first term. Thus, in this case the contribution to $\epsilon_{\Psi 2}$ should be subdominant, the findings of Sec.~\ref{sec:LISA} for non-spinning waveforms should apply.

In the generic case of double-spin precession, however, the above argument does not apply anymore as $\Omega_{\rm prec}$ varies itself on the precessional timescale and $\Omega_{\rm prec} / \dot{\omega}$ is not small in general. In fact, the cases where $\dot{\Omega}_{\rm prec} \sim \dot{\omega}$, with possible cancellations, are the ``catastrophes'' preventing the application of the SPA to the full time-domain precessing signal~\cite{KCY13}. We leave to future work the question of wether this situation could be challenging for the perturbative formalism, after taking into account the appropriate factors for the error estimates as in Sec.~\ref{subsec:lisafom}.

Note that, if we were instead to incorporate the full precession transfer function $\calT^{\ell}_{mm'}$ as an envelope function multiplying $A(f)$, then applying the criterion~\eqref{eq:deffom} for $\epsilon_{A1}$ would show that, even in the case of simple precession, the perturbative analysis would be challenged at both ends of the mass spectrum, for low-mass and high-mass systems.


\bibliography{references.bib}

\end{document}